\documentclass[12pt,graphicx,subfigure,axodraw,date]{article}
\usepackage{amsfonts}
\usepackage{amssymb}
\usepackage[usenames,dvipsnames]{color}
\usepackage{graphicx}
\usepackage{amsmath}
\usepackage{amsfonts}
\usepackage{amssymb}
\usepackage{cleveref}
\usepackage{float}
\restylefloat{table}
\usepackage{chngpage}
\def\rmuu{\gamma^{\mu}}
\def\rmud{\gamma_{\mu}}
\def\PL{{1-\gamma_5\over 2}}
\def\PR{{1+\gamma_5\over 2}}
\def\sinW2{\sin^2\theta_W}
\def\AEM{\alpha_{EM}}
\def\mul{M_{\tilde{u} L}^2}
\def\mur{M_{\tilde{u} R}^2}
\def\mdl{M_{\tilde{d} L}^2}
\def\mdr{M_{\tilde{d} R}^2}
\def\mz2{M_{z}^2}
\def\c2b{\cos 2\beta}
\def\au{A_u}
\def\ad{A_d}
\def\cob{\cot \beta}
\def\v#1{v_#1}
\def\tb{\tan\beta}
\def\epem{$e^+e^-$}
\def\KK{$K^0$-$\overline{K^0}$}
\def\wi{\omega_i}
\def\xj{\chi_j}
\def\Wmu{W_\mu}
\def\Wnu{W_\nu}
\def\m#1{{\tilde m}_#1}
\def\mH{m_H}
\def\mw#1{{\tilde m}_{\omega #1}}
\def\mx#1{{\tilde m}_{\chi^{0}_#1}}
\def\mc#1{{\tilde m}_{\chi^{+}_#1}}
\def\mwi{{\tilde m}_{\omega i}}
\def\mxi{{\tilde m}_{\chi^{0}_i}}
\def\mci{{\tilde m}_{\chi^{+}_i}}

\def\ch{{\tilde\chi^{+}_1}}
\def\c2{{\tilde\chi^{+}_2}}

\def\tt{{\tilde\theta}}

\def\tp{{\tilde\phi}}

\def\mz{M_z}
\def\sw{\sin\theta_W}
\def\cw{\cos\theta_W}
\def\cb{\cos\beta}
\def\sb{\sin\beta}
\def\rwi{r_{\omega i}}
\def\rxj{r_{\chi j}}
\def\rfp{r_f'}
\def\Kik{K_{ik}}
\def\Fq2{F_{2}(q^2)}
\def\f{\({\cal F}\)}
\def\d1{{\f(\tilde c;\tilde s;\tilde W)+ \f(\tilde c;\tilde \mu;\tilde W)}}
\def\tw{\tan\theta_W}
\def\sec2w{sec^2\theta_W}
\begin{document}
\baselineskip 18pt
\def\today{\ifcase\month\or
 January\or February\or March\or April\or May\or June\or
 July\or August\or September\or October\or November\or December\fi
 \space\number\day, \number\year}
\def\thebibliography#1{\section*{References\markboth
 {References}{References}}\list
 {[\arabic{enumi}]}{\settowidth\labelwidth{[#1]}
 \leftmargin\labelwidth
 \advance\leftmargin\labelsep
 \usecounter{enumi}}
 \def\newblock{\hskip .11em plus .33em minus .07em}
 \sloppy
 \sfcode`\.=1000\relax}
\let\endthebibliography=\endlist
\def\lsim{\ ^<\llap{$_\sim$}\ }
\def\gsim{\ ^>\llap{$_\sim$}\ }
\def\r2{\sqrt 2}
\def\beq{\begin{equation}}
\def\eeq{\end{equation}}
\def\beqn{\begin{eqnarray}}
\def\eeqn{\end{eqnarray}}
\def\rmuu{\gamma^{\mu}}
\def\rmud{\gamma_{\mu}}
\def\PL{{1-\gamma_5\over 2}}
\def\PR{{1+\gamma_5\over 2}}
\def\sinW2{\sin^2\theta_W}
\def\AEM{\alpha_{EM}}
\def\mul{M_{\tilde{u} L}^2}
\def\mur{M_{\tilde{u} R}^2}
\def\mdl{M_{\tilde{d} L}^2}
\def\mdr{M_{\tilde{d} R}^2}
\def\mz2{M_{z}^2}
\def\c2b{\cos 2\beta}
\def\au{A_u}
\def\ad{A_d}
\def\cob{\cot \beta}
\def\v#1{v_#1}
\def\tb{\tan\beta}
\def\epem{$e^+e^-$}
\def\KK{$K^0$-$\bar{K^0}$}
\def\wi{\omega_i}
\def\xj{\chi_j}
\def\Wmu{W_\mu}
\def\Wnu{W_\nu}
\def\m#1{{\tilde m}_#1}
\def\mH{m_H}
\def\mw#1{{\tilde m}_{\omega #1}}
\def\mx#1{{\tilde m}_{\chi^{0}_#1}}
\def\mc#1{{\tilde m}_{\chi^{+}_#1}}
\def\mwi{{\tilde m}_{\omega i}}
\def\mxi{{\tilde m}_{\chi^{0}_i}}
\def\mci{{\tilde m}_{\chi^{+}_i}}
\def\mz{M_z}
\def\sw{\sin\theta_W}
\def\cw{\cos\theta_W}
\def\cb{\cos\beta}
\def\sb{\sin\beta}
\def\rwi{r_{\omega i}}
\def\rxj{r_{\chi j}}
\def\rfp{r_f'}
\def\Kik{K_{ik}}
\def\Fq2{F_{2}(q^2)}
\def\f{\({\cal F}\)}
\def\d1{{\f(\tilde c;\tilde s;\tilde W)+ \f(\tilde c;\tilde \mu;\tilde W)}}
\def\tw{\tan\theta_W}
\def\sec2w{sec^2\theta_W}
\def\ch{{\tilde\chi^{+}_1}}
\def\c2{{\tilde\chi^{+}_2}}

\def\tt{{\tilde\theta}}

\def\tp{{\tilde\phi}}

\def\mz{M_z}
\def\sw{\sin\theta_W}
\def\cw{\cos\theta_W}
\def\cb{\cos\beta}
\def\sb{\sin\beta}
\def\rwi{r_{\omega i}}
\def\rxj{r_{\chi j}}
\def\rfp{r_f'}
\def\Kik{K_{ik}}
\def\Fq2{F_{2}(q^2)}
\def\f{\({\cal F}\)}
\def\d1{{\f(\tilde c;\tilde s;\tilde W)+ \f(\tilde c;\tilde \mu;\tilde W)}}


\def\tw{\tan\theta_W}
\def\sec2w{sec^2\theta_W}
\newcommand{\pn}[1]{{\color{blue}{#1}}}
\newcommand{\pr}[1]{{\color{red}{#1}}}

\begin{titlepage}

\begin{center}
{\large {\bf
The Higgs boson mass constraint and the
 CP even-CP odd Higgs boson mixing in an MSSM extension
 }}\\

\vspace{2cm}
\renewcommand{\thefootnote}
{\fnsymbol{footnote}}
  Tarek Ibrahim$^{a}$\footnote{Email: tibrahim@zewailcity.edu.eg},
   Pran Nath$^{b}$\footnote{Email: nath@neu.edu} and Anas Zorik$^{c}$$\footnote{Email: anas.zorik@alexU.edu.eg}$
\vskip 0.5 true cm
\end{center}

\date{Feb 14, 2015}

\noindent
{$^{a}$University of Science and Technology, Zewail City of Science and Technology,}\\
{ 6th of October City, Giza 12588, Egypt\footnote{Permanent address:  Department of  Physics, Faculty of Science,
University of Alexandria, Alexandria, Egypt}\\
}
{$^{b}$Department of Physics, Northeastern University,
Boston, MA 02115-5000, USA} \\
{$^{c}$Department of Physics, Faculty of Science, Alexandria University, Alexandria, Egypt}
\vskip 1.0 true cm

\centerline{\bf Abstract}
One loop contributions to the CP even-CP odd Higgs boson mixings arising from contributions
due to exchange of a vectorlike  multiplet are computed under the Higgs boson mass constraint.
 The vectorlike multiplet consists of
a fourth generation of quarks and a mirror generation. This sector brings in new CP phases
which can be large consistent with EDM constraints. In this work we compute the
contributions from the exchange of quarks and mirror quarks
$t_{4L}, t_{4R}, T_{L}, T_{R}$, and their scalar partners, the squarks and the mirror squarks.  The effect of their
contributions to the Higgs boson masses and mixings are computed and analyzed.
The possibility of measuring the effects of mixing of CP even and CP odd Higgs in experiment
is discussed. It is shown that the branching ratios of the Higgs bosons into fermion pairs are sensitive to new
physics and specifically to CP phases.\\


\medskip
\end{titlepage}

\medskip

{
\section{Introduction \label{sec1}}

{
One of the important phenomenon in MSSM is the observation that the CP even-CP odd Higgs bosons
can mix in the presence of an explicit CP violation
\cite{Pilaftsis:1998pe,Pilaftsis:1998dd,Pilaftsis:1999qt,Demir:1999hj,Choi:2000wz,Carena:2000yi,Ibrahim:2000qj,Carena:2001fw,Ibrahim:2002zk,Ellis:2004fs,FeynHiggs,Lee:2012wa,Carena:2015uoe}.
Such mixings give rise to effects which are observable at colliders. All of the early analyses, however,
were done in the era before the experimental observation of the light Higgs boson at 125 GeV by 
ATLAS ~\cite{Aad:2012tfa} and by CMS~\cite{Chatrchyan:2012ufa}.
It turns out that the Higgs boson mass constraint is rather stringent and severely limits the parameter space
of supersymmetry models. In this work we consider the effects of including a vectorlike multiplet in an 
MSSM extension. In this case the loop correction to the Higgs boson arises from two contributions: one
from the MSSM sector and the other from the vectorlike multiplet. It is shown that such an inclusion 
leads to significant enhancement of the CP even-CP odd mixing.
The explicit CP violation in the Higgs sector
can  be in conformity with the current limits on the EDM of quarks and leptons
due to either mass suppression~\cite{Nath:1991dn,Kizukuri:1992nj}
 in the sfermion sector  or via the cancellation mechanism~\cite{Ibrahim:1998je,Ibrahim:1997gj,Falk:1998pu,Ibrahim:1998je,Brhlik:1998zn,Ibrahim:1999af}. The neutral Higgs boson mixing is of great import since 
  the observation of such a mixing would be a direct indication of the existence of a new source of CP violation beyond what is observed in the Kaon and the B-meson system (for a review see~\cite{Ibrahim:2007fb}).
}

The outline of the rest of the paper is as follows:
In  section \ref{sec2} we describe the model and define notation.  Inclusion of the vectorlike
generation allowing for mixings between the vectorlike and the regular generations increases
the dimensionality of the quark mass matrices from three to five and increases the dimensionality
of the squark mass squared matrices from six to ten. In section \ref{sec3} the effect of
the vectorlike generation on the induced CP violation in the Higgs sector as a consequence of CP
violation in the matter sector including the vectorlike matter is discussed.  In section \ref{sec4}
a detailed computation of the corrections to the Higgs boson mass matrices is given.
A numerical analysis of the mixing of the CP even-CP odd sector is discussed in section\ref{sec5}.
A discussion of the constraints arising from the EDM of the quarks is also given in this
section. Conclusions are given in section \ref{sec6}. Further details of the  squark  mass
squared matrices including the vectorlike squarks are given in the Appendix.

}

\section{The Model and Notation\label{sec2}}
Here we briefly describe the model and further details are given in the appendix.  The model we consider is an
extension of MSSM with an additional vectorlike multiplet. Like MSSM the vectorlike extension is free of anomalies
and  vectorlike multiplets appear in a variety of settings which include grand unified models,
string and D brane models~\cite{vectorlike,Babu:2008ge,Liu:2009cc,Martin:2009bg}. Several analyses have recently
appeared which utilize vectorlike multiplets
\cite{Ibrahim:2008gg,Ibrahim:2010va,Ibrahim:2010hv,Ibrahim:2011im,Ibrahim:2012ds,Aboubrahim:2013gfa,Aboubrahim:2015zpa,Ibrahim:2015hva,Ibrahim:2014oia}

 Here we focus on the quark sector where the vectorlike multiplet consists of a
 fourth generation of quarks and their mirror quarks.
 Thus the quark sector of the extended MSSM model is given by where

\begin{align}
q_{iL}\equiv
 \left(\begin{matrix} t_{i L}\cr
 ~{b}_{iL}  \end{matrix} \right)  \sim \left(3,2,\frac{1}{6}\right) \ ;  ~~ ~t^c_{iL}\sim \left(3^*,1,-\frac{2}{3}\right)\ ;
 ~~~ b^c_{i L}\sim \left(3^*,1,\frac{1}{3}\right)\ ;
  ~~~i=1,2,3,4.
\label{2}
\end{align}

\begin{align}
Q^c\equiv
 \left(\begin{matrix} B_{ L}^c \cr
 T_L^c\end{matrix}\right)  \sim \left(3^*,2,-\frac{1}{6}\right)\ ;
~~  T_L \sim  \left(3,1,\frac{2}{3}\right)\ ;  ~~   B_L \sim \left(3^*,1,-\frac{1}{3}\right).
\label{3}
\end{align}
The numbers in the braces show  the properties  under $SU(3)_C\times SU(2)_L\times U(1)_Y$
where the first two entries label the representations for $SU(3)_C$ and $SU(2)_L$ and the last one
gives the value of the hypercharge normalized so  that $Q=T_3+Y$.
We allow the mixing of the vectorlike generation with the first three generations. Specifically we will focus
on the mixings of the mirrors in the vectorlike generation with the first three generations. Here we display some relevant features.  In the up quark sector we
choose a basis as follows

\begin{gather}
\bar\xi_R^T= \left(\begin{matrix}\bar t_{ R} & \bar T_R & \bar c_{ R}
&\bar u_{R} &\bar t_{4R} \end{matrix}\right),~~
\xi_L^T= \left(\begin{matrix} t_{ L} &  T_L &  c_{ L}
& u_{ L} &\bar t_{4L}\end{matrix}\right)\,.
\label{basis-xi}
\end{gather}
and we write the mass term  so that

\beq
-{\cal L}^u_m= \bar\xi_R^T (M_u) \xi_L
+\text{h.c.},
\eeq
The superpotential (as shown in the appendix) of the theory  leads to the up-quark  mass matrix $M_u$
which is given by

\beqn
M_u=
 \left(\begin{matrix} y'_1 v_2/\sqrt{2} & h_5 & 0 & 0&0 \cr
 -h_3 & y_2 v_1/\sqrt{2} & -h_3' & -h_3''&-h_6 \cr
0&h_5'&y_3' v_2/\sqrt{2} & 0 &0\cr
0 & h_5'' & 0 & y_4' v_2/\sqrt{2}&0 \cr
0&h_8&0&0&y_5'v_2/\sqrt{2}\end{matrix}\right)\
\eeqn
This mass matrix is not hermitian and a  bi-unitary transformation is needed  to diagonalize it.
Thus one has
\beq
D^{u \dagger}_R (M_u) D^u_L=\text{diag}(m_{u_1},m_{u_2},m_{u_3}, m_{u_4},  m_{u_5} ).
\label{7a}
\eeq
Under the bi-unitary transformations the basis vectors transform so that
\beqn
 \left(\begin{matrix} t_{R}\cr
 T_{ R} \cr
c_{R} \cr
u_{R} \cr
t_{4R}
\end{matrix}\right)=D^{u}_R \left(\begin{matrix} u_{1_R}\cr
 u_{2_R}  \cr
u_{3_R} \cr
u_{4_R}\cr
u_{5_R}
\end{matrix}\right), \  \
\left(\begin{matrix} t_{L}\cr
 T_{ L} \cr
c_{L} \cr
u_{L}\cr
t_{4L}
\end{matrix} \right)=D^{u}_L \left(\begin{matrix} u_{1_L}\cr
 u_{2_L} \cr
u_{3_L} \cr
u_{4_L}\cr
u_{5_L}
\end{matrix}\right) \ .
\label{8}
\eeqn

A similar analysis can be carried out for the down quarks. Here we choose the basis set as
\begin{gather}
\bar\eta_R^T= \left(\begin{matrix}\bar{b}_R & \bar B_R & \bar{s}_R
&\bar{d}_R
&\bar{b}_{4R}
\end{matrix}\right),
~~\eta_L^T= \left(\begin{matrix} {b_ L} &  B_L &  {s_ L}
& {d_ L}
&{b_{4L}}
 \end{matrix}\right)\,.
\label{basis-eta}
\end{gather}

In this basis  the down quark mass terms are given by
\beq
-{\cal L}^d_m=
\bar\eta_R^T(M_{d}) \eta_L
+\text{h.c.},
\eeq
where using the interactions of   $M_d$ has the following form
\beqn
M_d=\left(\begin{matrix} y_1 v_1/\sqrt{2} & h_4 & 0 & 0  & 0\cr
 h_3 & y'_2 v_2/\sqrt{2} & h_3' & h_3'' &h_6\cr
0&h_4'&y_3 v_1/\sqrt{2} & 0&0 \cr
0 & h_4'' & 0 & y_4 v_1/\sqrt{2}&0\cr
0& h_7 & 0 &0 &y_5 v_1/\sqrt{2}
\end{matrix} \right)\ .
\label{7bb}
\eeqn
In general $h_3, h_4, h_5, h_3', h_4',h_5',  h_3'', h_4'',h_5'', h_6, h_7, h_8$ can be complex and we define their phases
so that

\beqn
h_k= |h_k| e^{i\chi_k}, ~~h_k'= |h_k'| e^{i\chi_k'}, ~~~h_k''= |h_k''| e^{i\chi_k''}\
\label{mix}
\eeqn
The squark sector of the model contains a variety of terms including F -type, D-type, soft  as well as
mixings terms involving squarks and mirror squarks. The details of these contributions to squark mass
square matrices are  discussed in the appendix.

{
\section{Computation of correction to the  Higgs boson mass \label{sec3}}

In MSSM the
 Higgs sector at the one loop level is described by the
scalar potential
\beq
V(H_1,H_2)=V_0+\Delta V\nonumber\\
\eeq
In our analysis we use the renormalization group improved
effective potential where
\beqn
V_0=m_1^2 |H_1|^2+m_2^2|H_2|^2 +(m_3^2 H_1.H_2 + H.C.)\nonumber\\
+\frac{(g_2^2+g_1^2)}{8}|H_1|^4+
\frac{(g_2^2+g_1^2)}{8}|H_2|^4
-\frac{g_2^2}{2}|H_1.H_2|^2
+\frac{(g_2^2-g_1^2)}{4}|H_1|^2|H_2|^2
\eeqn
where $m_1^2=m_{H_1}^2+|\mu|^2,~~~~m_2^2=m_{H_2}^2+|\mu|^2,
~~~~m_3^2=|\mu B|$
and $m_{H_{1,2}}$ and $B$ are the soft SUSY breaking parameters,
and $\Delta V$ is the one loop correction to the effective
potential and is given by

\beq
\Delta V=\frac{1}{64\pi^2}
 Str(M^4(H_1,H_2)(log\frac{M^2(H_1,H_2)}{Q^2}-\frac{3}{2}))
\eeq
where $Str=\sum_i C_i (2J_i+1)(-1)^{2J_i}$
where the sum runs over all particles with spin $J_i$
and $C_i(2J_i+1)$ counts the degrees of freedom of the particle i,
and Q is the running scale. In the evaluation
of $\Delta V$ one should include the contributions of all of the
fields that enter in MSSM. This includes the Standard Model fields
and their superpartners, the sfermions, the higgsinos and the
gauginos. The one loop corrections
to the effective potential make significant contributions
to the minimization conditions.

 It is well known that  the presence of  CP violating effect in the one loop effective
potential   induce CP violating phase in the Higgs VEV
through the minimization of the effective potential. One can
parametrize this effect by the CP phase $\theta_H$ where
\beqn
(H_1)= \left(\begin{matrix} H_1^0\cr
 H_1^-  \end{matrix}  \right)
 =
\left(\begin{matrix} \frac{1}{\sqrt 2}(v_1+\phi_1+i\psi_1)\cr
             H_1^-\end{matrix}\right)
    \eeqn
     \beqn
 (H_2)= \left(\begin{matrix}H_2^+\cr
             H_2^0\end{matrix}\right)
=e^{i\theta_H} \left(\begin{matrix}H_2^+ \cr
            \frac{1}{\sqrt 2} (v_2+\phi_2+i\psi_2)\end{matrix}\right)
\eeqn

The non-vanishing of the phase $\theta_H$ can be seen by
looking at the minimization of the effective potential.
For the present case with the inclusion of CP violating effects
  the variations with respect to
the fields $\phi_1, \phi_2, \psi_1,\psi_2$ give the following

\beq
-\frac{1}{v_1}(\frac{\partial \Delta V}{\partial \phi_1})_0=
m_1^2+\frac{g_2^2+g_1^2}{8}(v_1^2-v_2^2)+m_3^2 \tan\beta \cos\theta_H
\eeq

\beq
-\frac{1}{v_2}(\frac{\partial \Delta V}{\partial \phi_2})_0=
m_2^2-\frac{g_2^2+g_1^2}{8}(v_1^2-v_2^2)+m_3^2 cot\beta \cos\theta_H
\eeq

\beq
\frac{1}{v_1}(\frac{\partial \Delta V}{\partial \psi_2})_0=
m_3^2 \sin\theta_H= \frac{1}{v_2}
(\frac{\partial \Delta V}{\partial \psi_1})_0
\eeq
where the subscript 0 means that the quantities are evaluated
at the point $\phi_1=\phi_2=\psi_1=\psi_2=0$.

The masses $M$ to be included in the $\Delta V$  analysis are the masses of three MSSM quark and their squark partners along with the masses of the generations in the vectorlike sector of the theory. In this case the phase $\theta_H$ is determined by
\beqn
m^2_3 \sin \theta_H=\frac{1}{2}\beta_{h_t} |\mu| |A_t| \sin\gamma_t  f_1 ({M^2_{\tilde{u}_1}},{M^2_{\tilde{u}_3}})
+\frac{1}{2}\beta_{h_u} |\mu| |A_u| \sin\gamma_u  f_1 ({M^2_{\tilde{u}_7}},{M^2_{\tilde{u}_8}})\nonumber\\
+\frac{1}{2}\beta_{h_c} |\mu| |A_c| \sin\gamma_c  f_1 ({M^2_{\tilde{u}_5}},{M^2_{\tilde{u}_6}})
+\frac{1}{2}\beta_{h_{4t}} |\mu| |A_{4t}| \sin\gamma_{4t}  f_1 ({M^2_{\tilde{u}_9}},{M^2_{\tilde{u}_{10}}})\nonumber\\
+\frac{1}{2}\beta_{h_T} |\mu| |A_T| \sin\gamma_T  f_1 ({M^2_{\tilde{u}_2}},{M^2_{\tilde{u}_4}})+
\frac{1}{2}\beta_{h_b} |\mu| |A_b| \sin\gamma_b  f_1 ({M^2_{\tilde{d}_1}},{M^2_{\tilde{d}_3}})\nonumber\\
+\frac{1}{2}\beta_{h_d} |\mu| |A_d| \sin\gamma_d  f_1 ({M^2_{\tilde{d}_7}},{M^2_{\tilde{d}_8}})
+\frac{1}{2}\beta_{h_s} |\mu| |A_s| \sin\gamma_s  f_1 ({M^2_{\tilde{d}_5}},{M^2_{\tilde{d}_6}})\nonumber\\
+\frac{1}{2}\beta_{h_{4b}} |\mu| |A_{4b}| \sin\gamma_{4b}  f_1 ({M^2_{\tilde{d}_9}},{M^2_{\tilde{d}_{10}}})
+\frac{1}{2}\beta_{h_B} |\mu| |A_B| \sin\gamma_B  f_1 ({M^2_{\tilde{d}_2}},{M^2_{\tilde{d}_4}})
\eeqn
where
\beqn
f_1(x,y)=-2+\log(\frac{xy}{Q^2})+\frac{y+x}{y-x}\log\frac{y}{x}\nonumber\\
\beta_{h_q}=\frac{3h^2_q}{16\pi^2},~~\gamma_q=\theta_{\mu}+\alpha_{A_q}
\eeqn
To construct the mass squared matrix of the Higgs scalars
we need to compute the quantities
\beq
M_{ab}^2=(\frac{\partial^2 V}{\partial \Phi_a\partial\Phi_b})_0
\eeq
where $\Phi_a$ (a=1-4) are defined by
\beq
\{\Phi_a\}= \{\phi_1,\phi_2, \psi_1,\psi_2\}
\eeq
and as already specified the subscript 0  means that we set
$\phi_1=\phi_2=\psi_1=\psi_2=0$
after the evaluation of the mass matrix.
The tree and loop contributions to $M_{ab}^2$ are given by

\beq
M_{ab}^2= M_{ab}^{2(0)}+ \Delta M_{ab}^2
\eeq
where  $M_{ab}^{2(0)}$ are the contributions at the tree level and
 $\Delta M_{ab}^2$ are the loop contributions where

\beq
\Delta M_{ab}^2=
\frac{1}{32\pi^2}
Str(\frac{\partial M^2}{\partial \Phi_a}\frac{\partial M^2}{\partial\Phi_b}
log\frac{M^2}{Q^2}+M^2 \frac{\partial^2 M^2}{\partial \Phi_a\partial \Phi_b}
log\frac{M^2}{eQ^2})_0
\eeq
 where e=2.718.
Computation of the $4\times 4$ Higgs  mass matrix in the basis
of Eq.(22) gives

\beq
\left(\begin{matrix}M_Z^2c_{\beta}^2+M_A^2s_{\beta}^2+\Delta_{11} &
-(M_Z^2+M_A^2)s_{\beta}c_{\beta}+\Delta_{12} &\Delta_{13}s_{\beta}&\Delta_{13}
 c_{\beta}\cr
-(M_Z^2+M_A^2)s_{\beta}c_{\beta}+\Delta_{12} &
M_Z^2s_{\beta}^2+M_A^2c_{\beta}^2+\Delta_{22} & \Delta_{23} s_{\beta}
&\Delta_{23} c_{\beta}\cr
\Delta{13} s_{\beta} & \Delta_{23} s_{\beta}&(M_A^2+\Delta_{33})s_{\beta}^2 &
(M_A^2+\Delta_{33})s_{\beta}c_{\beta}\cr
\Delta_{13} c_{\beta} &\Delta_{23} c_{\beta} &(M_A^2+\Delta_{33})s_{\beta}c_{\beta} &
(M_A^2+\Delta_{33})c_{\beta}^2\end{matrix}\right)
\eeq
where $(c_{\beta}, s_{\beta})=(\cos\beta, \sin\beta)$.
In the above the explicit Q dependence has been absorbed in $m_A^2$
which is given by
\beqn
m_A^2
=\frac{1}{\sin\beta\cos\beta}
[-m_3^2\cos\theta_H +\frac{1}{2}\beta_{h_t} |\mu| |A_t| \cos\gamma_t  f_1 ({M^2_{\tilde{u}_1}},{M^2_{\tilde{u}_3}})
+\frac{1}{2}\beta_{h_u} |\mu| |A_u| \cos\gamma_u  f_1 ({M^2_{\tilde{u}_7}},{M^2_{\tilde{u}_8}})\nonumber\\
+\frac{1}{2}\beta_{h_c} |\mu| |A_c| \cos\gamma_c  f_1 ({M^2_{\tilde{u}_5}},{M^2_{\tilde{u}_6}})
+\frac{1}{2}\beta_{h_{4t}} |\mu| |A_{4t}| \cos\gamma_{4t}  f_1 ({M^2_{\tilde{u}_9}},{M^2_{\tilde{u}_{10}}})\nonumber\\
+\frac{1}{2}\beta_{h_T} |\mu| |A_T| \cos\gamma_T  f_1 ({M^2_{\tilde{u}_2}},{M^2_{\tilde{u}_4}})+
\frac{1}{2}\beta_{h_b} |\mu| |A_b| \cos\gamma_b  f_1 ({M^2_{\tilde{d}_1}},{M^2_{\tilde{d}_3}})\nonumber\\
+\frac{1}{2}\beta_{h_d} |\mu| |A_d| \cos\gamma_d  f_1 ({M^2_{\tilde{d}_7}},{M^2_{\tilde{d}_8}})
+\frac{1}{2}\beta_{h_s} |\mu| |A_s| \cos\gamma_s  f_1 ({M^2_{\tilde{d}_5}},{M^2_{\tilde{d}_6}})\nonumber\\
+\frac{1}{2}\beta_{h_{4b}} |\mu| |A_{4b}| \cos\gamma_{4b}  f_1 ({M^2_{\tilde{d}_9}},{M^2_{\tilde{d}_{10}}})
+\frac{1}{2}\beta_{h_B} |\mu| |A_B| \cos\gamma_B  f_1 ({M^2_{\tilde{d}_2}},{M^2_{\tilde{d}_4}})]
 \eeqn
The first term in the second
 brace on the right hand side of the above equation is the tree term,
while the rest ten terms are coming from the three generations of MSSM (six terms) and four terms from the vectorlike multiplet.
One may reduce the $4\times 4$ matrix of the Higgs matrix by introducing a new basis $\{\phi_1, \phi_2, \psi_{1D}, \psi_{2D}\}$ where
\beqn
\psi_{1D}=\sin\beta \psi_1 + \cos\beta \psi_2\nonumber\\
\psi_{2D}=-\cos\beta\psi_1 +\sin\beta \psi_2
\eeqn
In this basis the field $\psi_{2D}$ decouples from the other three fields as a Goldstone field with a zero mass eigen value.
The Higgs mass$^2$ matrix of the remaining three fields are given by
\beq
M^2_{Higgs}=
\left(\begin{matrix}M_Z^2c_{\beta}^2+M_A^2s_{\beta}^2+\Delta_{11} &
-(M_Z^2+M_A^2)s_{\beta}c_{\beta}+\Delta_{12} &\Delta_{13}\cr
-(M_Z^2+M_A^2)s_{\beta}c_{\beta}+\Delta_{12} &
M_Z^2s_{\beta}^2+M_A^2c_{\beta}^2+\Delta_{22} & \Delta_{23} \cr
\Delta_{13}  &\Delta_{23} &(M_A^2+\Delta_{33}) \end{matrix}\right)
\eeq

\section{Computation of Corrections $\Delta_{ij}$
 to the Higgs boson mass squared matrix\label{sec4}}
We consider the exchange contribution from the quarks/mirror quarks and from the squarks/mirror squarks in the susy standard model enriched with the vectorlike generation.

\beqn
\Delta V (u,\tilde{u},d,\tilde{d})= \frac{1}{64\pi^2}
\left(
 \sum_{a=1}^{10} 6 M^4_{\tilde{u}_a} (\log\frac{M^2_{\tilde{u}_a}}{Q^2}-\frac{3}{2}) - 12\sum_{q=u,c,t,{t_4},T}  m^4_{q} (\log\frac{m^2_{q}}{Q^2}-\frac{3}{2})\right)\nonumber\\
+ \frac{1}{64\pi^2}
\left(
\sum_{a=1}^{10} 6 M^4_{\tilde{d}_a} (\log\frac{M^2_{\tilde{d}_a}}{Q^2}-\frac{3}{2}) - 12\sum_{q=d,s,b,{b_4},B}  m^4_{q} (\log\frac{m^2_{q}}{Q^2}-\frac{3}{2})\right)
\eeqn
Note that in the supersymmetric limit, quark masses would be equal to the squark masses and the loop corrections vanish.

Using the above loop corrections we can calculate the corrections to the different Higgs mass$^2$ elements as
\beq
\Delta_{ij}=\Delta_{ij\tilde{q}_u}+\Delta_{ij\tilde{q}_d}
\eeq
where
\beqn
\Delta_{ij\tilde{q}_u}= \Delta_{ij\tilde{t}}+\Delta_{ij\tilde{c}}+\Delta_{ij\tilde{u}}+\Delta_{ij\tilde{t}_4}
+\Delta_{ij\tilde{T}}\nonumber\\
\Delta_{ij\tilde{q}_d}= \Delta_{ij\tilde{b}}+\Delta_{ij\tilde{s}}+\Delta_{ij\tilde{d}}+\Delta_{ij\tilde{b}_4}
+\Delta_{ij\tilde{B}}
\eeqn

For the up quarks/squarks we have the contributions
\beqn
\Delta_{11\tilde{q}}= -2\beta_{hq} m^2_q |\mu|^2 \frac{(|A_q|\cos\gamma_q-|\mu|\cot\beta)^2}{({M^2_{\tilde{u}_i}}-{M^2_{\tilde{u}_j}})^2}
 f_2 ({M^2_{\tilde{u}_i}},{M^2_{\tilde{u}_j}}) \nonumber\\
\Delta_{22\tilde{q}}= -2\beta_{hq} m^2_q |A_q|^2 \frac{(|A_q|-|\mu|\cot\beta \cos\gamma_q)^2}{({M^2_{\tilde{u}_i}}-{M^2_{\tilde{u}_j}})^2}
 f_2 ({M^2_{\tilde{u}_i}},{M^2_{\tilde{u}_j}})+\nonumber\\
2\beta_{hq} m^2_q \log(\frac{M^2_{\tilde{u}_i}
M^2_{\tilde{u}_j}
}{m^4_q}) +
4\beta_{hq} m^2_q |A_q| \frac{(|A_q|-|\mu|\cot\beta \cos\gamma_q)}{({M^2_{\tilde{u}_i}}-{M^2_{\tilde{u}_j}})}
\log(\frac{M^2_{\tilde{u}_i}}{M^2_{\tilde{u}_j}})\nonumber\\
\Delta_{12\tilde{q}}=-2\beta_{hq} m^2_q |\mu| \frac{(|A_q|\cos\gamma_q-|\mu|\cot\beta)}{({M^2_{\tilde{u}_i}}-{M^2_{\tilde{u}_j}})}
\log(\frac{M^2_{\tilde{u}_i}}{M^2_{\tilde{u}_j}})+\nonumber\\
2\beta_{hq} m^2_q |\mu||A_q| \frac{(|A_q|\cos\gamma_q-|\mu|\cot\beta)(|A_q|-|\mu|\cot\beta\cos\gamma_q)}{({M^2_{\tilde{u}_i}}-{M^2_{\tilde{u}_j}})^2}
 f_2 ({M^2_{\tilde{u}_i}},{M^2_{\tilde{u}_j}})\nonumber\\
\Delta_{13\tilde{q}}=-2\beta_{hq} m^2_q |\mu|^2 |A_q|\sin\gamma_q \frac{(|\mu|\cot\beta-|A_q|\cos\gamma_q)}{\sin\beta({M^2_{\tilde{u}_i}}-{M^2_{\tilde{u}_j}})^2}  f_2 ({M^2_{\tilde{u}_i}},{M^2_{\tilde{u}_j}}) \nonumber\\
\Delta_{23\tilde{q}}=-2\beta_{hq} m^2_q |\mu| |A_q|^2\sin\gamma_q \frac{(|A_q|-|\mu|\cot\beta\cos\gamma_q)}{\sin\beta({M^2_{\tilde{u}_i}}-{M^2_{\tilde{u}_j}})^2}  f_2 ({M^2_{\tilde{u}_i}},{M^2_{\tilde{u}_j}}) \nonumber\\
+2\beta_{hq}\frac{m^2_q|\mu||A_q|\sin\gamma_q}{\sin\beta({M^2_{\tilde{u}_i}}-{M^2_{\tilde{u}_j}})}\log(\frac{M^2_{\tilde{u}_i}}{M^2_{\tilde{u}_j}})\nonumber\\
\Delta_{33\tilde{q}}=-2\beta_{hq}\frac{m^2_q|\mu|^2|A_q|^2\sin^2\gamma_q}{\sin^2\beta({M^2_{\tilde{u}_i}}-{M^2_{\tilde{u}_j}})^2}f_2 ({M^2_{\tilde{u}_i}},{M^2_{\tilde{u}_j}})
\eeqn

where $(i,j)=(1,3)$ for $q=t$, $(i,j)=(7,8)$ for $q=u$, $(i,j)=(5,6)$ for $q=c$, $(i,j)=(9,10)$ for $q=t_4$ and
\beqn
f_2(x,y)= -2+\frac{y+x}{y-x}\log\frac{y}{x}
\eeqn
For the mirror $q=T$ contribution is given by
\beqn
\Delta_{11\tilde{T}}= -2\beta_{hT} m^2_T |A_T|^2 \frac{(|A_T|-|\mu|\tan\beta \cos\gamma_T)^2}{({M^2_{\tilde{u}_2}}-{M^2_{\tilde{u}_4}})^2}
 f_2 ({M^2_{\tilde{u}_2}},{M^2_{\tilde{u}_4}})+\nonumber\\
2\beta_{hT} m^2_T \log(\frac{M^2_{\tilde{u}_2}
M^2_{\tilde{u}_4}
}{m^4_T}) +
4\beta_{hT} m^2_T |A_T| \frac{(|A_T|-|\mu|\tan\beta \cos\gamma_T)}{({M^2_{\tilde{u}_2}}-{M^2_{\tilde{u}_4}})}
\log(\frac{M^2_{\tilde{u}_2}}{M^2_{\tilde{u}_4}})\nonumber\\
\Delta_{22\tilde{T}}= -2\beta_{hT} m^2_T |\mu|^2 \frac{(|A_T|\cos\gamma_T-|\mu|\tan\beta)^2}{({M^2_{\tilde{u}_2}}-{M^2_{\tilde{u}_4}})^2}
 f_2 ({M^2_{\tilde{u}_2}},{M^2_{\tilde{u}_4}}) \nonumber\\
\Delta_{12\tilde{T}}=-2\beta_{hT} m^2_T |\mu| \frac{(|A_T|\cos\gamma_T-|\mu|\tan\beta)}{({M^2_{\tilde{u}_2}}-{M^2_{\tilde{u}_4}})}
\log(\frac{M^2_{\tilde{u}_2}}{M^2_{\tilde{u}_4}})+\nonumber\\
2\beta_{hT} m^2_T |\mu||A_T| \frac{(|A_T|\cos\gamma_T-|\mu|\tan\beta)(|A_T|-|\mu|\tan\beta\cos\gamma_T)}{({M^2_{\tilde{u}_2}}-{M^2_{\tilde{u}_4}})^2}
 f_2 ({M^2_{\tilde{u}_2}},{M^2_{\tilde{u}_4}})\nonumber\\
\Delta_{13\tilde{T}}=-2\beta_{hT} m^2_T |\mu| |A_T|^2\sin\gamma_T \frac{(|A_T|-|\mu|\tan\beta\cos\gamma_T)}{\cos\beta({M^2_{\tilde{u}_2}}-{M^2_{\tilde{u}_4}})^2}  f_2 ({M^2_{\tilde{u}_2}},{M^2_{\tilde{u}_4}}) \nonumber\\
+2\beta_{hT}\frac{m^2_T|\mu||A_T|\sin\gamma_T}{\cos\beta({M^2_{\tilde{u}_2}}-{M^2_{\tilde{u}_4}})}\log(\frac{M^2_{\tilde{u}_2}}{M^2_{\tilde{u}_4}})\nonumber\\
\Delta_{23\tilde{T}}=-2\beta_{hT} m^2_T |\mu|^2 |A_T|\sin\gamma_T \frac{(|\mu|\tan\beta-|A_T|\cos\gamma_T)}{\cos\beta({M^2_{\tilde{u}_2}}-{M^2_{\tilde{u}_4}})^2}  f_2 ({M^2_{\tilde{u}_2}},{M^2_{\tilde{u}_4}}) \nonumber\\
\Delta_{33\tilde{T}}=-2\beta_{hT}\frac{m^2_T|\mu|^2|A_T|^2\sin^2\gamma_T}{\cos^2\beta({M^2_{\tilde{u}_2}}-{M^2_{\tilde{u}_4}})^2}f_2 ({M^2_{\tilde{u}_2}},{M^2_{\tilde{u}_4}})
\eeqn

For the down quarks/squarks we have the contributions
\beqn
\Delta_{11\tilde{q}}= -2\beta_{hq} m^2_q |A_q|^2 \frac{(|A_q|-|\mu|\tan\beta \cos\gamma_q)^2}{({M^2_{\tilde{d}_i}}-{M^2_{\tilde{d}_j}})^2}
 f_2 ({M^2_{\tilde{d}_i}},{M^2_{\tilde{d}_j}})+\nonumber\\
2\beta_{hq} m^2_q \log(\frac{M^2_{\tilde{d}_i}
M^2_{\tilde{d}_j}
}{m^4_q}) +
4\beta_{hq} m^2_q |A_q| \frac{(|A_q|-|\mu|\tan\beta \cos\gamma_q)}{({M^2_{\tilde{d}_i}}-{M^2_{\tilde{d}_j}})}
\log(\frac{M^2_{\tilde{d}_i}}{M^2_{\tilde{d}_j}})\nonumber\\
\Delta_{22\tilde{q}}= -2\beta_{hq} m^2_q |\mu|^2 \frac{(|A_q|\cos\gamma_q-|\mu|\tan\beta)^2}{({M^2_{\tilde{d}_i}}-{M^2_{\tilde{d}_j}})^2}
 f_2 ({M^2_{\tilde{d}_i}},{M^2_{\tilde{d}_j}}) \nonumber\\
\Delta_{12\tilde{q}}=-2\beta_{hq} m^2_q |\mu| \frac{(|A_q|\cos\gamma_q-|\mu|\tan\beta)}{({M^2_{\tilde{d}_i}}-{M^2_{\tilde{d}_j}})}
\log(\frac{M^2_{\tilde{d}_i}}{M^2_{\tilde{d}_j}})+\nonumber\\
2\beta_{hq} m^2_q |\mu||A_q| \frac{(|A_q|\cos\gamma_q-|\mu|\tan\beta)(|A_q|-|\mu|\tan\beta\cos\gamma_q)}{({M^2_{\tilde{d}_i}}-{M^2_{\tilde{d}_j}})^2}
 f_2 ({M^2_{\tilde{d}_i}},{M^2_{\tilde{d}_j}})\nonumber\\
\Delta_{13\tilde{q}}=-2\beta_{hq} m^2_q |\mu| |A_q|^2\sin\gamma_q \frac{(|A_q|-|\mu|\tan\beta\cos\gamma_q)}{\cos\beta({M^2_{\tilde{d}_i}}-{M^2_{\tilde{d}_j}})^2}  f_2 ({M^2_{\tilde{d}_i}},{M^2_{\tilde{d}_j}}) \nonumber\\
+2\beta_{hq}\frac{m^2_q|\mu||A_q|\sin\gamma_q}{\cos\beta({M^2_{\tilde{d}_i}}-{M^2_{\tilde{d}_j}})}\log(\frac{M^2_{\tilde{d}_i}}{M^2_{\tilde{d}_j}})\nonumber\\
\Delta_{23\tilde{q}}=-2\beta_{hq} m^2_q |\mu|^2 |A_q|\sin\gamma_q \frac{(|\mu|\tan\beta-|A_q|\cos\gamma_q)}{\cos\beta({M^2_{\tilde{d}_i}}-{M^2_{\tilde{d}_j}})^2}  f_2 ({M^2_{\tilde{d}_i}},{M^2_{\tilde{d}_j}}) \nonumber\\
\Delta_{33\tilde{q}}=-2\beta_{hq}\frac{m^2_q|\mu|^2|A_q|^2\sin^2\gamma_q}{\cos^2\beta({M^2_{\tilde{d}_i}}-{M^2_{\tilde{d}_j}})^2}f_2 ({M^2_{\tilde{d}_i}},{M^2_{\tilde{d}_j}})
\eeqn
where $(i,j)=(1,3)$ for $q=b$, $(i,j)=(7,8)$ for $q=d$, $(i,j)=(5,6)$ for $q=s$ and  $(i,j)=(9,10)$ for $q=b_4$.

Finally the contribution of the mirror $B$ is given by
\beqn
\Delta_{11\tilde{B}}= -2\beta_{hB} m^2_B |\mu|^2 \frac{(|A_B|\cos\gamma_B-|\mu|\cot\beta)^2}{({M^2_{\tilde{d}_2}}-{M^2_{\tilde{d}_4}})^2}
 f_2 ({M^2_{\tilde{d}_2}},{M^2_{\tilde{d}_4}}) \nonumber\\
\Delta_{22\tilde{B}}= -2\beta_{hB} m^2_B |A_B|^2 \frac{(|A_B|-|\mu|\cot\beta \cos\gamma_q)^2}{({M^2_{\tilde{d}_2}}-{M^2_{\tilde{d}_4}})^2}
 f_2 ({M^2_{\tilde{d}_2}},{M^2_{\tilde{d}_4}})+\nonumber\\
2\beta_{hB} m^2_B \log(\frac{M^2_{\tilde{d}_2}
M^2_{\tilde{d}_4}
}{m^4_B}) +
4\beta_{hB} m^2_B |A_B| \frac{(|A_B|-|\mu|\cot\beta \cos\gamma_B)}{({M^2_{\tilde{d}_2}}-{M^2_{\tilde{d}_4}})}
\log(\frac{M^2_{\tilde{d}_2}}{M^2_{\tilde{d}_4}})\nonumber\\
\Delta_{12\tilde{B}}=-2\beta_{hB} m^2_B |\mu| \frac{(|A_B|\cos\gamma_B-|\mu|\cot\beta)}{({M^2_{\tilde{d}_2}}-{M^2_{\tilde{d}_4}})}
\log(\frac{M^2_{\tilde{d}_2}}{M^2_{\tilde{d}_4}})+\nonumber\\
2\beta_{hB} m^2_B |\mu||A_B| \frac{(|A_B|\cos\gamma_B-|\mu|\cot\beta)(|A_B|-|\mu|\cot\beta\cos\gamma_B)}{({M^2_{\tilde{d}_2}}-{M^2_{\tilde{d}_4}})^2}
 f_2 ({M^2_{\tilde{d}_2}},{M^2_{\tilde{d}_4}})\nonumber\\
\Delta_{13\tilde{B}}=-2\beta_{hB} m^2_B |\mu|^2 |A_B|\sin\gamma_B \frac{(|\mu|\cot\beta-|A_q|\cos\gamma_B)}{\sin\beta({M^2_{\tilde{d}_2}}-{M^2_{\tilde{d}_4}})^2}  f_2 ({M^2_{\tilde{d}_2}},{M^2_{\tilde{d}_4}}) \nonumber\\
\Delta_{23\tilde{B}}=-2\beta_{hB} m^2_B |\mu| |A_B|^2\sin\gamma_B \frac{(|A_B|-|\mu|\cot\beta\cos\gamma_B)}{\sin\beta({M^2_{\tilde{d}_2}}-{M^2_{\tilde{d}_4}})^2}  f_2 ({M^2_{\tilde{d}_2}},{M^2_{\tilde{d}_4}}) \nonumber\\
+2\beta_{hB}\frac{m^2_B|\mu||A_B|\sin\gamma_B}{\sin\beta({M^2_{\tilde{d}_2}}-{M^2_{\tilde{d}_4}})}\log(\frac{M^2_{\tilde{d}_2}}{M^2_{\tilde{d}_4}})\nonumber\\
\Delta_{33\tilde{B}}=-2\beta_{hB}\frac{m^2_B|\mu|^2|A_B|^2\sin^2\gamma_B}{\sin^2\beta({M^2_{\tilde{d}_2}}-{M^2_{\tilde{d}_4}})^2}f_2 ({M^2_{\tilde{d}_2}},{M^2_{\tilde{d}_4}})
\eeqn
The Yukawa couplings and quark masses in the $\Delta_{ij}$ elements are defined as follows
\beqn
h_{t_4}=y'_5,~h_t=y'_1,~h_c=y'_3,~h_u=y'_4,h_T=y_2\nonumber\\
h_{b_4}=y_5,~h_b=y_1,~h_s=y_3,~h_d=y_4,h_B=y'_2\nonumber\\
m^2_T=\frac{v^2_1 |y_2|^2}{2},~m^2_{t_4}=\frac{v^2_2 |y'_5|^2}{2},~m^2_u=\frac{v^2_2 |y'_4|^2}{2}\nonumber\\
m^2_c=\frac{v^2_2 |y'_3|^2}{2},~m^2_t=\frac{v^2_2 |y'_1|^2}{2},~m^2_B=\frac{v^2_2 |y'_2|^2}{2}\nonumber\\
m^2_{b_4}=\frac{v^2_1 |y_5|^2}{2},~m^2_d=\frac{v^2_1 |y_4|^2}{2},~m^2_s=\frac{v^2_1 |y_3|^2}{2},~m^2_b=\frac{v^2_1 |y_1|^2}{2}
\eeqn
The mass eigen values of the squark mass$^2$ matrices $M^2_{\tilde{q}_i}$ are defined in the appendix.

\section{Numerical Analysis\label{sec5}}
We present now a numerical analysis of the CP even-CP odd mixings of the Higgs bosons. The mixings arise from
the Higgs boson mass squared matrix which as discussed above will be $3\times 3$.
In the preceding section this mass squared matrix has been computed in the basis $\phi_1, \phi_2, \psi_{1D}$ as
explained in the text of the previous section. The Higgs mass squared matrix computed in section \ref{sec4} is a
 real symmetric $3\times3$ matrix and can be diagonalized by an orthogonal transformations so that

\beq
D M^2 D^T = diag(M^2_{H1}, M^2_{H2},M^2_{H3})
\eeq
Here  the $H_1$ is the lightest field and the remaining two fields $H_2, H_3$
are typically significantly heavier than $H_1$.
We can investigate the CP structure of the two heavy fields through the estimate of the eigen vectors of the Higgs mass$^2$ matrix.
\beqn
H_2 = D_{21} \phi_1 +D_{22} \phi_2 + D_{23} \psi_{1D}\nonumber\\
H_3 = D_{31} \phi_1 +D_{32} \phi_2 + D_{33} \psi_{1D}
\label{mixings}
\eeqn
The percentage of CP odd part of $H_2$ is defined to be $|D_{23}|^2 \times 100$ and its CP even part is defined to be
 $(|D_{22}|^2 +|D_{21}|^2) \times 100$. The same definitions  apply to the other neutral  heavy Higgs $H_3$.
{The  CP even-CP odd Higgs mixing depends directly on CP phases. On the other hand CP phases also
generate EDM for the quarks and for the neutron. The current experimental limit on the EDM of the neutron is~\cite{Baker:2006ts}
$|d_n|< 2.9\times 10^{-26} e{\rm cm} (90\% \rm{CL})$.
We note that the combinations of the phases that enter in the EDM of the quarks are not the same
that enter in the CP even-CP odd Higgs mixings. Thus significant CP even-CP odd Higgs mixings can occur
while at the same time the  EDM constraint
can be satisfied.}\\

We present now a numerical analysis of the $CP$ structure of the two heavy physical fields $H_2$ and $H_3$. We order the eigen values so that in the limit of no mixing between the $CP$ even and the $CP$ odd states one has
($M_{H1}$, $M_{H2}$, $M_{H3}$) tends to ($m_h$, $m_H$, $m_A$) where $m_h$ is the mass of the light $CP$ even state, $m_H$ the mass of the heavy $CP$ even and $m_A$ is the mass of the $CP$ odd Higgs in MSSM when all $CP$ phases are set to zero.
In the squark sector we assume $m^{u^2}_0=M^2_{\tilde T}=M^2_{\tilde t_1}=M^2_{\tilde t_2}=M^2_{\tilde t_3}$ and $m^{d^2}_0=M^2_{\tilde 1 L}=M^2_{\tilde B}=M^2_{\tilde b_1}=M^2_{\tilde Q}=M^2_{\tilde 2 L}=M^2_{\tilde b_2}=M^2_{\tilde 3 L}=M^2_{\tilde b_3}$. {{and}} $m^u_0=m^d_0=m_0$. Additionally the trilinear couplings are chosen {so that}: $A^u_0=A_t=A_T=A_c=A_u=A_{4t}$ and $A^d_0=A_b=A_B=A_s=A_d=A_{4b}$.\\

{
 One expects the CP even-CP odd mixing to be a very sensitive function of the CP phases.
 We study this sensitivity for the case of MSSM first.
In Fig.~\ref{fig1}
we exhibit this dependence as a function of $\theta_{\mu}$.
The left panel exhibits the CP even and CP odd components of the Higgs boson $H_2$
while the right panel exhibits the CP even and CP odd components of the Higgs boson $H_3$.
In figure \ref{fig2} we exhibit this dependence for the case of $\alpha_{A_0}$ where ($\alpha_{A_0}=\alpha_{A_0^u}=\alpha_{A_0^d}$).}
Next  let us suppose that not all the loop correction to the light Higgs boson mass arises from
 the MSSM sector. Rather there are two components to this correction, one that arises from
 MSSM while the other arises from exchange of a vectorlike quark multiplet. In this case
 the vectorlike multiplet brings in new sources of CP violation which can contribute to the
 CP even-CP odd Higgs mixings.  We give an illustration of this in table \ref{table:1}
 and table \ref{table:2}. Table \ref{table:1} gives the contribution to the Higgs mass from
 the MSSM sector alone which is a few GeV smaller than the desired value. The deficit is
 made up by exchange of a vectorlike multiplet.
  The contributions of the MSSM and of
 the vectorlike multiplet together are exhibited in table {2} which gives the
 Higgs mass consistent with the experimental value within a small error corridor
 of $\pm 2$ GeV.
{{Comparison of tables 1 and 2, especially the last three lines, shows that the CP even-CP odd mixing for the case of table 2 is very different from the case of table 1. Thus for $H_2(H_3)$, the CP odd (even) component is as much as 10\% for the case when the vector multiplet is included  where without the inclusion of the vector multiplet
the even-odd mixing was vanishing. Thus inclusion of the 
vectorlike multiplet in the analysis has a strong effect on the CP even-CP odd mixing.
   }}

\begin{table}[H]
\begin{center}
\begin{tabular}{l  c  c  c  c  c  c  c  c  c }
\hline\hline
    & $M_{H1}$ & $CP$even& $CP$odd& $M_{H2}$ & $CP$even& $CP$odd& $ M_{H3}$  & $CP$even  & $CP$odd  \\
\cline{2-10}
 (1)& $118.02$ & $99.99$ & $0.01$ & $501.57$ & $93.96$ & $6.04$ & $499.56$ & $6.05$ & $93.95$ \\
 (2)& $116.76$ & $ 100 $ & $0.00$ & $500.44$ & $95.46$ & $4.54$ & $499.88$ & $4.54$ & $95.46$ \\
 (3)& $117.21$ & $ 100 $ & $0.00$ & $500.22$ & $97.57$ & $2.43$ & $499.95$ & $2.43$ & $97.57$ \\
 (4)& $117.36$ & $ 100 $ & $0.00$ & $500.14$ & $ 100 $ & $0.00$ & $ 500 $  & $0.00$ & $ 100 $ \\
 (5)& $119.53$ & $ 100 $ & $0.00$ & $500.10$ & $ 100 $ & $0.00$ & $ 500 $  & $0.00$ & $ 100 $ \\
 (6)& $119.82$ & $ 100 $ & $0.00$ & $500.07$ & $ 100 $ & $0.00$ & $ 500 $  & $0.00$ & $ 100 $ \\
\hline\hline
\end{tabular}
\caption{An exhibition of the $CP$ structure of the $H_1$, $H_2$ and $H_3$ fields for the case without the contributions of the vectorlike generation. The analysis is for six benchmark points (1), (2), (3), (4), (5) and (6).
 Benchmark (1): $\tan\beta=5$, $m_0=m^u_0=m^d_0=2300$, $|\mu|=800$, $|A^u_0|=8500$, $|A^d_0|=9500$, $\theta_{\mu}=0.9$,
  $\alpha_{A^u_0}=0.5$, $\alpha_{A^d_0}=1.5$.
 Benchmark (2): $\tan\beta=10$, $m_0=m^u_0=m^d_0=2000$, $|\mu|=380$, $|A^u_0|=7400$, $|A^d_0|=8300$, $\theta_{\mu}=0.4$, $\alpha_{A^u_0}=1.2$, $\alpha_{A^d_0}=1.3$.
 Benchmark (3): $\tan\beta=15$, $m_0=m^u_0=m^d_0=2300$, $|\mu|=300$, $|A^u_0|=8600$, $|A^d_0|=8000$, $\theta_{\mu}=0.9$, $\alpha_{A^u_0}=3.5$, $\alpha_{A^d_0}=2.2$.
 Benchmark (4): $\tan\beta=20$, $m_0=m^u_0=m^d_0=2100$, $|\mu|=200$, $|A^u_0|=7800$, $|A^d_0|=7000$, $\theta_{\mu}=1.7$, $\alpha_{A^u_0}=1.4$, $\alpha_{A^d_0}=1$.
 Benchmark (5): $\tan\beta=25$, $m_0=m^u_0=m^d_0=2500$, $|\mu|=260$, $|A^u_0|=9350$, $|A^d_0|=3500$, $\theta_{\mu}=2.2$, $\alpha_{A^u_0}=1$, $\alpha_{A^d_0}=3.2$.
 Benchmark (6): $\tan\beta=30$, $m_0=m^u_0=m^d_0=2400$, $|\mu|=200$, $|A^u_0|=8950$, $|A^d_0|=1000$, $\theta_{\mu}=2.37$, $\alpha_{A^u_0}=0.9$, $\alpha_{A^d_0}=2.8$.
  The common parameters are: $m_A=500$, $|h_3|=1.58$, $|h'_3|=6.34\times10^{-2}$, $|h''_3|=1.97\times10^{-2}$, $|h_4|=4.42$, $|h'_4|=5.07$, $|h''_4|=12.87$, $|h_5|=6.6$, $|h'_5|=2.67$, $|h''_5|=1.86\times10^{-1}$, $|h_6|=1000$, $|h_7|=1000$, $|h_8|=1000$, $\chi_3=2\times10^{-2}$, $\chi'_3=1\times10^{-3}$, $\chi''_3=4\times10^{-3}$, $\chi_4=7\times10^{-3}$, $\chi'_4=\chi''_4=1\times10^{-3}$, $\chi_5=9\times10^{-3}$, $\chi'_5=5\times10^{-3}$, $\chi''_5=2\times10^{-3}$, $\chi_6=\chi_7=\chi_8=5\times10^{-3}$. All masses are in GeV and all phases in rad.}
\label{table:1}
\end{center}
\end{table}

\begin{table}[H]
\begin{center}
\begin{tabular}{l  c  c  c  c  c  c  c  c  c }
\hline\hline
    & $M_{H1}$ & $CP$even& $CP$odd& $M_{H2}$ & $CP$even& $CP$odd& $ M_{H3}$  & $CP$even  & $CP$odd  \\
\cline{2-10}
 (1)& $124.08$ & $99.98$ & $0.02$ & $504.80$ & $91.68$ & $8.32$ & $497.46$ & $8.33$ & $91.67$ \\
 (2)& $124.54$ & $99.98$ & $0.02$ & $523.51$ & $90.71$ & $9.29$ & $486.87$ & $9.30$ & $90.70$ \\
 (3)& $124.17$ & $99.99$ & $0.01$ & $533.10$ & $92.69$ & $7.31$ & $486.07$ & $7.32$ & $92.68$ \\
 (4)& $124.06$ & $ 100 $ & $0.00$ & $539.99$ & $94.79$ & $5.21$ & $494.94$ & $5.21$ & $94.79$ \\
 (5)& $123.99$ & $ 100 $ & $0.00$ & $514.14$ & $89.28$ &$10.72$ & $492.61$ &$10.72$ & $89.28$ \\
 (6)& $124.71$ & $ 100 $ & $0.00$ & $539.94$ & $94.35$ & $5.65$ & $495.41$ & $5.66$ & $94.34$ \\
\hline\hline
\end{tabular}
\caption{An exhibition of the $CP$ structure of the $H_1$, $H_2$ and $H_3$ fields for the case with the contributions of the vectorlike generation. The analysis is for six benchmark points corresponding to the parameter space of table~\ref{table:1}. The Yukawa couplings are:
 (1): $h_T=1.5$, $h_B=0.4$, $h_{t_4}=0.6$, $h_{b_4}=1.5$;
 (2): $h_T=2.9$, $h_B=0.4$, $h_{t_4}=0.5$, $h_{b_4}=2.9$;
 (3): $h_T=4.3$, $h_B=0.4$, $h_{t_4}=0.5$, $h_{b_4}=4.3$;
 (4): $h_T=5.8$, $h_B=0.4$, $h_{t_4}=0.5$, $h_{b_4}=5.8$;
(5): $h_T=7.2$, $h_B=0.4$, $h_{t_4}=0.5$, $h_{b_4}=7.2$;
(6): $h_T=8.6$, $h_B=0.4$, $h_{t_4}=0.5$, $h_{b_4}=8.6$.
 Masses for the vectorlike quarks are gotten by diagonalization of the matrices of Eqs.~(5) and
(10) and are given as follows: mirror up quark  $m_{t'}$ = 980.14,
  mirror down quark  mass  $m_{b'}$ = 1062.63,  fourth generation up quark mass $m^{\rm up}_{4}$ = 1025.14,
  fourth generation down quark mass $m^{\rm down}_{4}$ = 937.64. All masses are in GeV.
 The inputs from the MSSM sector are listed in table \ref{table:4}.
  }
\label{table:2}
\end{center}
\end{table}

The MSSM sector inputs of the six benchmark points in table \ref{table:1} and table \ref{table:2}.

\begin{table}[H]
\begin{center}
\begin{tabular}{l c  c  c  c  c  c  c  c  c}
 \hline\hline
 (case)& $\tan\beta$ & $|\mu|$ & $\theta_{\mu}$ & $m_0$ & $|A^u_0|$ & $|A^d_0|$ & $\alpha_{A^u_0}$ & $\alpha_{A^d_0}$ \\
 \hline
 (1)   &     5       &   800   &       0.9      &  2300 &    8500   &   9500    &       0.5        &      1.5         \\
 (2)   &     10      &   380   &       0.4      &  2000 &    7400   &   8300    &       1.2        &      1.3         \\
 (3)   &     15      &   300   &       0.9      &  2300 &    8600   &   8000    &       3.5        &      2.2         \\
 (4)   &     20      &   200   &       1.7      &  2100 &    7800   &   7000    &       1.4        &      1           \\
 (5)   &     25      &   260   &       2.2      &  2500 &    9350   &   3500    &       1          &      3.2         \\
 (6)   &     30      &   200   &       2.37     &  2400 &    8950   &   1000    &       0.9        &      2.8         \\
 \hline\hline
\end{tabular}
\caption{The inputs of the six benchmark points of table~\ref{table:1}.}
\label{table:4}
\end{center}
\end{table}

We give now a more detailed analysis of CP even-CP odd mixing  for the case with inclusion of 
 the vectorlike multiplet. Specifically we discuss 
three illustrative benchmark points of table \ref{table:2}.
In figure \ref{fig6} we exhibit this dependence as a function of $\theta_{\mu}$.
The left panel exhibits the CP even and CP odd components of the Higgs boson $H_2$
while the right panel exhibits the CP even and CP odd components of the Higgs boson $H_3$.
One finds that the mixing can be very substantial for a significant parameter range of $\theta_\mu$.
A similar analysis is presented in figure \ref{fig7} for the case of $\alpha_{A_0^u}$ dependence.
The $\alpha_{A_0^d}$ dependence is very similar to that for $\alpha_{A_0^u}$
 and is not exhibited.
Figure \ref{fig10} exhibits the
dependence of the CP even-CP mixing for $H_2$ and $H_3$ as a function of $m_0$.
In Fig. \ref{fig11} we give an analysis of the sensitivity of the masses for the boson $H_1, H_2, H_3$
as a function of $\theta_\mu$ and a similar analysis as a function of $\alpha_{A_0^u}$ is given
in Fig. \ref{fig12}. One finds only a mild sensitivity of the light Higgs $H_1$ mass but much larger
sensitivity of the masses of $H_2$ and $H_3$ on the CP phases. This is consistent with the significant
CP even -CP odd mixing among the two heavy neutral Higgs.


\begin{figure}[H]
\begin{center}
{\rotatebox{0}{\resizebox*{6.0cm}{!}{\includegraphics{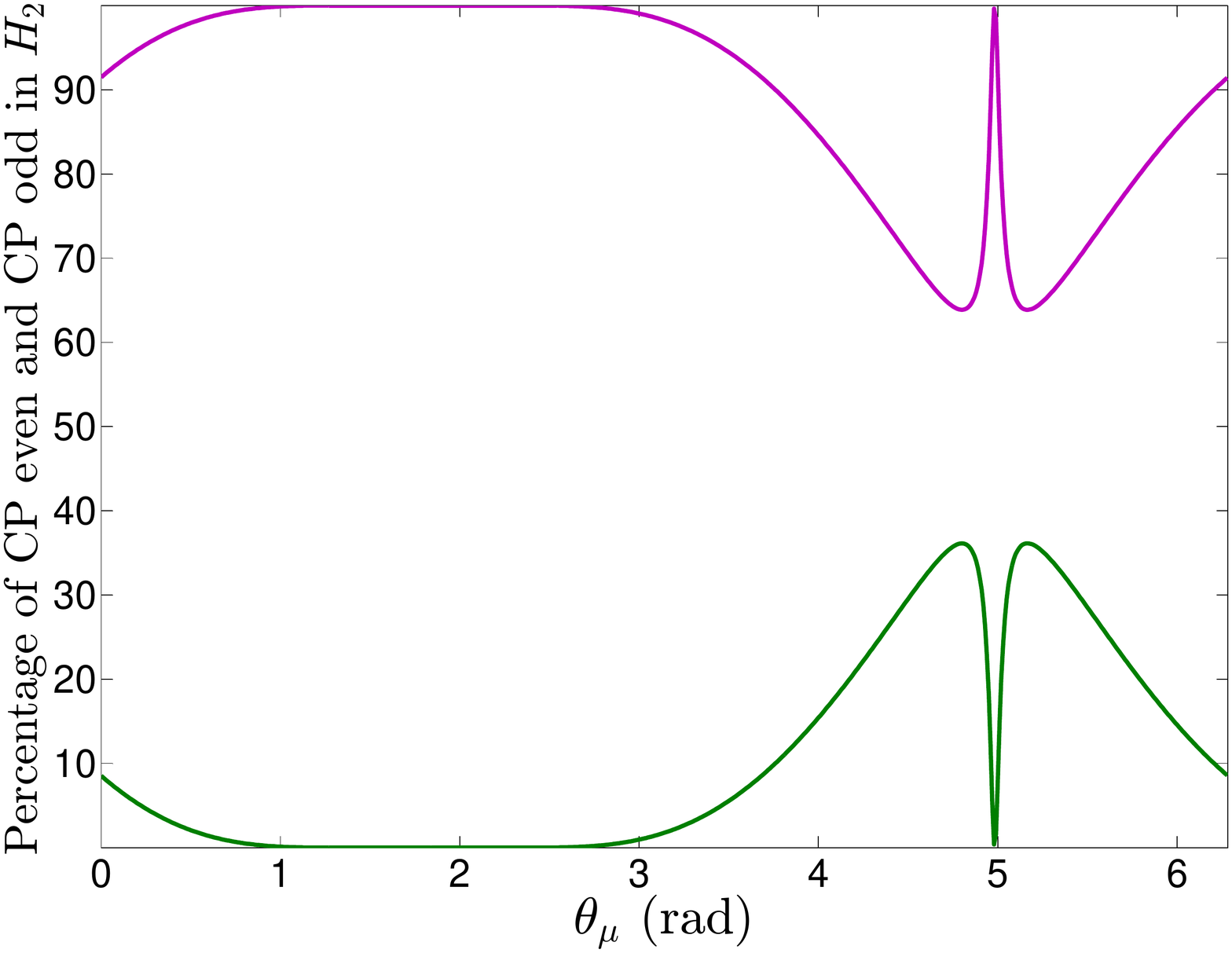}}\hglue5mm}}
{\rotatebox{0}{\resizebox*{6.0cm}{!}{\includegraphics{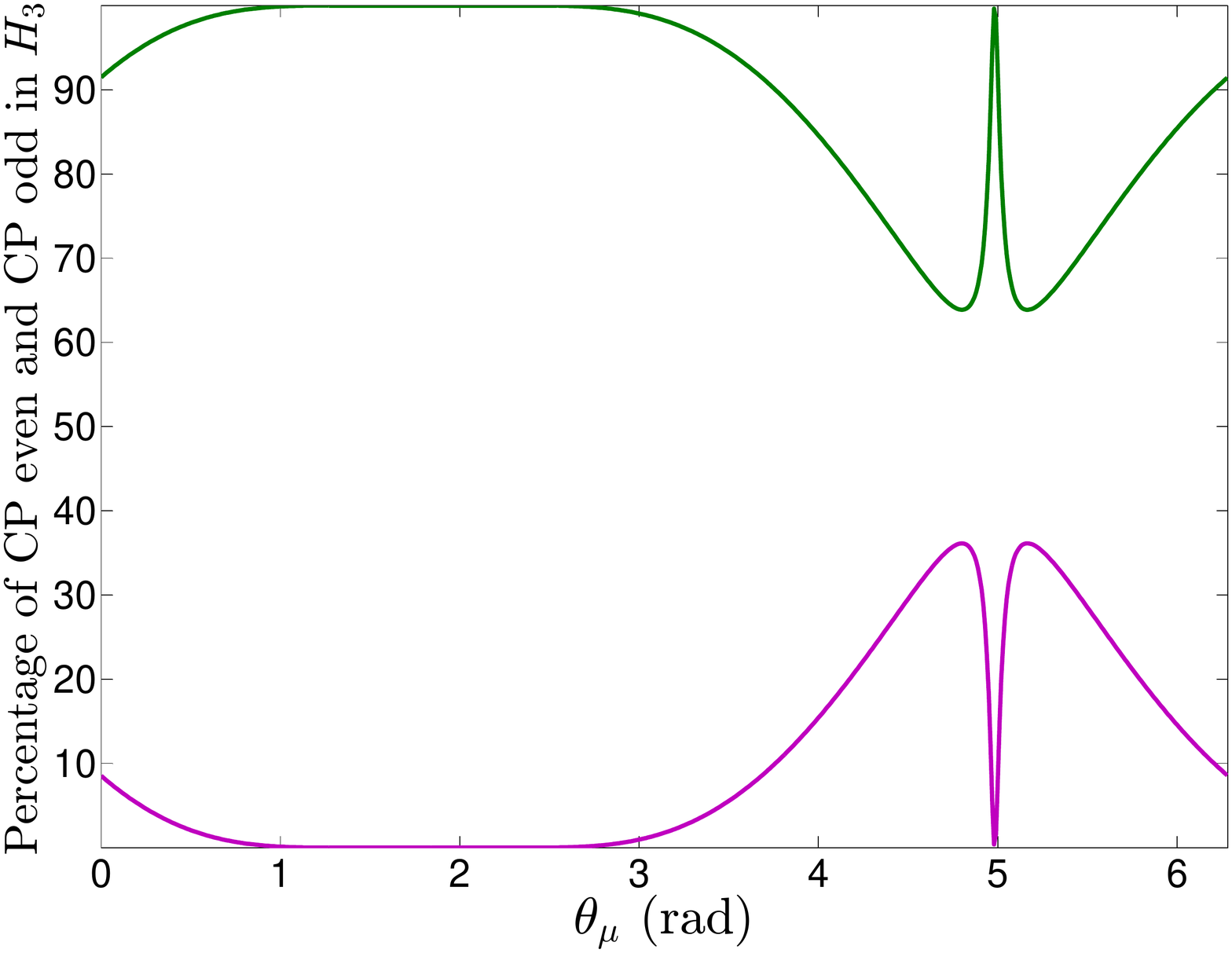}}\hglue5mm}}
\caption{Left panel:   Variation of the $CP$ even component of $H_2$ (upper curve) and the $CP$ odd
component of $H_2$ (lower curve) without including  the contributions of the vectorlike generation versus $\theta_{\mu}$.
The input parameters are: $\tan\beta=20$,$m_A=500$, $m_0=m^u_0=m^d_0=2400$, $|\mu|=300$, $|A^u_0|=|A^d_0|=8750$,
$\alpha_{A^u_0}=\alpha_{A^d_0}=1.3$, $|h_3|=1.58$, $|h'_3|=6.34\times10^{-2}$, $|h''_3|=1.97\times10^{-2}$, $|h_4|=4.42$, $|h'_4|=5.07$, $|h''_4|=12.87$, $|h_5|=6.6$, $|h'_5|=2.67$, $|h''_5|=1.86\times10^{-1}$, $|h_6|=1000$, $|h_7|=1000$, $|h_8|=1000$, $\chi_3=2\times10^{-2}$, $\chi'_3=1\times10^{-3}$, $\chi''_3=4\times10^{-3}$, $\chi_4=7\times10^{-3}$, $\chi'_4=\chi''_4=1\times10^{-3}$, $\chi_5=9\times10^{-3}$, $\chi'_5=5\times10^{-3}$, $\chi''_5=2\times10^{-3}$, $\chi_6=\chi_7=\chi_8=5\times10^{-3}$. Right panel: Variation of the $CP$ even component of $H_3$ (lower curve) and the $CP$ odd component of $H_3$ (upper curve) without including the contributions of the vectorlike generation versus $\theta_{\mu}$ for the same inputs as left panel. All masses are in GeV and all phases in rad.}
\label{fig1}
\end{center}
\end{figure}


\begin{figure}[H]
\begin{center}
{\rotatebox{0}{\resizebox*{6.0cm}{!}{\includegraphics{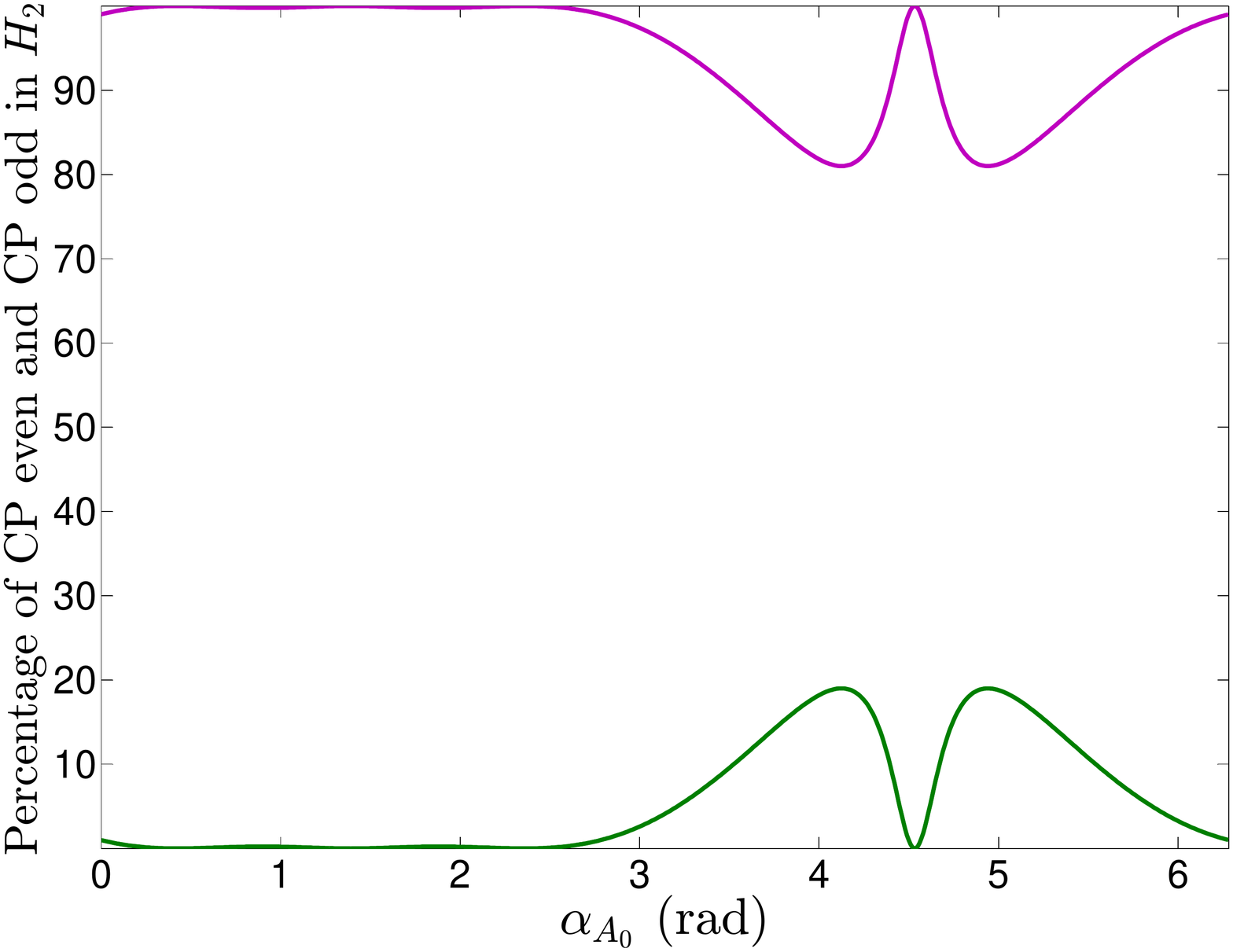}}\hglue5mm}}
{\rotatebox{0}{\resizebox*{6.0cm}{!}{\includegraphics{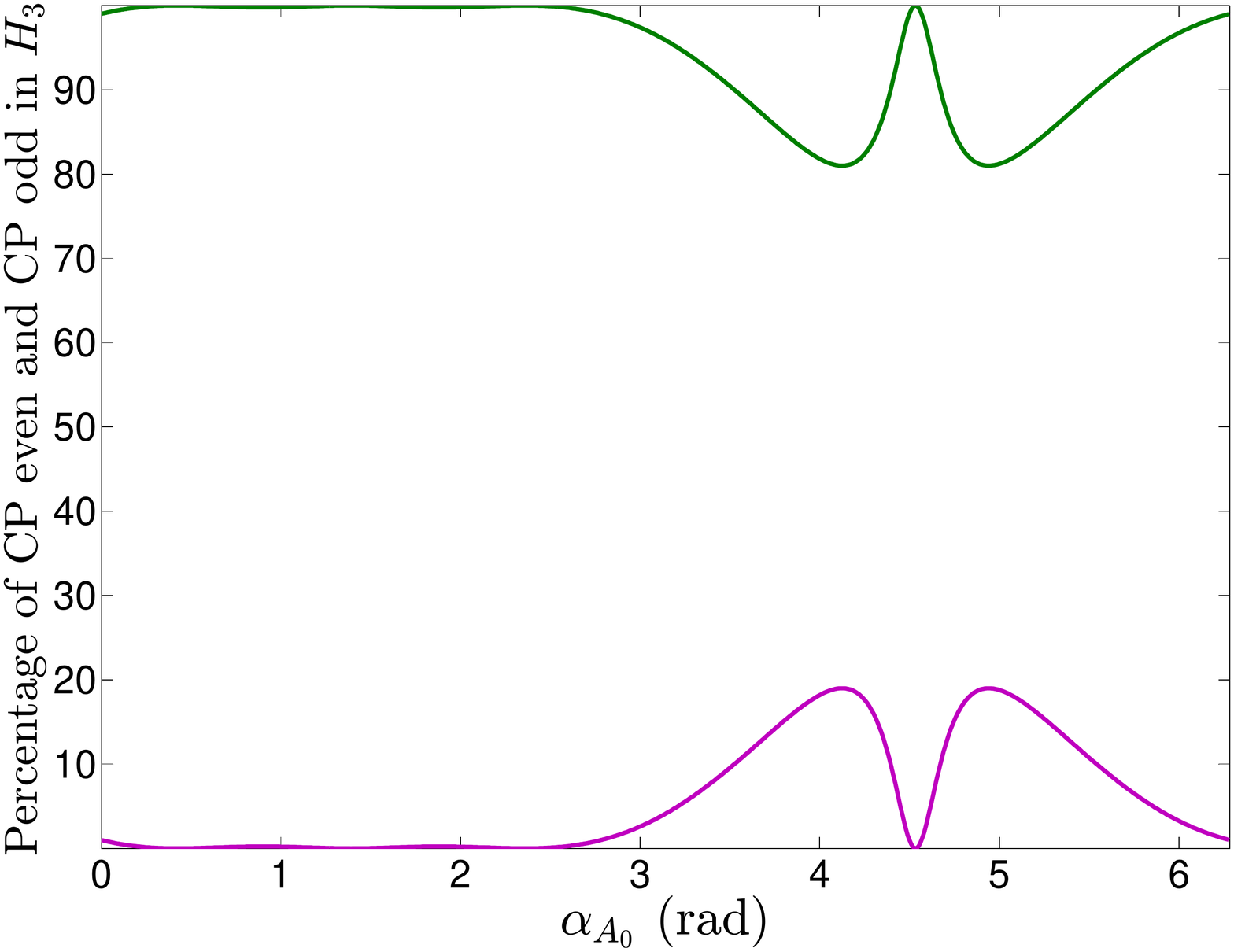}}\hglue5mm}}
\caption{Left panel: Variation of the $CP$ even component of $H_2$ (upper curve) and the $CP$ odd
component of $H_2$ (lower curve) without including the contributions of the vectorlike generation versus $\alpha_{A_0}$ ($\alpha_{A_0}=\alpha_{A^u_0}=\alpha_{A^d_0}$).The input parameters are: $\tan\beta=30$,$m_A=500$, $m_0=m^u_0=m^d_0=2200$, $|\mu|=180$, $|A^u_0|=|A^d_0|=8000$, $\theta_{\mu}=1.75$, $|h_3|=1.58$, $|h'_3|=6.34\times10^{-2}$, $|h''_3|=1.97\times10^{-2}$, $|h_4|=4.42$, $|h'_4|=5.07$, $|h''_4|=12.87$, $|h_5|=6.6$, $|h'_5|=2.67$, $|h''_5|=1.86\times10^{-1}$, $|h_6|=1000$, $|h_7|=1000$, $|h_8|=1000$, $\chi_3=2\times10^{-2}$, $\chi'_3=1\times10^{-3}$, $\chi''_3=4\times10^{-3}$, $\chi_4=7\times10^{-3}$, $\chi'_4=\chi''_4=1\times10^{-3}$, $\chi_5=9\times10^{-3}$, $\chi'_5=5\times10^{-3}$, $\chi''_5=2\times10^{-3}$, $\chi_6=\chi_7=\chi_8=5\times10^{-3}$.
Right panel: Variation of the $CP$ even component of $H_3$ (lower curve) and the $CP$ odd component of $H_3$ (upper curve)
without including the contributions of the vectorlike generation versus $\alpha_{A_0}$ for the same inputs as left panel.}
\label{fig2}
\end{center}
\end{figure}

\begin{figure}[H]
\begin{center}
{\rotatebox{0}{\resizebox*{6.0cm}{!}{\includegraphics{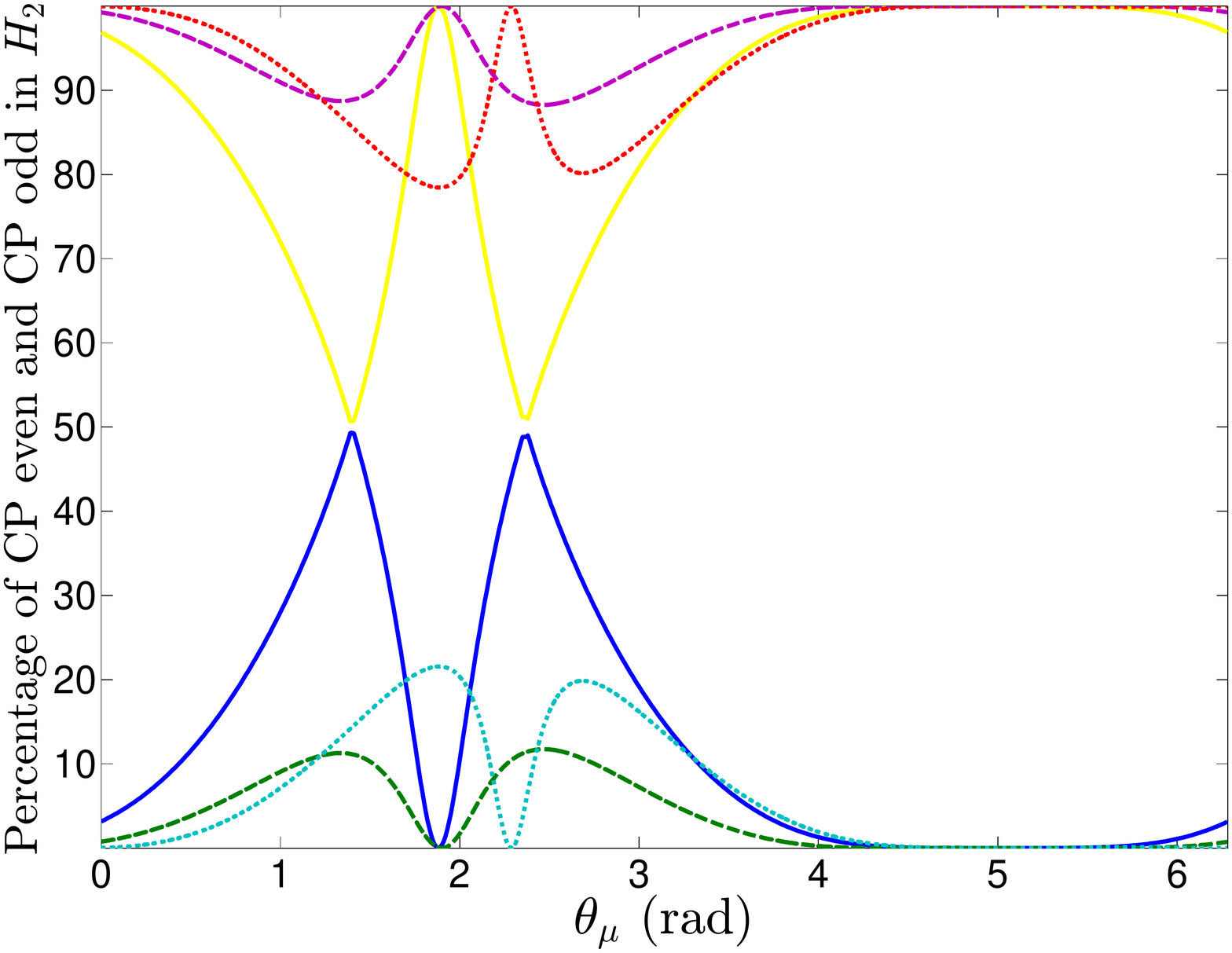}}\hglue5mm}}
{\rotatebox{0}{\resizebox*{6.0cm}{!}{\includegraphics{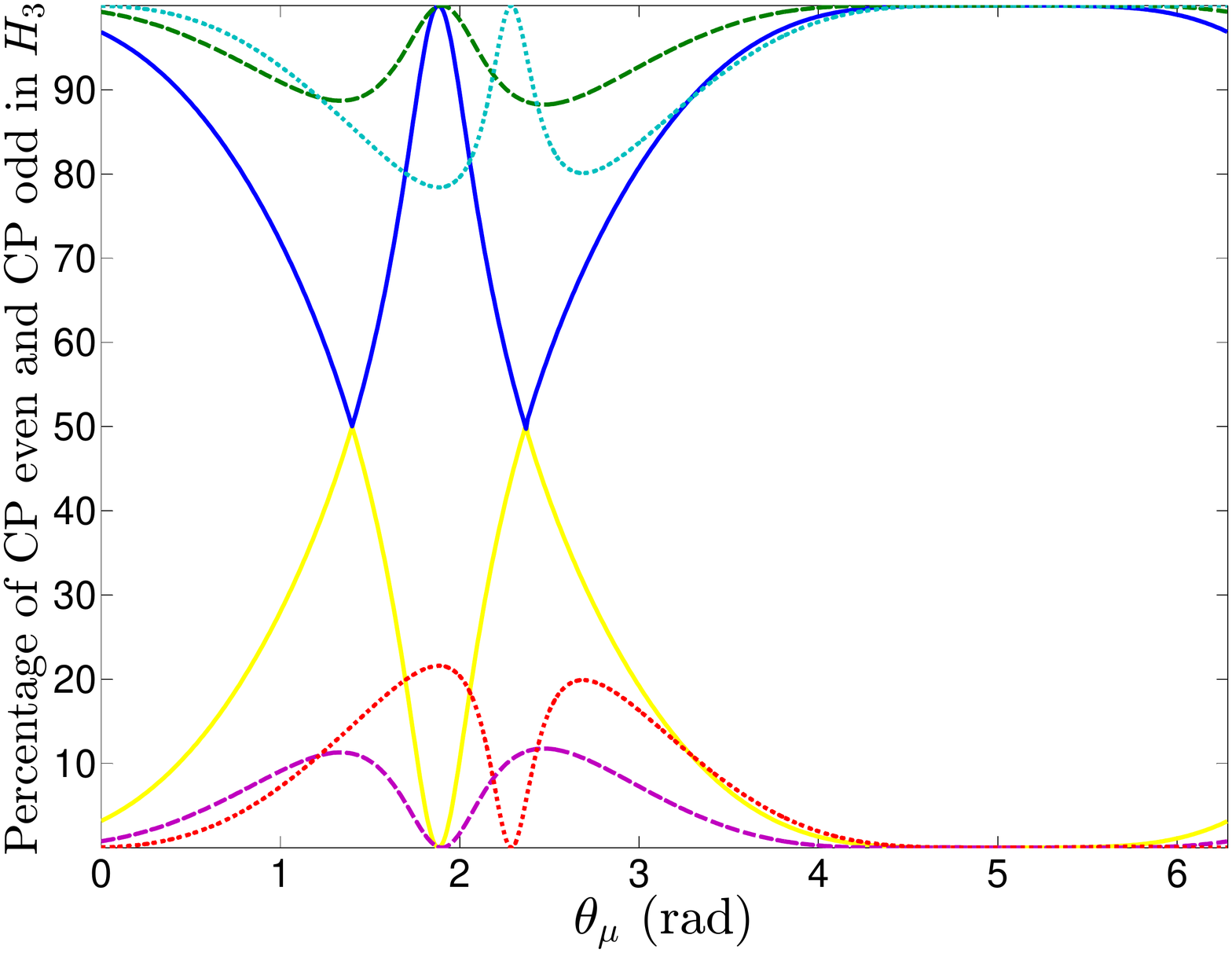}}\hglue5mm}}
\caption{Left panel: Variation of the $CP$ even component of $H_2$ (upper curves) and the $CP$ odd
component of $H_2$ (lower curves) including  the contributions of the vectorlike generation versus $\theta_{\mu}$.
The input for the solid curves is $\tan\beta=10$, $m_0=m^u_0=m^d_0=2000$, $|\mu|=380$, $|A^u_0|=7400$, $|A^d_0|=8300$, $\alpha_{A^u_0}=1.2$, $\alpha_{A^d_0}=1.3$, $h_T=2.9$, $h_B=0.4$, $h_{t_4}=0.5$, $h_{b_4}=2.9$ (Point 2).
The input for the dashed curves is $\tan\beta=20$, $m_0=m^u_0=m^d_0=2100$, $|\mu|=200$, $|A^u_0|=7800$, $|A^d_0|=7000$, $\alpha_{A^u_0}=1.4$, $\alpha_{A^d_0}=1$, $h_T=5.8$, $h_B=0.4$, $h_{t_4}=0.5$, $h_{b_4}=5.8$ (Point 4).
The input for the dotted curves is $\tan\beta=30$, $m_0=m^u_0=m^d_0=2400$, $|\mu|=200$, $|A^u_0|=8950$, $|A^d_0|=1000$, $\alpha_{A^u_0}=0.9$, $\alpha_{A^d_0}=2.8$, $h_T=8.6$, $h_B=0.4$, $h_{t_4}=0.5$, $h_{b_4}=8.6$ (Point 6).
The common parameters are: $m_A=500$, $|h_3|=1.58$, $|h'_3|=6.34\times10^{-2}$, $|h''_3|=1.97\times10^{-2}$, $|h_4|=4.42$, $|h'_4|=5.07$, $|h''_4|=12.87$, $|h_5|=6.6$, $|h'_5|=2.67$, $|h''_5|=1.86\times10^{-1}$, $|h_6|=1000$, $|h_7|=1000$, $|h_8|=1000$, $\chi_3=2\times10^{-2}$, $\chi'_3=1\times10^{-3}$, $\chi''_3=4\times10^{-3}$, $\chi_4=7\times10^{-3}$, $\chi'_4=\chi''_4=1\times10^{-3}$, $\chi_5=9\times10^{-3}$, $\chi'_5=5\times10^{-3}$, $\chi''_5=2\times10^{-3}$, $\chi_6=\chi_7=\chi_8=5\times10^{-3}$. Right panel: Variation of the $CP$ even component of $H_3$
(lower curves) and the $CP$ odd component of $H_3$ (upper curves)  including   the contributions of the vectorlike generation versus $\theta_{\mu}$ for the same inputs as left panel. All masses are in GeV and all phases in rad.}
\label{fig6}
\end{center}
\end{figure}

\begin{figure}[H]
\begin{center}
{\rotatebox{0}{\resizebox*{6.0cm}{!}{\includegraphics{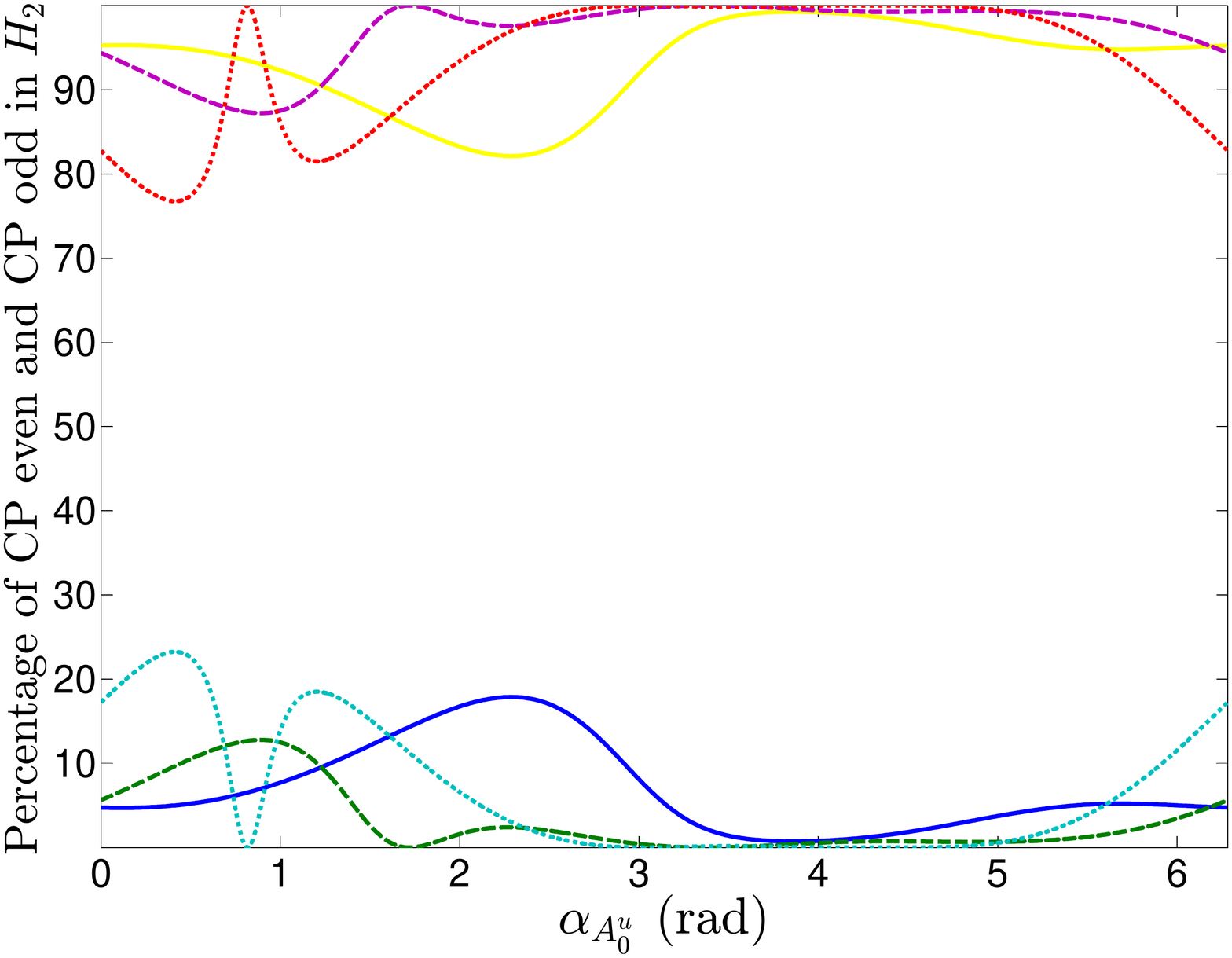}}\hglue5mm}}
{\rotatebox{0}{\resizebox*{6.0cm}{!}{\includegraphics{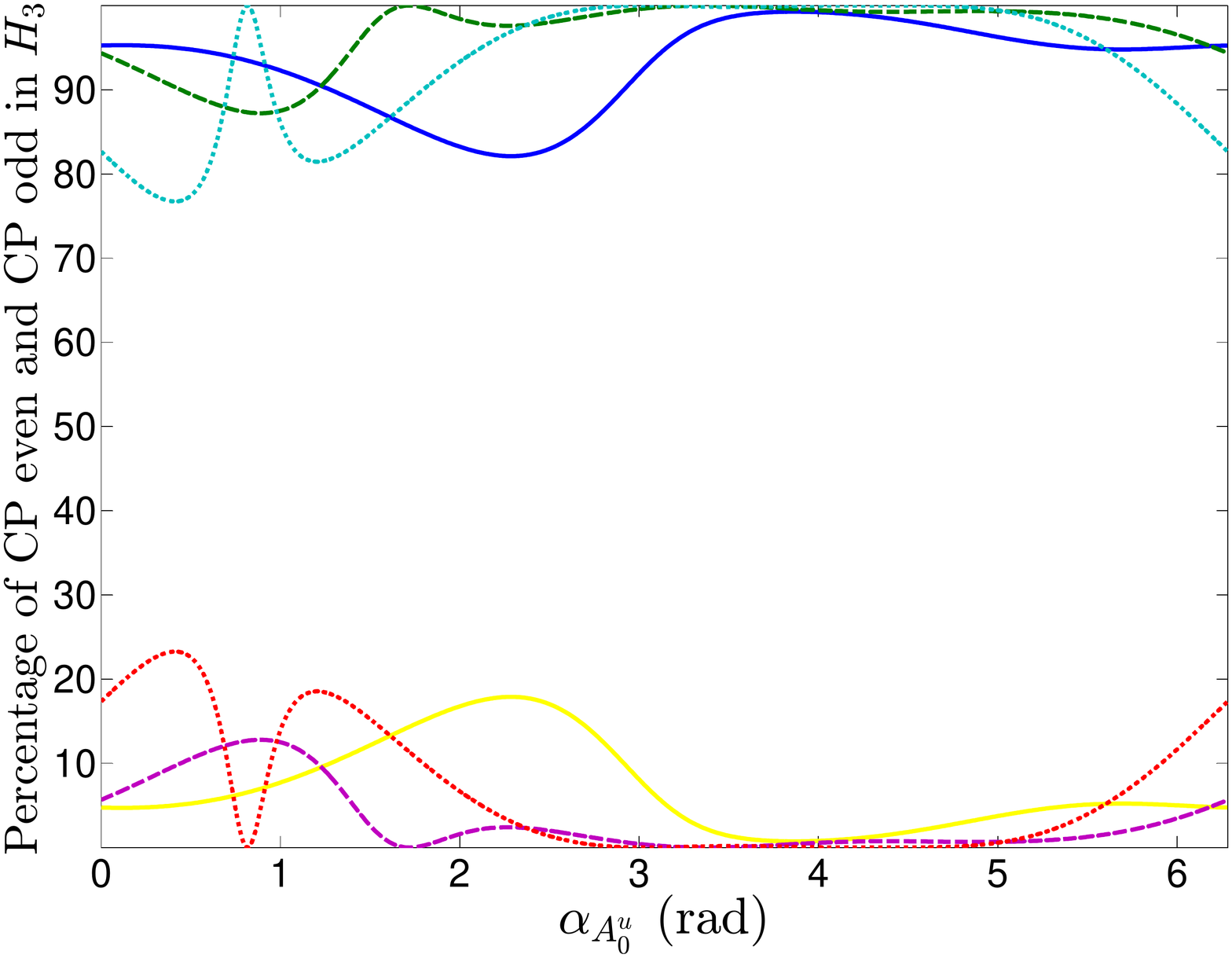}}\hglue5mm}}
\caption{Left panel: Variation of the $CP$ even component of $H_2$ (upper curves) and the $CP$ odd
component of $H_2$ (lower curves)  including   the contributions of the vectorlike generation versus $\alpha_{A_0^u}$.
Right panel: Variation of the $CP$ even component of $H_3$ (lower curves) and the $CP$ odd component of $H_3$ (upper curves)
 including
 the contributions of the vectorlike generation versus $\alpha_{A_0^u}$ for the same inputs as figure~\ref{fig6}.}
\label{fig7}
\end{center}
\end{figure}

\begin{figure}[H]
\begin{center}
{\rotatebox{0}{\resizebox*{6.0cm}{!}{\includegraphics{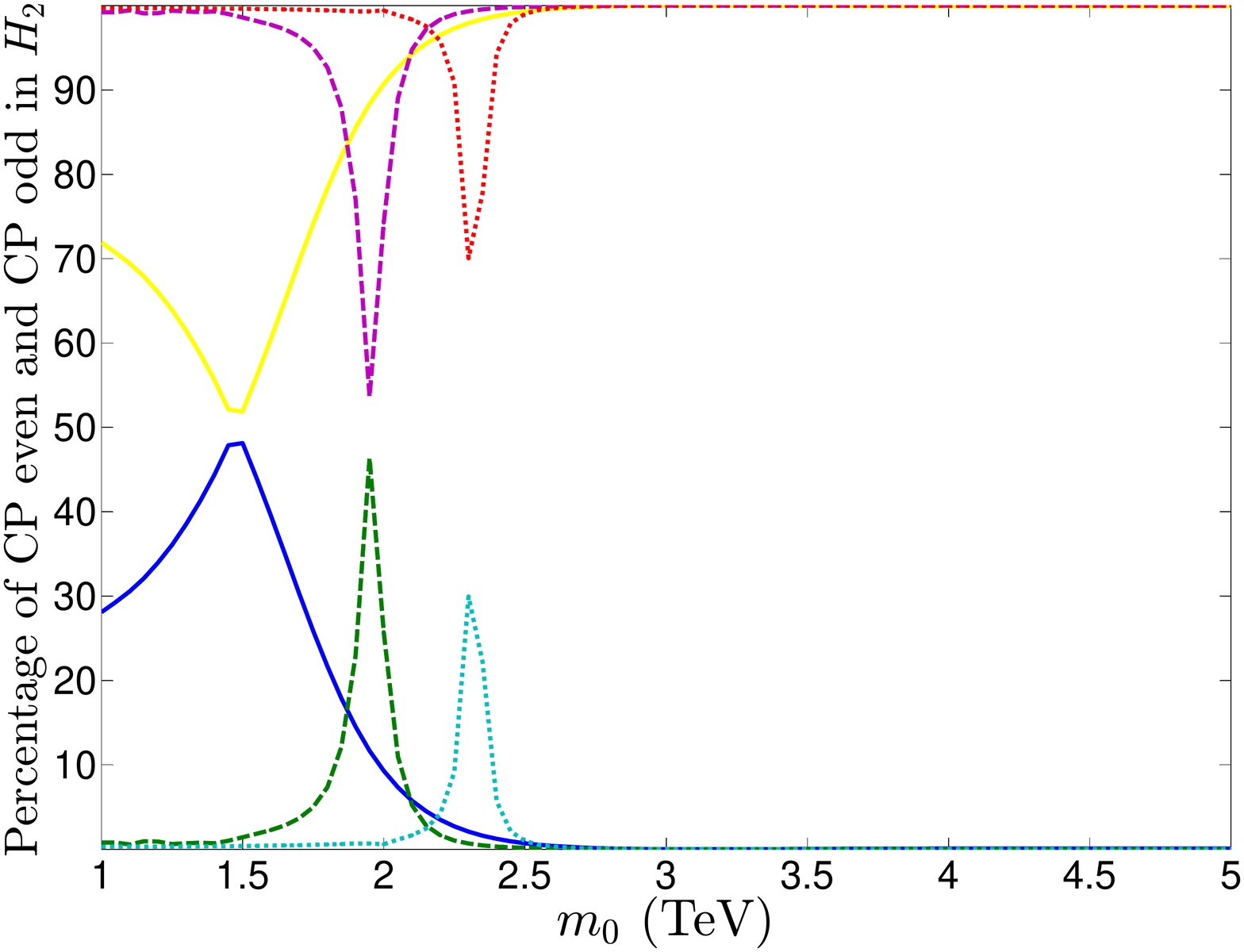}}\hglue5mm}}
{\rotatebox{0}{\resizebox*{6.0cm}{!}{\includegraphics{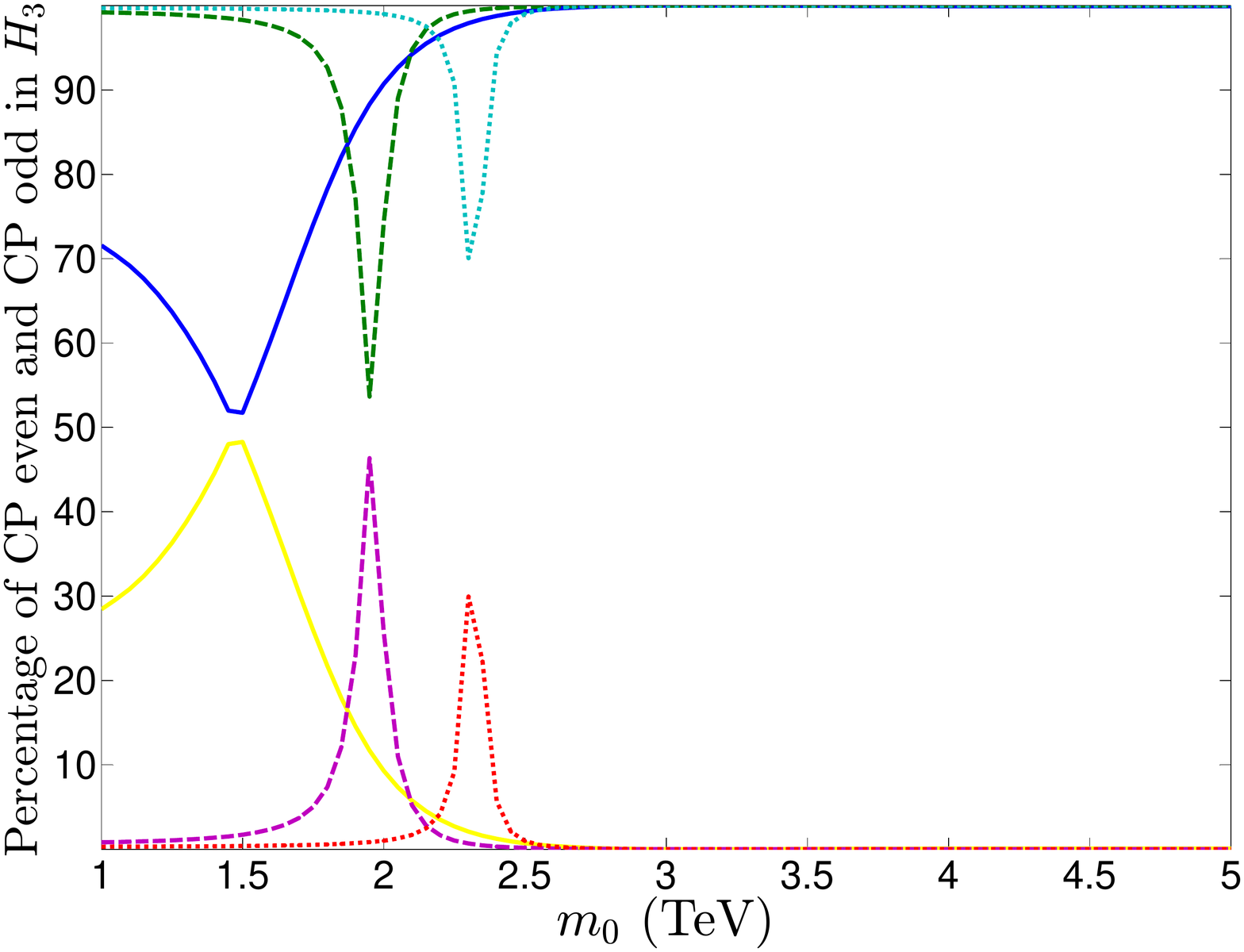}}\hglue5mm}}
\caption{Left panel: Variation of the $CP$ even component of $H_2$ (upper curves) and the $CP$ odd
component of $H_2$ (lower curves)  including   the contributions of the vectorlike generation versus $m_0$.
Right panel: Variation of the $CP$ even component of $H_3$ (lower curves) and the $CP$ odd component of $H_3$ (upper curves)  including  the contributions of the vectorlike generation versus $m_0$ for the same inputs as figure~\ref{fig6}.}
\label{fig10}
\end{center}
\end{figure}


\begin{figure}[H]
\begin{center}
{\rotatebox{0}{\resizebox*{4.7cm}{!}{\includegraphics{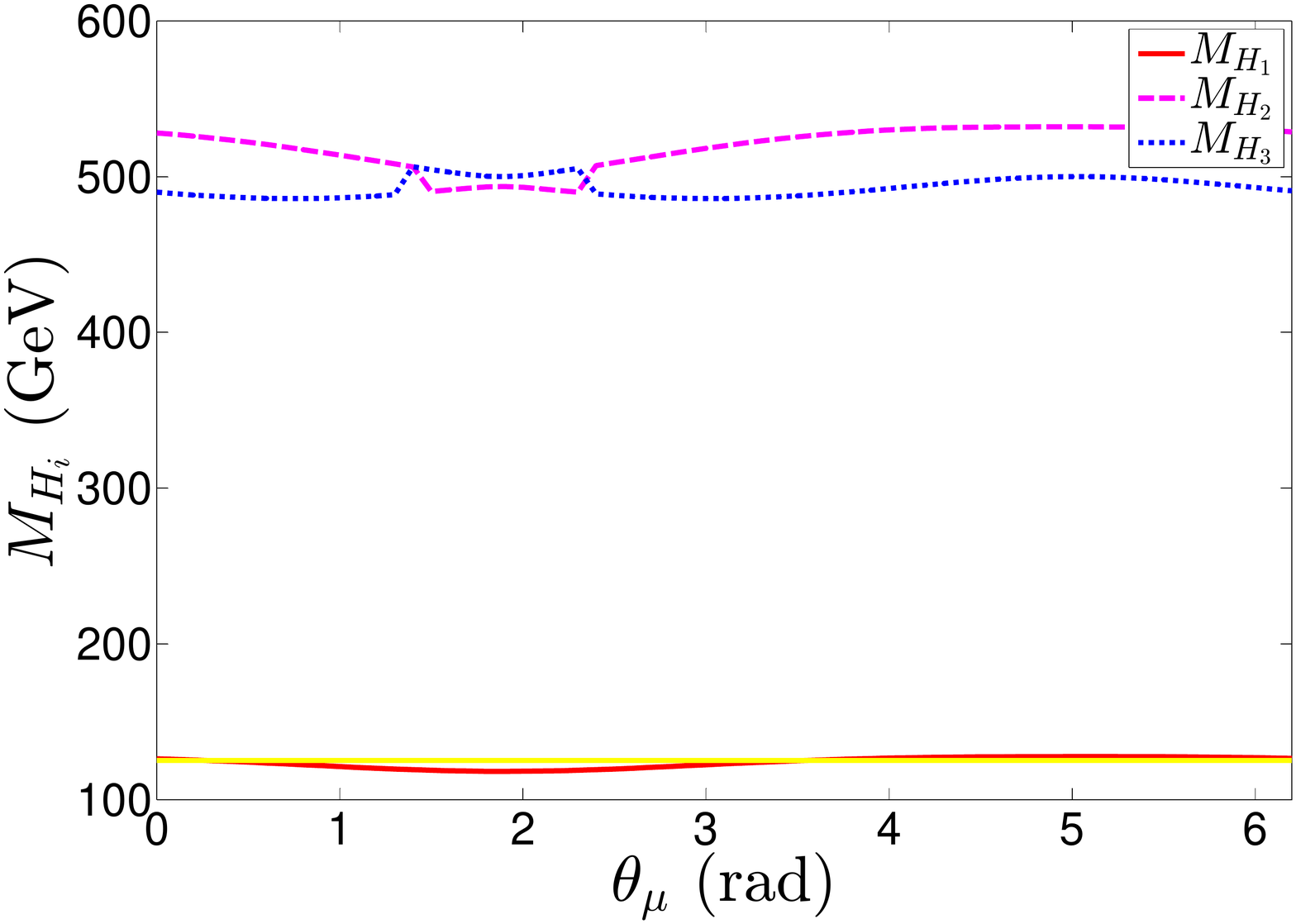}}\hglue5mm}}
{\rotatebox{0}{\resizebox*{4.7cm}{!}{\includegraphics{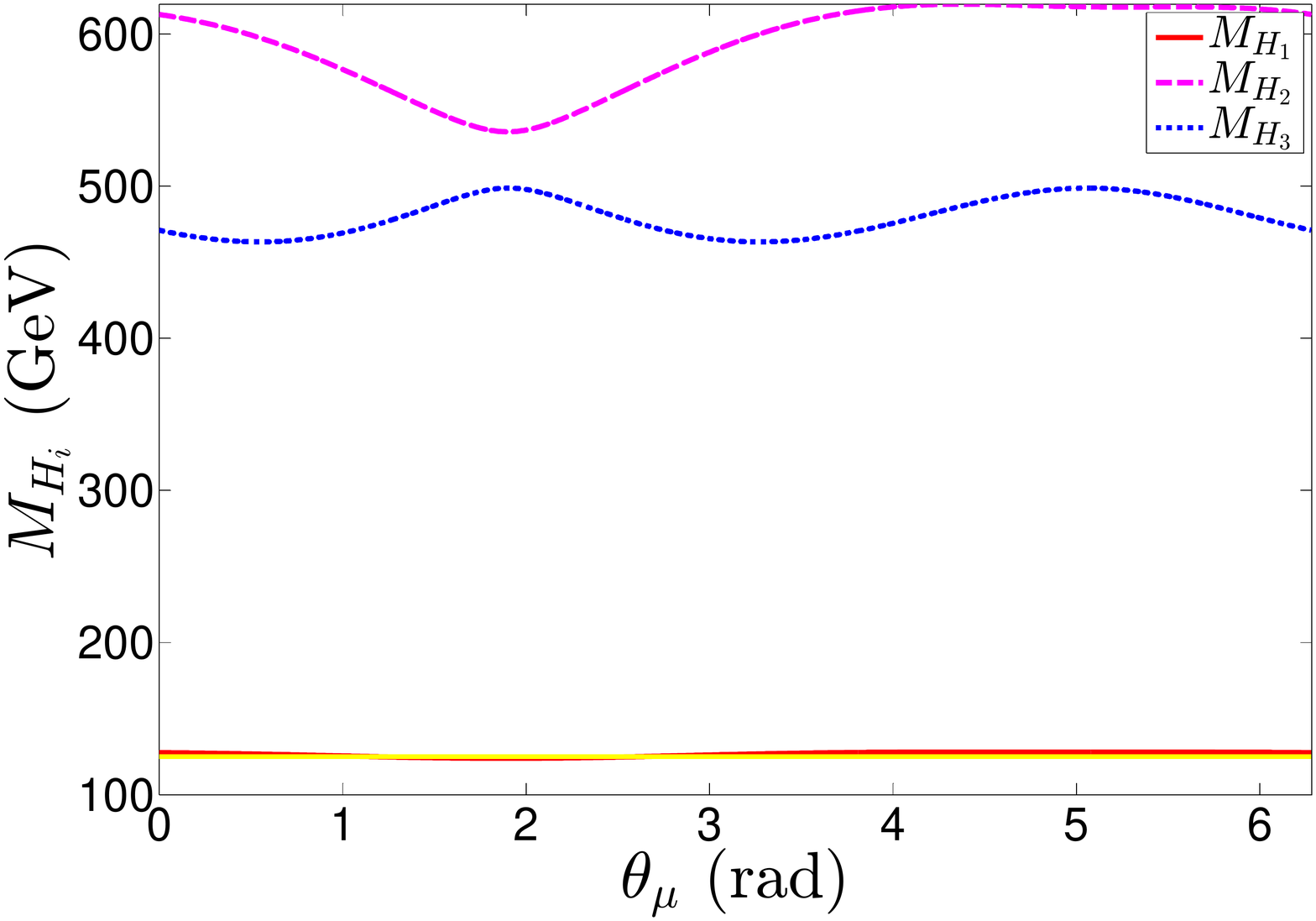}}\hglue5mm}}
{\rotatebox{0}{\resizebox*{4.7cm}{!}{\includegraphics{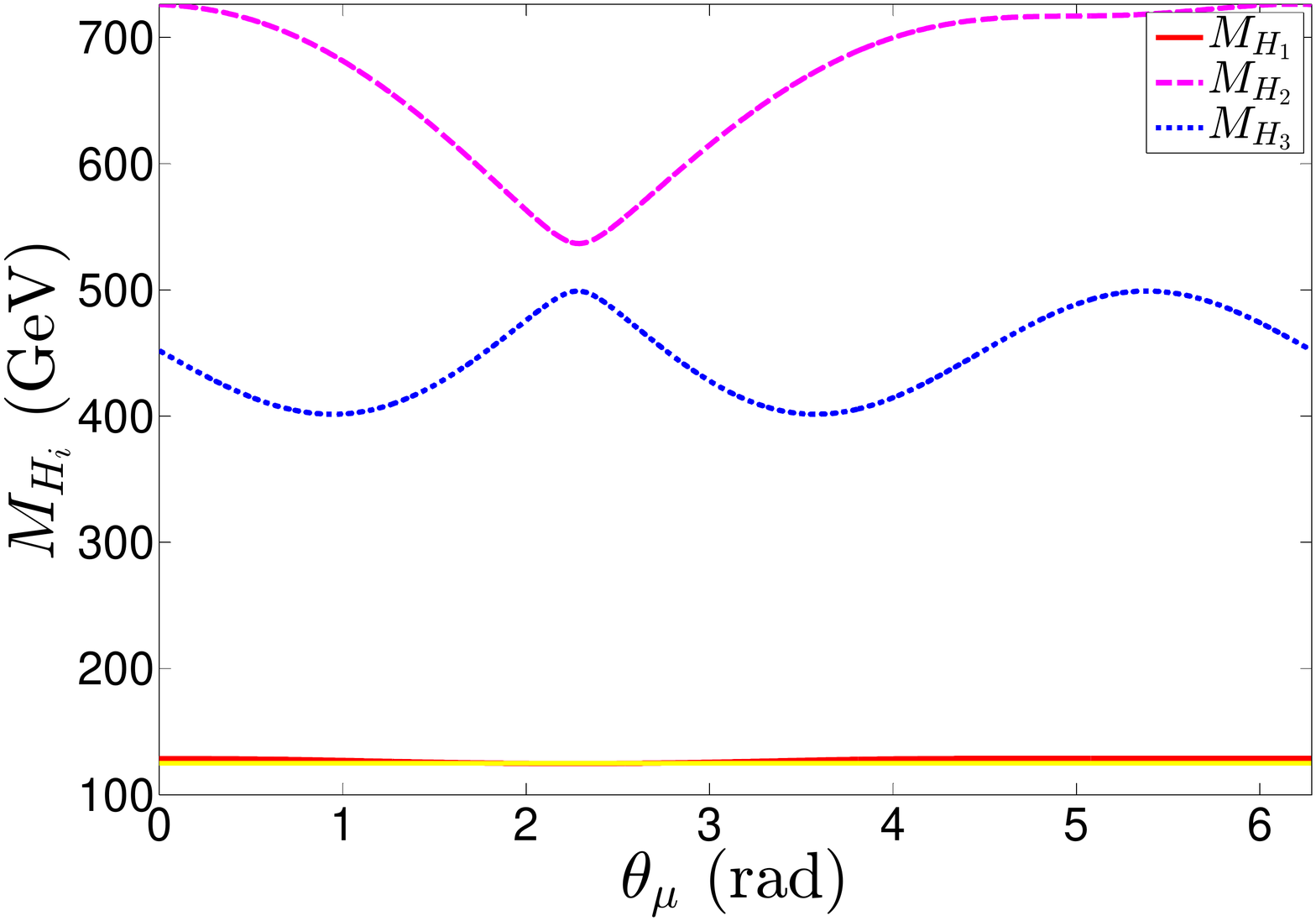}}\hglue5mm}}
\caption{Left panel: Variation of the $M_{H_1}$ (solid curve), $M_{H_2}$ (dashed curve) and $M_{H_3}$ (dotted curve) versus $\theta_{\mu}$ for $\tan\beta=10$, $m_0=m^u_0=m^d_0=2000$, $|\mu|=380$, $|A^u_0|=7400$, $|A^d_0|=8300$, $\alpha_{A^u_0}=1.2$, $\alpha_{A^d_0}=1.3$, $h_T=2.9$, $h_B=0.4$, $h_{t_4}=0.5$, $h_{b_4}=2.9$.
Middle panel: Variation of the $M_{H_1}$ (solid curve), $M_{H_2}$ (dashed curve) and $M_{H_3}$ (dotted curve) versus $\theta_{\mu}$ for $\tan\beta=20$, $m_0=m^u_0=m^d_0=2100$, $|\mu|=200$, $|A^u_0|=7800$, $|A^d_0|=7000$, $\alpha_{A^u_0}=1.4$, $\alpha_{A^d_0}=1$, $h_T=5.8$, $h_B=0.4$, $h_{t_4}=0.5$, $h_{b_4}=5.8$.
Right panel: Variation of the $M_{H_1}$ (solid curve), $M_{H_2}$ (dashed curve) and $M_{H_3}$ (dotted curve) versus $\theta_{\mu}$ for $\tan\beta=30$, $m_0=m^u_0=m^d_0=2400$, $|\mu|=200$, $|A^u_0|=8950$, $|A^d_0|=1000$, $\alpha_{A^u_0}=0.9$, $\alpha_{A^d_0}=2.8$, $h_T=8.6$, $h_B=0.4$, $h_{t_4}=0.5$, $h_{b_4}=8.6$.
The common parameters are: $m_A=500$, $|h_3|=1.58$, $|h'_3|=6.34\times10^{-2}$, $|h''_3|=1.97\times10^{-2}$, $|h_4|=4.42$, $|h'_4|=5.07$, $|h''_4|=12.87$, $|h_5|=6.6$, $|h'_5|=2.67$, $|h''_5|=1.86\times10^{-1}$, $|h_6|=1000$, $|h_7|=1000$, $|h_8|=1000$, $\chi_3=2\times10^{-2}$, $\chi'_3=1\times10^{-3}$, $\chi''_3=4\times10^{-3}$, $\chi_4=7\times10^{-3}$, $\chi'_4=\chi''_4=1\times10^{-3}$, $\chi_5=9\times10^{-3}$, $\chi'_5=5\times10^{-3}$, $\chi''_5=2\times10^{-3}$, $\chi_6=\chi_7=\chi_8=5\times10^{-3}$. All masses are in GeV and all phases in rad.}
\label{fig11}
\end{center}
\end{figure}

\begin{figure}[H]
\begin{center}
{\rotatebox{0}{\resizebox*{4.7cm}{!}{\includegraphics{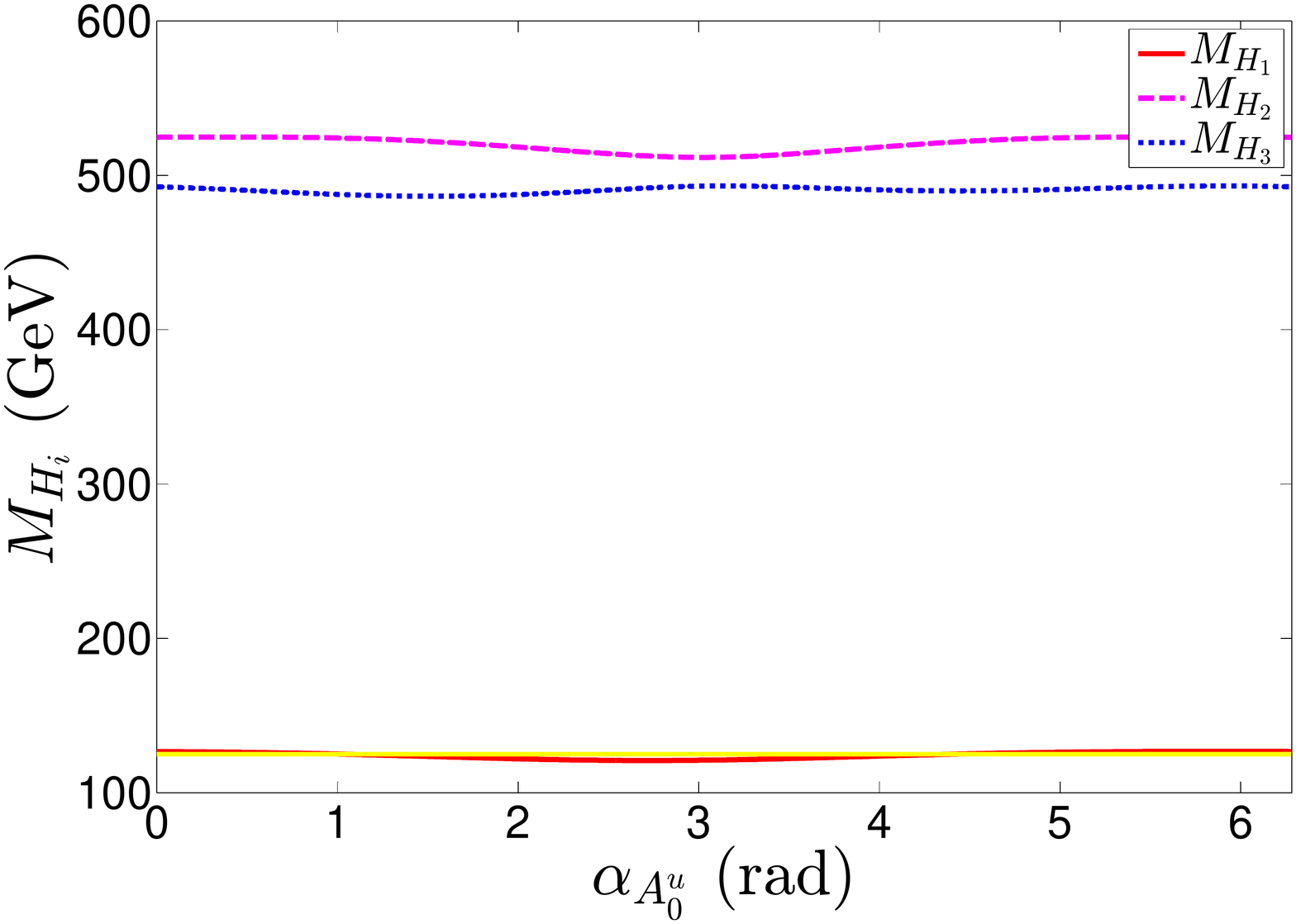}}\hglue5mm}}
{\rotatebox{0}{\resizebox*{4.7cm}{!}{\includegraphics{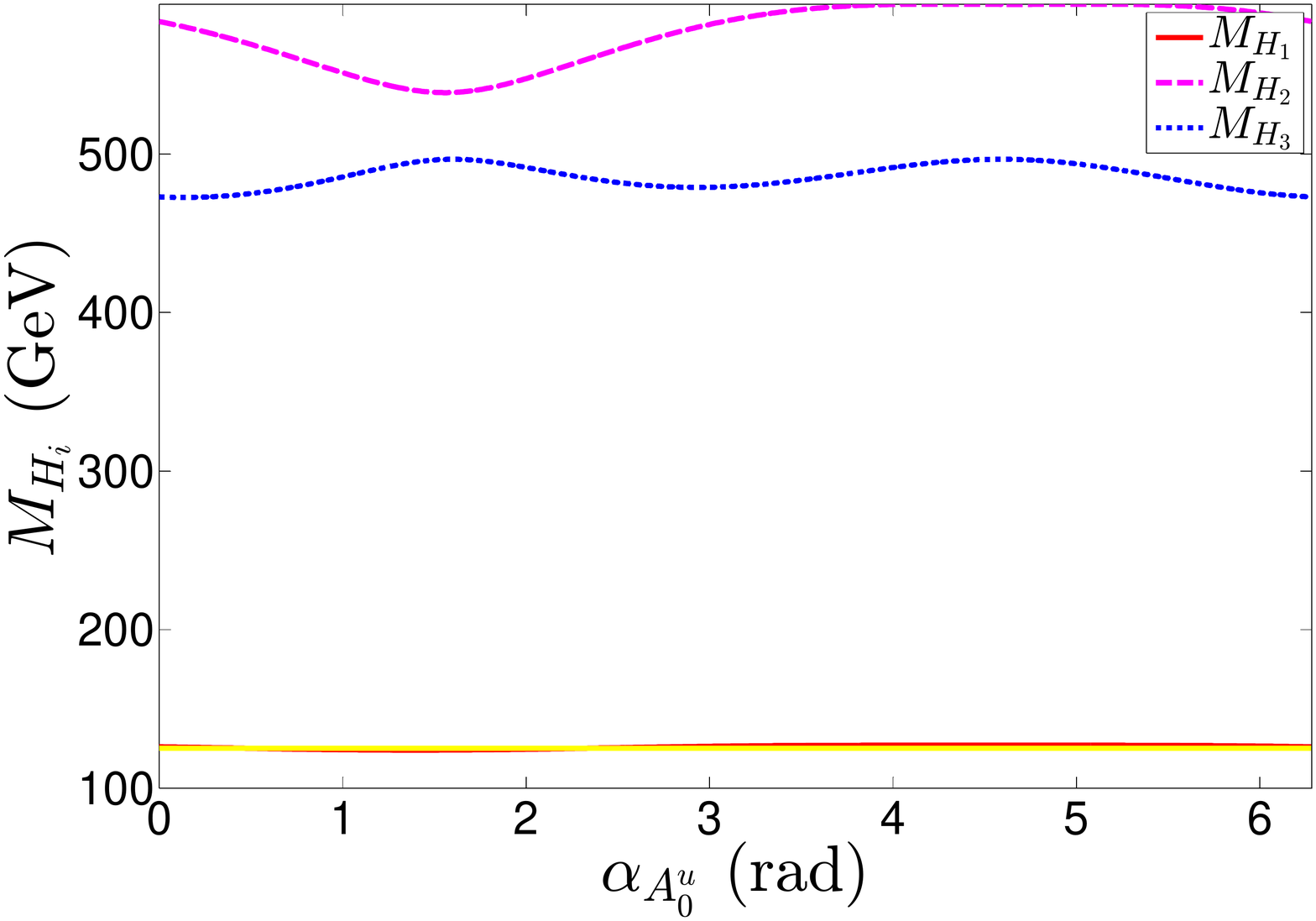}}\hglue5mm}}
{\rotatebox{0}{\resizebox*{4.7cm}{!}{\includegraphics{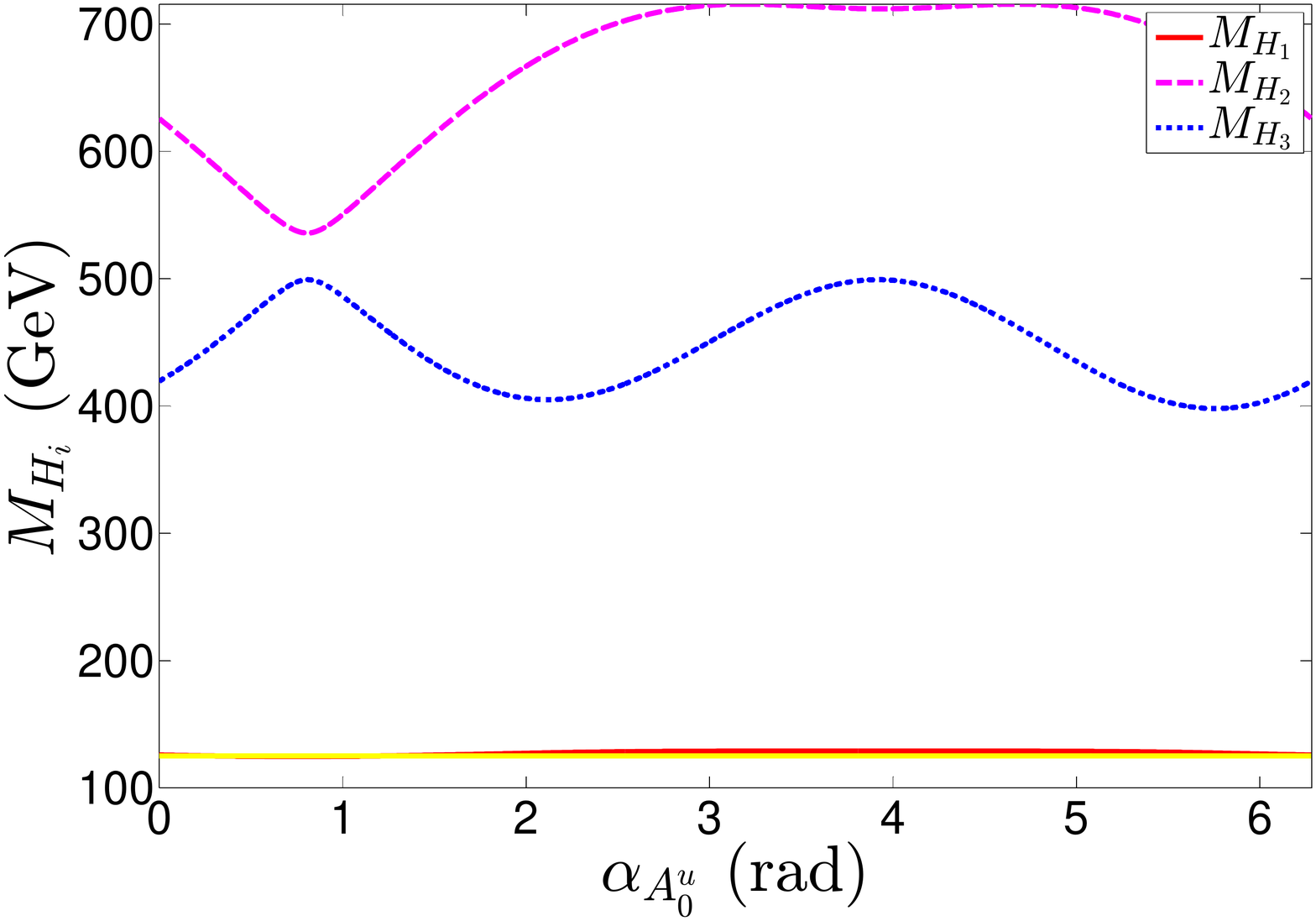}}\hglue5mm}}
\caption{Left panel: Variation of the $M_{H_1}$ (solid curve), $M_{H_2}$ (dashed curve) and $M_{H_3}$ (dotted curve) versus $\alpha_{A^u_0}$
 for $\tan\beta=10$, $m_0=m^u_0=m^d_0=2000$, $|\mu|=380$, $|A^u_0|=7400$, $|A^d_0|=8300$, $\theta_{\mu}=0.4$, $\alpha_{A^d_0}=1.3$, $h_T=2.9$, $h_B=0.4$, $h_{t_4}=0.5$, $h_{b_4}=2.9$.
Middle panel: Variation of the $M_{H_1}$ (solid curve), $M_{H_2}$ (dashed curve) and $M_{H_3}$ (dotted curve) versus $\alpha_{A^u_0}$
 for $\tan\beta=20$, $m_0=m^u_0=m^d_0=2100$, $|\mu|=200$, $|A^u_0|=7800$, $|A^d_0|=7000$, $\theta_{\mu}=1.7$, $\alpha_{A^d_0}=1$, $h_T=5.8$, $h_B=0.4$, $h_{t_4}=0.5$, $h_{b_4}=5.8$.
Right panel: Variation of the $M_{H_1}$ (solid curve), $M_{H_2}$ (dashed curve) and $M_{H_3}$ (dotted curve) versus $\alpha_{A^u_0}$
 for $\tan\beta=30$, $m_0=m^u_0=m^d_0=2400$, $|\mu|=200$, $|A^u_0|=8950$, $|A^d_0|=1000$, $\theta_{\mu}=2.37$, $\alpha_{A^d_0}=2.8$, $h_T=8.6$, $h_B=0.4$, $h_{t_4}=0.5$, $h_{b_4}=8.6$.
The common parameters are: $m_A=500$, $|h_3|=1.58$, $|h'_3|=6.34\times10^{-2}$, $|h''_3|=1.97\times10^{-2}$, $|h_4|=4.42$, $|h'_4|=5.07$, $|h''_4|=12.87$, $|h_5|=6.6$, $|h'_5|=2.67$, $|h''_5|=1.86\times10^{-1}$, $|h_6|=1000$, $|h_7|=1000$, $|h_8|=1000$, $\chi_3=2\times10^{-2}$, $\chi'_3=1\times10^{-3}$, $\chi''_3=4\times10^{-3}$, $\chi_4=7\times10^{-3}$, $\chi'_4=\chi''_4=1\times10^{-3}$, $\chi_5=9\times10^{-3}$, $\chi'_5=5\times10^{-3}$, $\chi''_5=2\times10^{-3}$, $\chi_6=\chi_7=\chi_8=5\times10^{-3}$. All masses are in GeV and all phases in rad.}
\label{fig12}
\end{center}
\end{figure}

\subsection{Decays of the Higgs bosons to fermion pairs}
Decays of the Higgs bosons are important channels which allow for  tests of new physics beyond the standard
model. A convenient ratio for this purpose is $R_{if}$ defined by~\cite{Demir:1999hj}

\begin{align}
R_{i f}&=\frac{\Gamma(H_{i}\rightarrow \bar{f} f)}  {\Gamma(H_{i}\rightarrow \bar{f}f)_0}\nonumber\\
&=\frac{{(D_{i k})}^{2}(1-x_f^2)^{3/2}+
f^{2}{(D_{i 3})}^{2}(1- x_f^{2})^{1/2}}{{(D_{i k
}(0))}^{2}(1- x_{f0}^2)^{3/2}+
f^{2}{(D_{i 3}(0))}^{2}(1-x_{f0}^2)^{1/2}}
\end{align}
where $x_f^2= 4m_f^2/M_{H_i}^2$,  $x_{f0}^2= 4m_{f}^2/M_{H_i}(0)^2$,
where $k=2 (1)$ and  ${f}=\cos\beta (\sin\beta)$ for $u$-type quarks ($d$-type quarks and charged
leptons). The argument 0 in D and in the subscript of $x_f$  in the denominator indicates that
$\theta_\mu+ \alpha_{A_0}=0$.
For the case when there is no contribution from the vectorlike multiplet
the ratio between the decay widths of the higgs into quark pairs is exhibited in table \ref{table:5}
for the model point 3 in table~\ref{table:1}. As a comparison we exhibit the same ratios for the
case when a vectorlike multiplet is included again for model point 3 of table \ref{table:6}.
One finds significant differences between the two tables for certain decay width ratios
which points to the significant contribution from the vectorlike multiplet to the ratio.
We now study the CP phase dependence for the case with contributions from the vectorlike
multiplet are included.  In Fig. \ref{fig19} we give the dependence of $R_{1b}$ and $R_{1c}$ on the
$\theta_\mu$ and $\alpha_{A_0}=\alpha_{A_0^u} = \alpha_{A_0^d}$. One finds a large sensitivity of the
ratio to the CP phases. A similar analysis for $R_{2b}, R_{2c}$ is given in Fig. \ref{fig21} and for $R_{3b}, R_{3c}$ in
Fig. \ref{fig23}.

\begin{table}[H]
\begin{center}
\begin{tabular}{l  c  c  c  c  c  c}
 \hline\hline
 $R_{if}$ &  $b$  &  $s$  &  $d$  &  $t$  &  $c$  &  $u$   \\
 \hline
 $i=1$    & 1.125 & 1.125 & 1.125 &  ...  & 0.999 & 0.999  \\
 $i=2$    & 0.999 & 0.999 & 0.999 & 1.154 & 1.133 & 1.133  \\
 $i=3$    & 1     & 1     & 1     & 0.984 & 0.992 & 0.992  \\
 \hline\hline
\end{tabular}
\caption{An exhibition of the ratio between the decay widths of the higgs scalars into quark pairs for the case without the contributions of vectorlike multiplet. The parameter space corresponding to point 3 in table~\ref{table:1}.}
\label{table:5}
\end{center}
\end{table}
\begin{table}[H]
\begin{center}
\begin{tabular}{l  c  c  c  c  c  c}
 \hline\hline
 $R_{if}$ &  $b$  &  $s$  &  $d$  &  $t$  &  $c$  &  $u$   \\
 \hline
 $i=1$    & 2.069 & 2.07  & 2.07  &  ...  & 0.997 & 0.997  \\
 $i=2$    & 0.998 & 0.998 & 0.998 & 1.701 & 1.861 & 1.861  \\
 $i=3$    & 0.999 & 0.999 & 0.999 & 1.001 & 1.134 & 1.134  \\
 \hline\hline
\end{tabular}
\caption{An exhibition of the ratio between the decay widths of the higgs scalars into quark pairs for the case with the contributions of vectorlike multiplet. The parameter space corresponding to point 3 in table~\ref{table:1}.}
\label{table:6}
\end{center}
\end{table}

\begin{figure}[H]
\begin{center}
{\rotatebox{0}{\resizebox*{3.5cm}{!}{\includegraphics{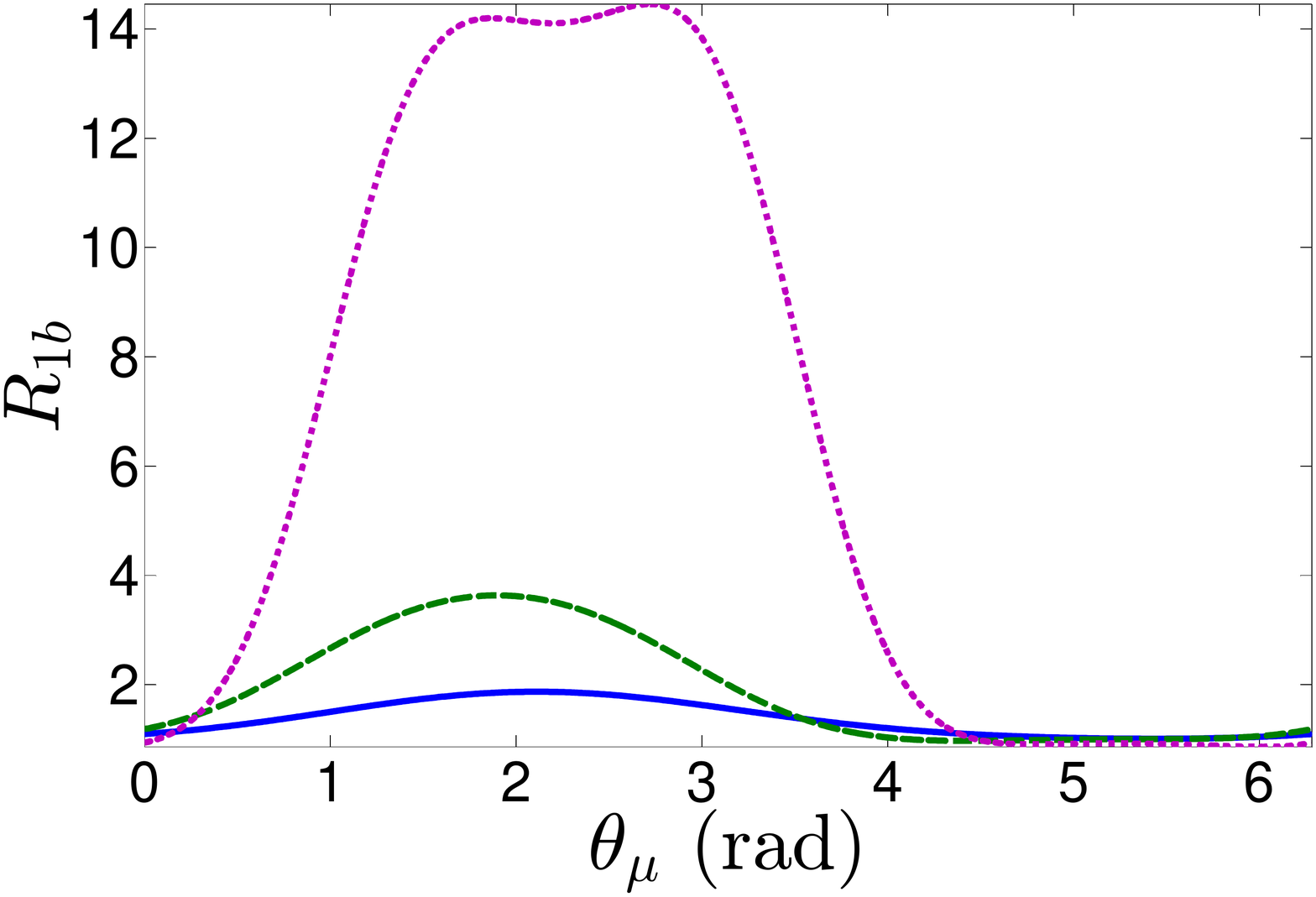}}\hglue5mm}}
{\rotatebox{0}{\resizebox*{3.5cm}{!}{\includegraphics{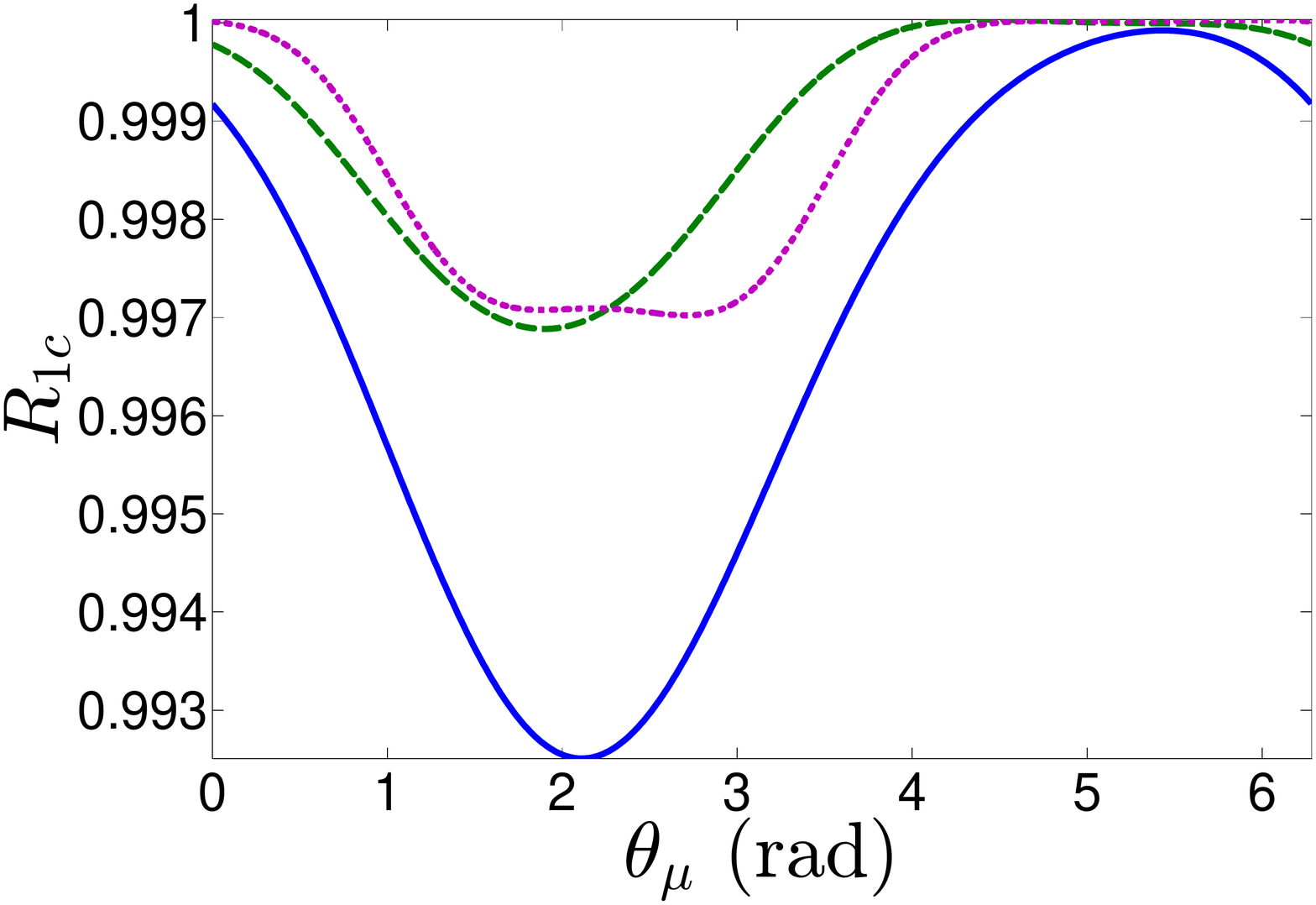}}\hglue5mm}}
{\rotatebox{0}{\resizebox*{3.5cm}{!}{\includegraphics{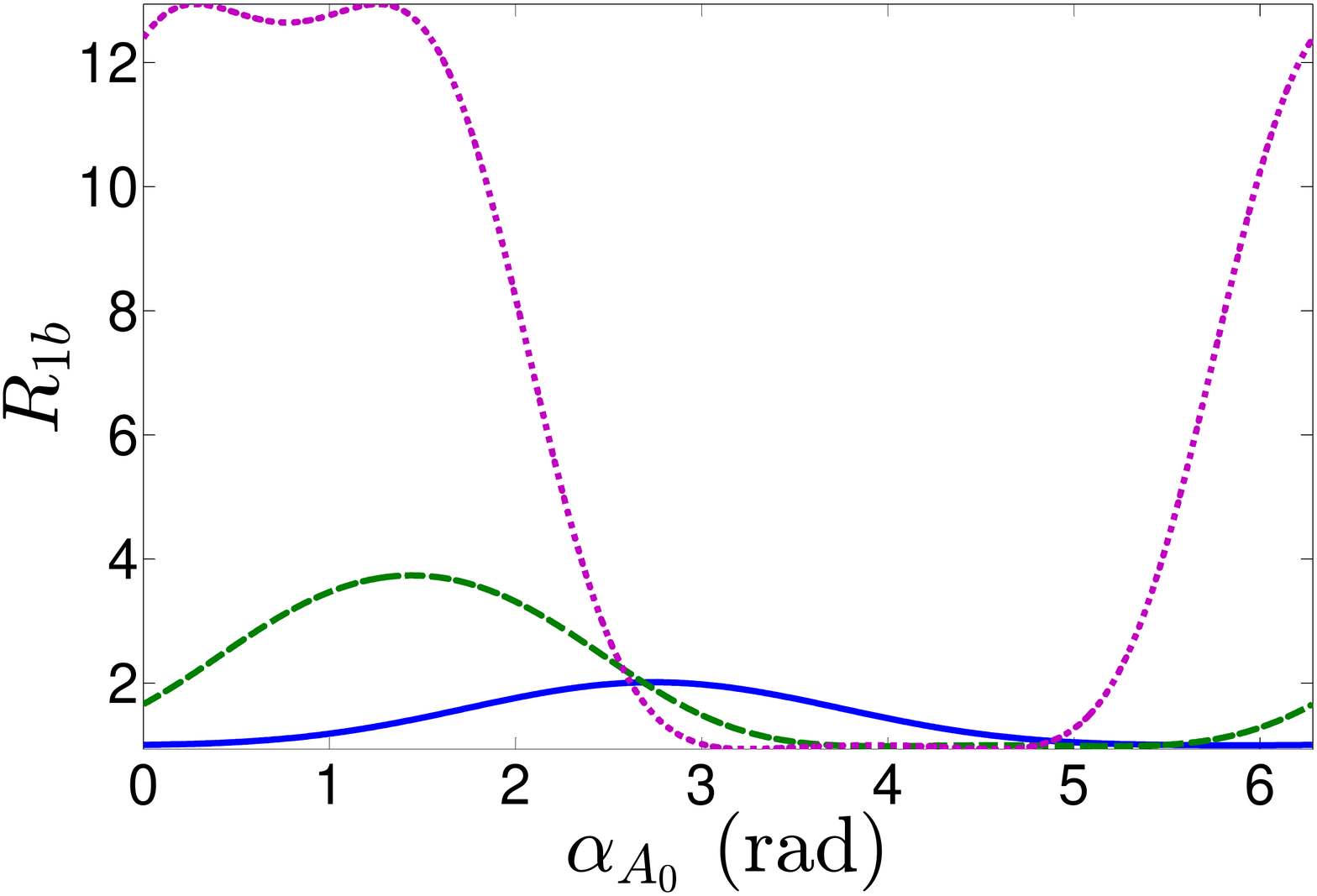}}\hglue5mm}}
{\rotatebox{0}{\resizebox*{3.5cm}{!}{\includegraphics{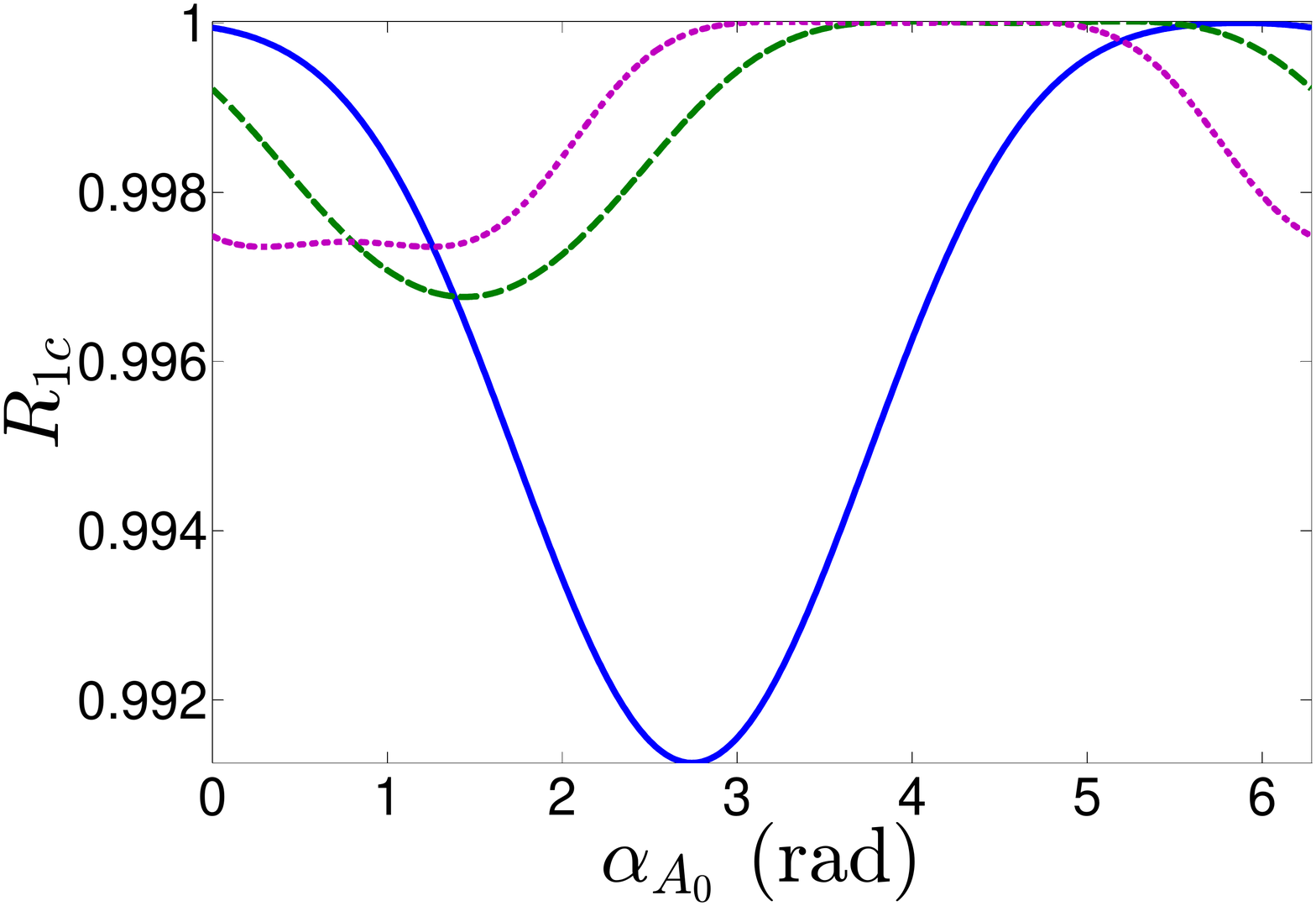}}\hglue5mm}}
\caption{Left panel: Variation of the $R_{1b}$ versus $\theta_{\mu}$ for the case with the contributions of the vectorlike generation.
Second  left panel: Variation of the $R_{1c}$ versus $\theta_{\mu}$ for the case with the contributions of the vectorlike generation.
The input corresponding to point 2 (solid curve), point 4 (dashed curve) and point 6 (dotted curve) in table~\ref{table:4}.
Second right panel: Variation of the $R_{1b}$ versus $\alpha_{A_0}$ ($\alpha_{A_0}=\alpha_{A^u_0}=\alpha_{A^d_0}$) for the case with the contributions of the vectorlike generation.
Right panel: Variation of the $R_{1c}$ versus $\alpha_{A_0}$  for the case with the contributions of the vectorlike generation.
The input corresponding to point 2 (solid curve), point 4 (dashed curve) and point 6 (dotted curve) in table~\ref{table:4}.}
\label{fig19}
\end{center}
\end{figure}
\begin{figure}[H]
\begin{center}
{\rotatebox{0}{\resizebox*{3.5cm}{!}{\includegraphics{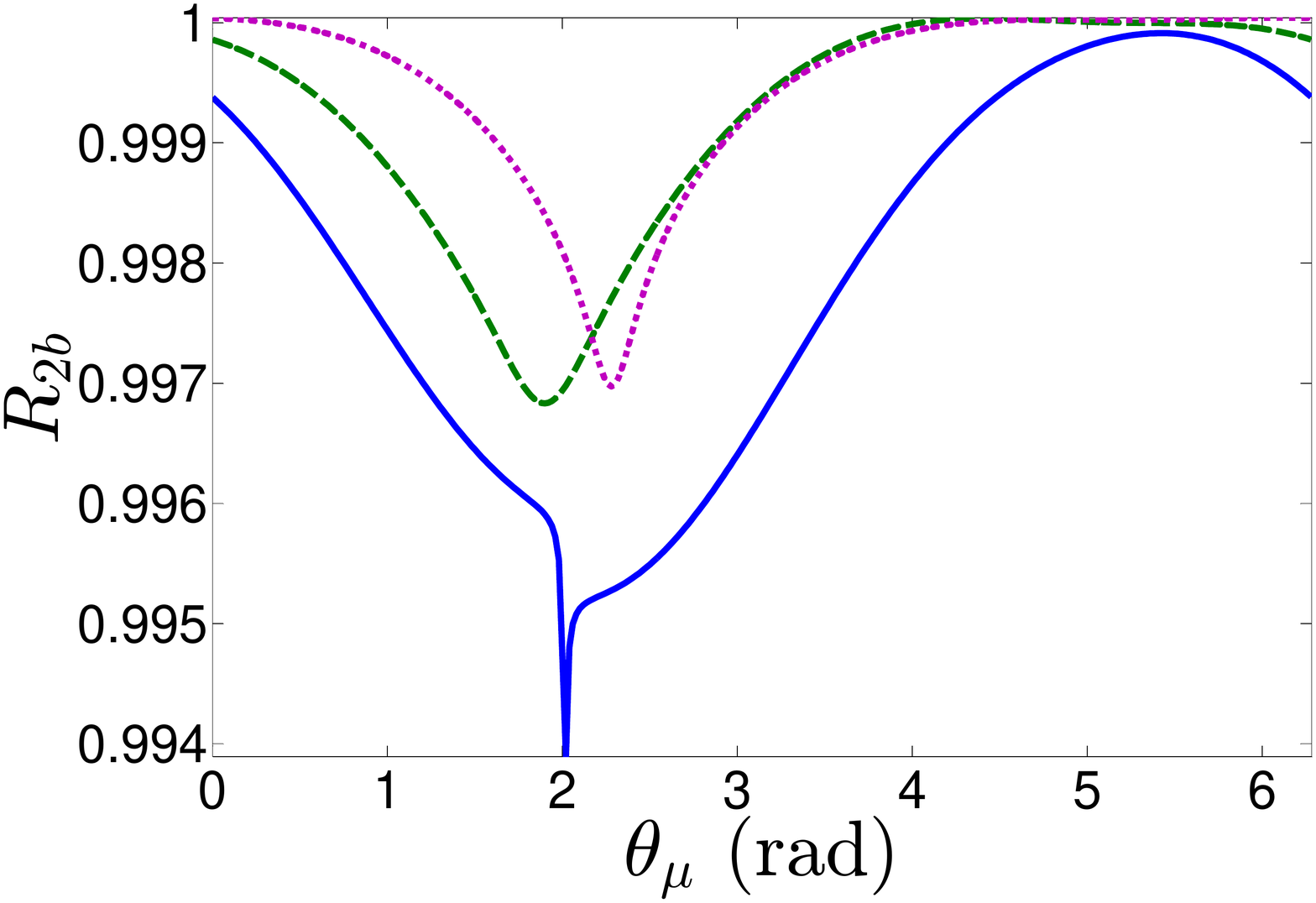}}\hglue5mm}}
{\rotatebox{0}{\resizebox*{3.5cm}{!}{\includegraphics{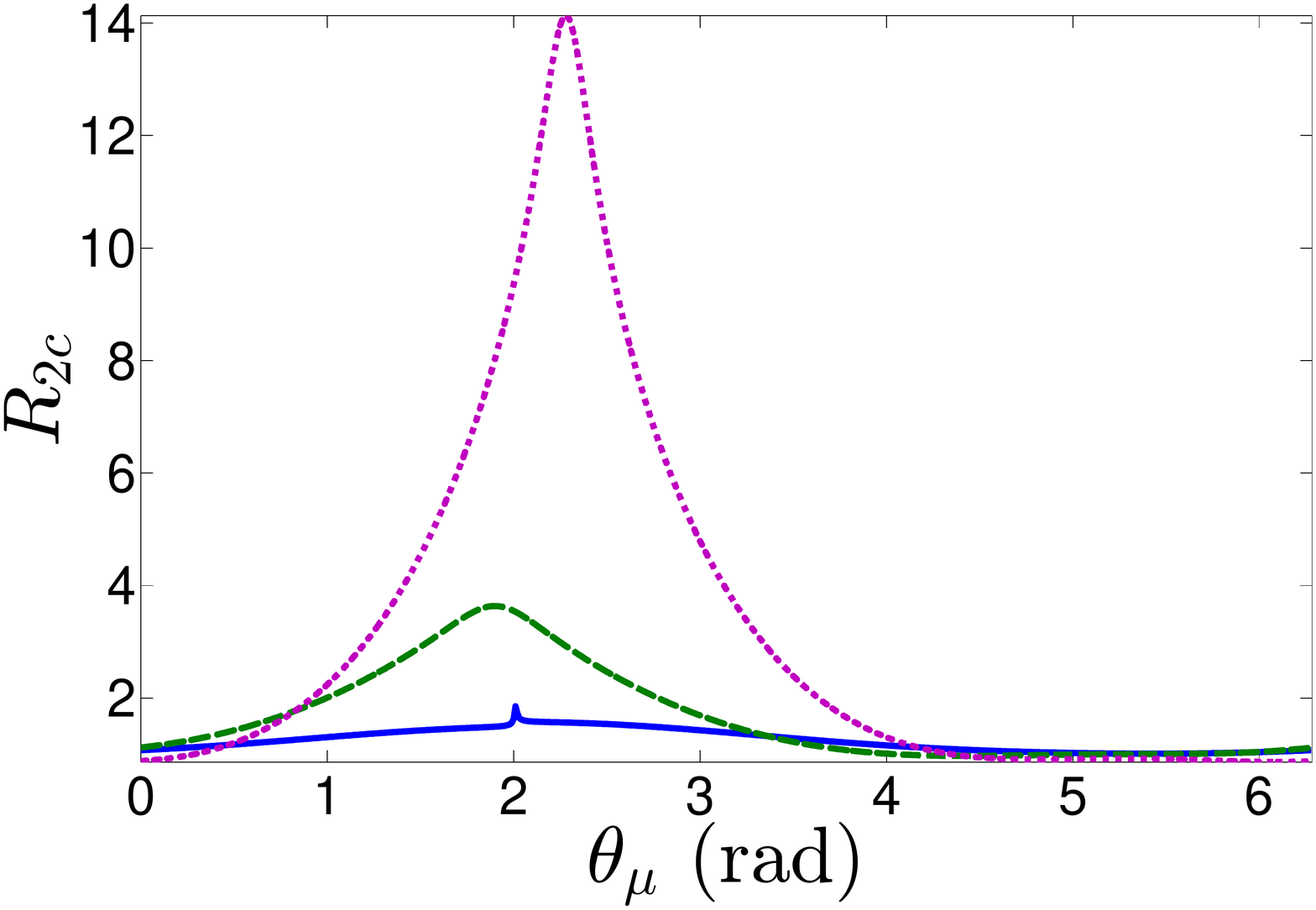}}\hglue5mm}}
{\rotatebox{0}{\resizebox*{3.5cm}{!}{\includegraphics{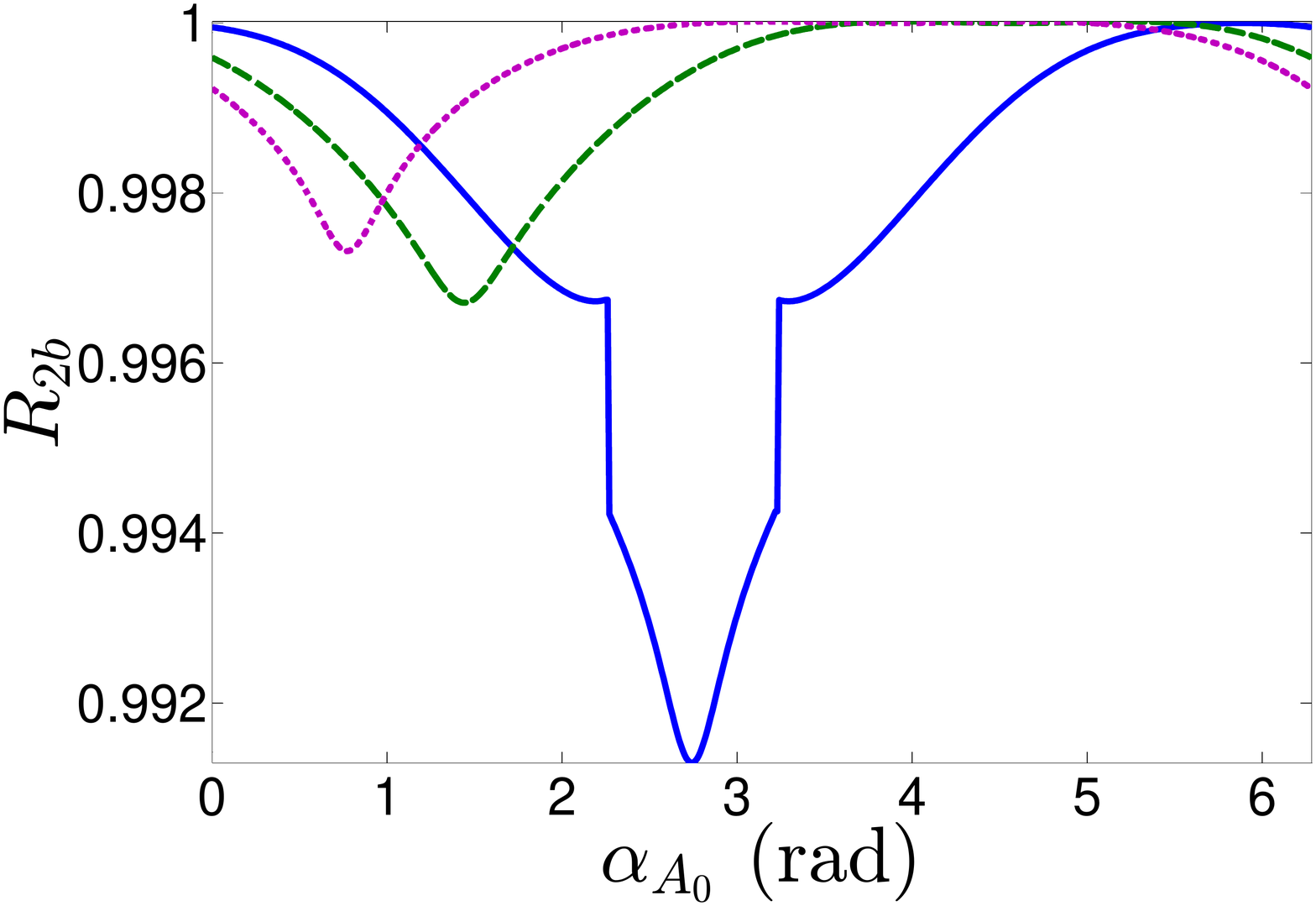}}\hglue5mm}}
{\rotatebox{0}{\resizebox*{3.5cm}{!}{\includegraphics{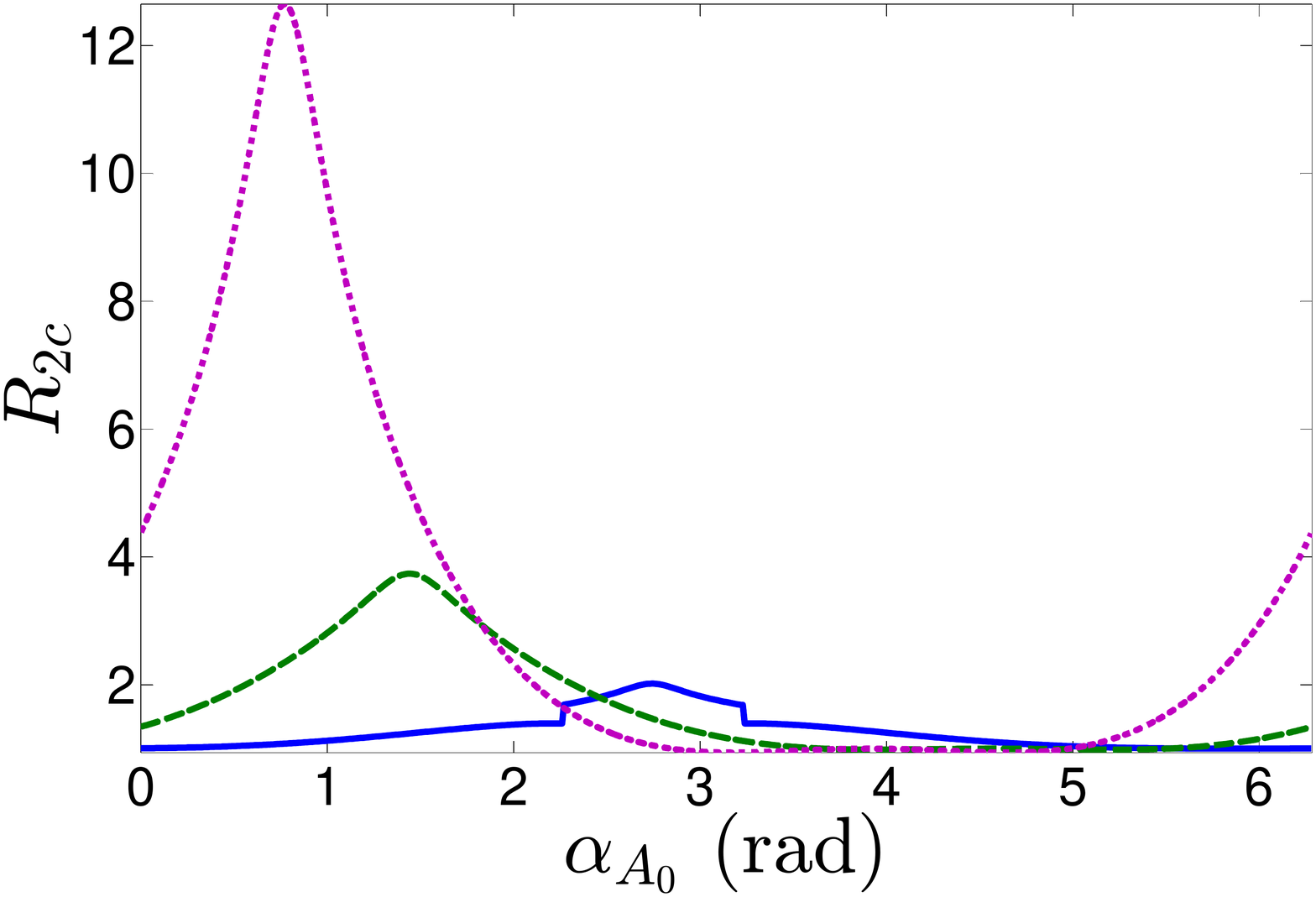}}\hglue5mm}}
\caption{Left panel: Variation of the $R_{2b}$ versus $\theta_{\mu}$ for the case with the contributions of the vectorlike generation.
Second left panel: Variation of the $R_{2c}$ versus $\theta_{\mu}$ for the case with the contributions of the vectorlike generation.
The input corresponding to point 2 (solid curve), point 4 (dashed curve) and point 6 (dotted curve) in table~\ref{table:4}.
Second right panel: Variation of the $R_{2b}$ versus $\alpha_{A_0}$ ($\alpha_{A_0}=\alpha_{A^u_0}=\alpha_{A^d_0}$) for the case with the contributions of the vectorlike generation.
Right panel: Variation of the $R_{2c}$ versus $\alpha_{A_0}$  for the case with the contributions of the vectorlike generation.
The input corresponding to point 2 (solid curve), point 4 (dashed curve) and point 6 (dotted curve) in table~\ref{table:4}.}
\label{fig21}
\end{center}
\end{figure}
\begin{figure}[H]
\begin{center}
{\rotatebox{0}{\resizebox*{3.5cm}{!}{\includegraphics{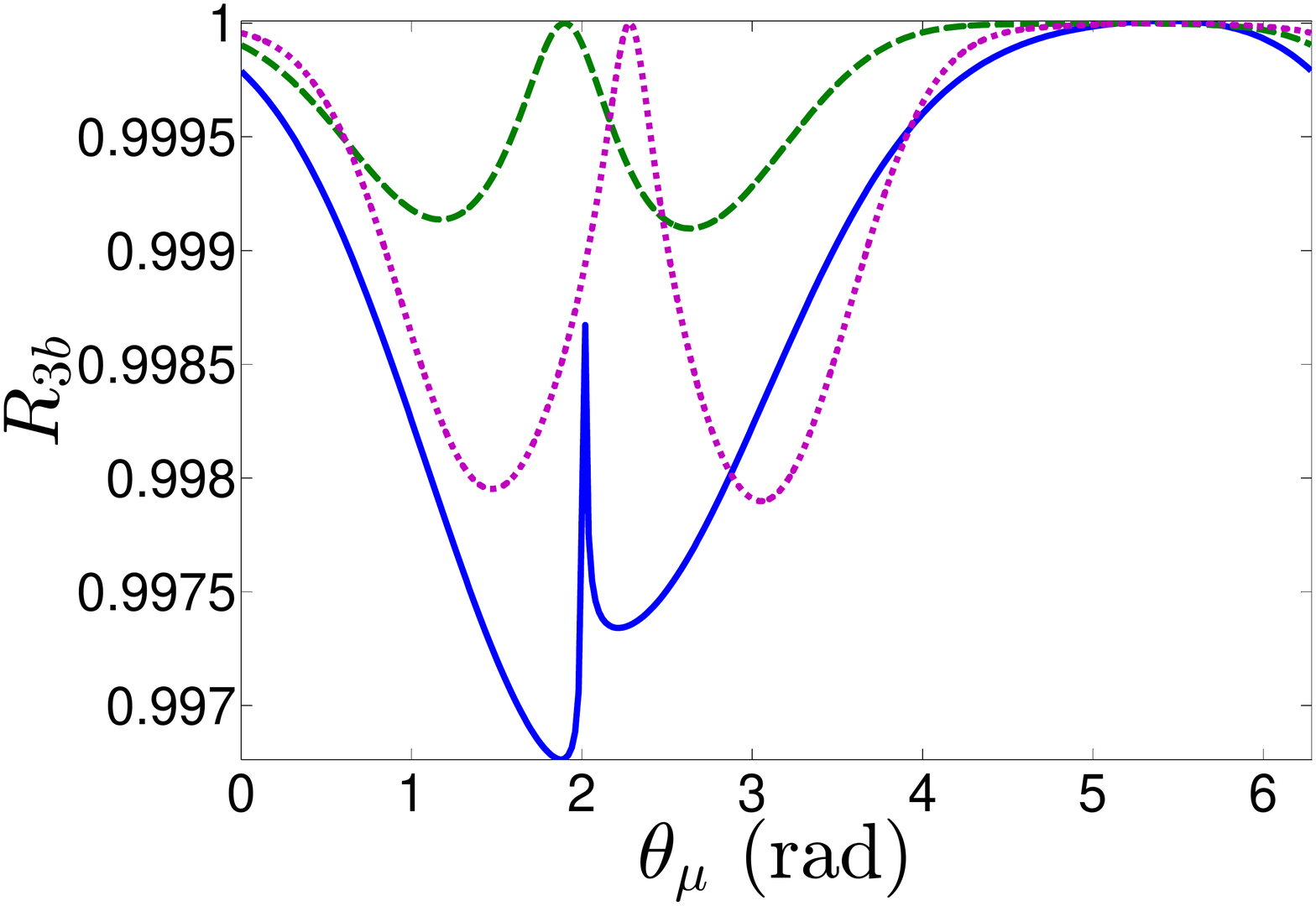}}\hglue5mm}}
{\rotatebox{0}{\resizebox*{3.5cm}{!}{\includegraphics{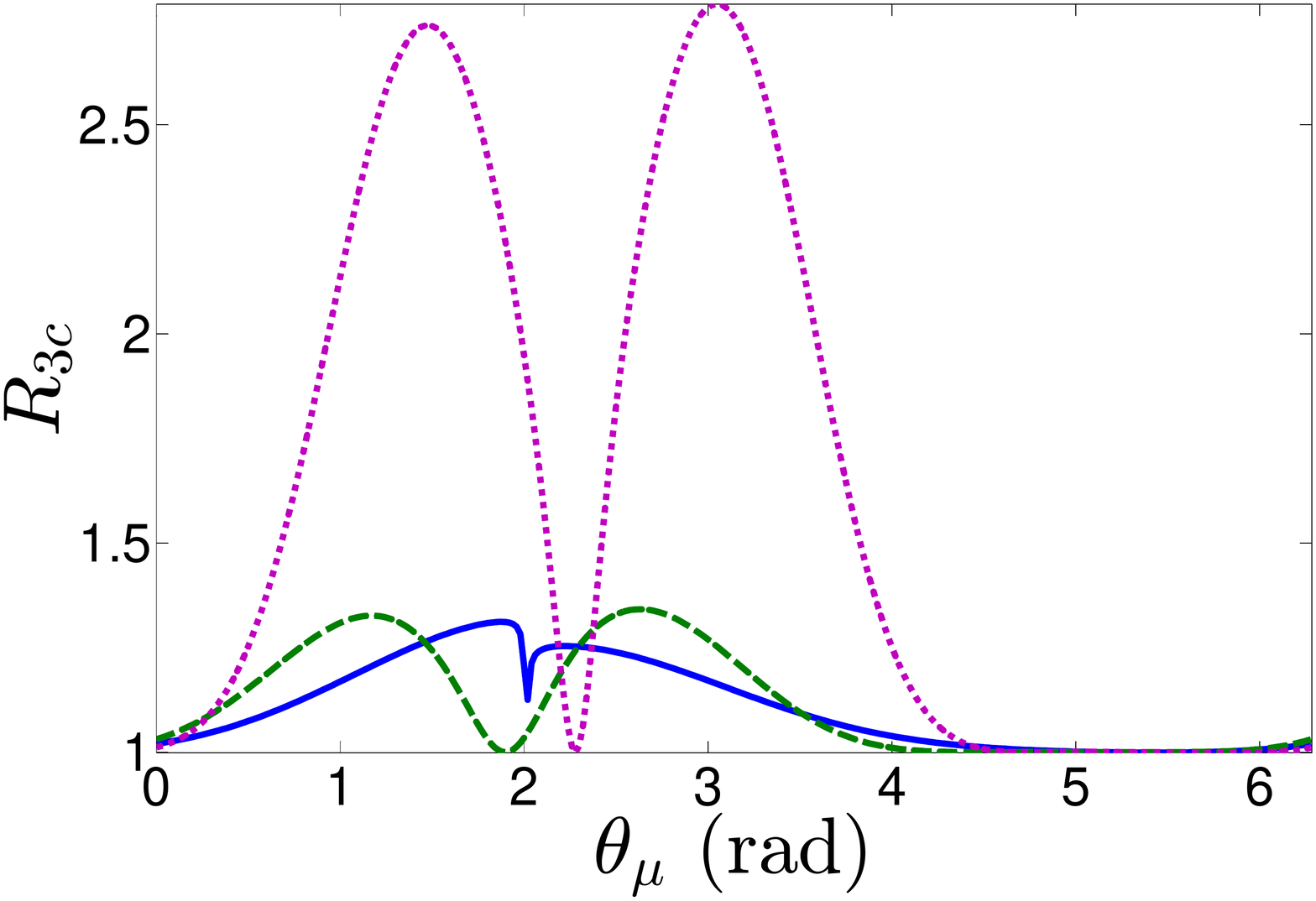}}\hglue5mm}}
{\rotatebox{0}{\resizebox*{3.5cm}{!}{\includegraphics{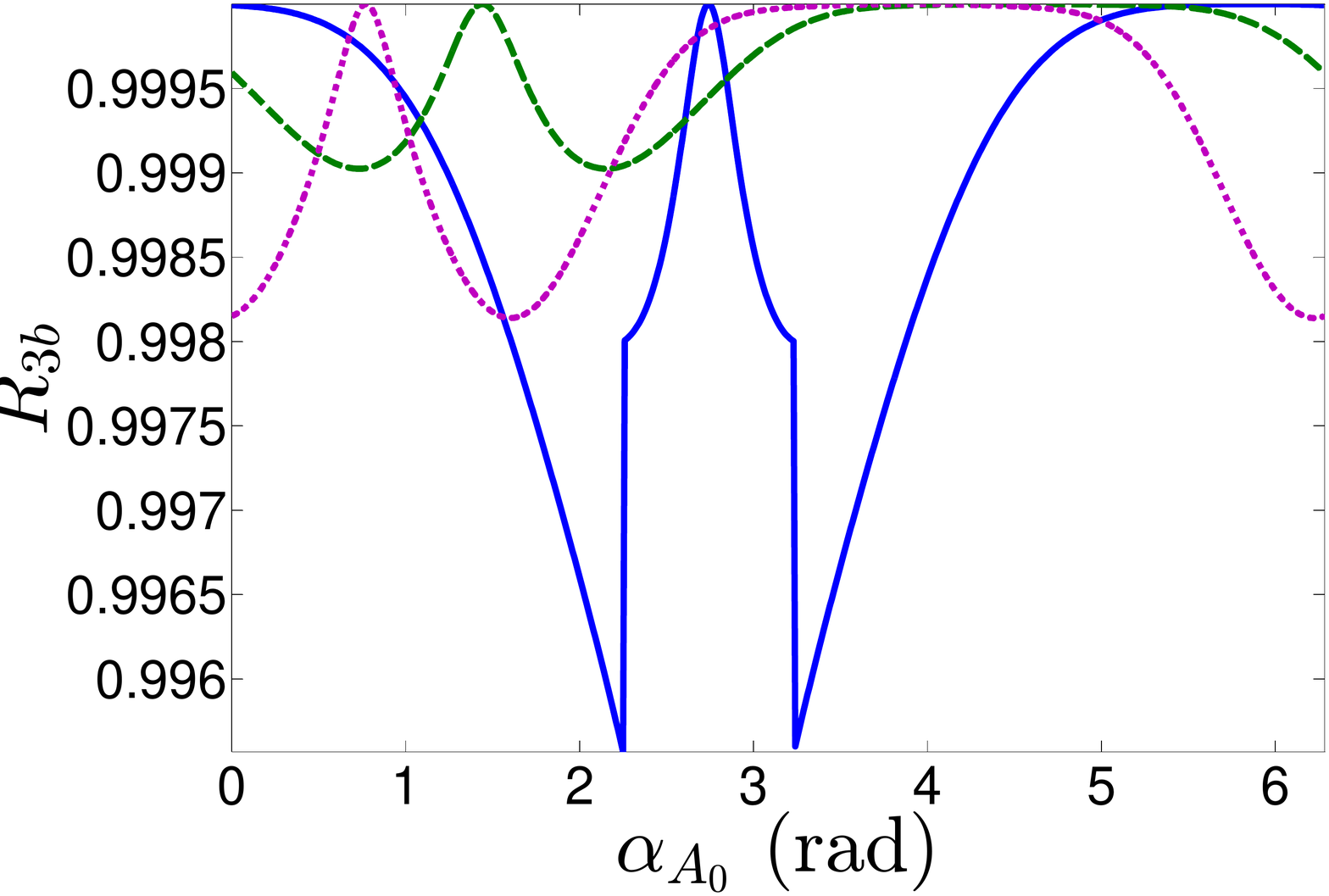}}\hglue5mm}}
{\rotatebox{0}{\resizebox*{3.5cm}{!}{\includegraphics{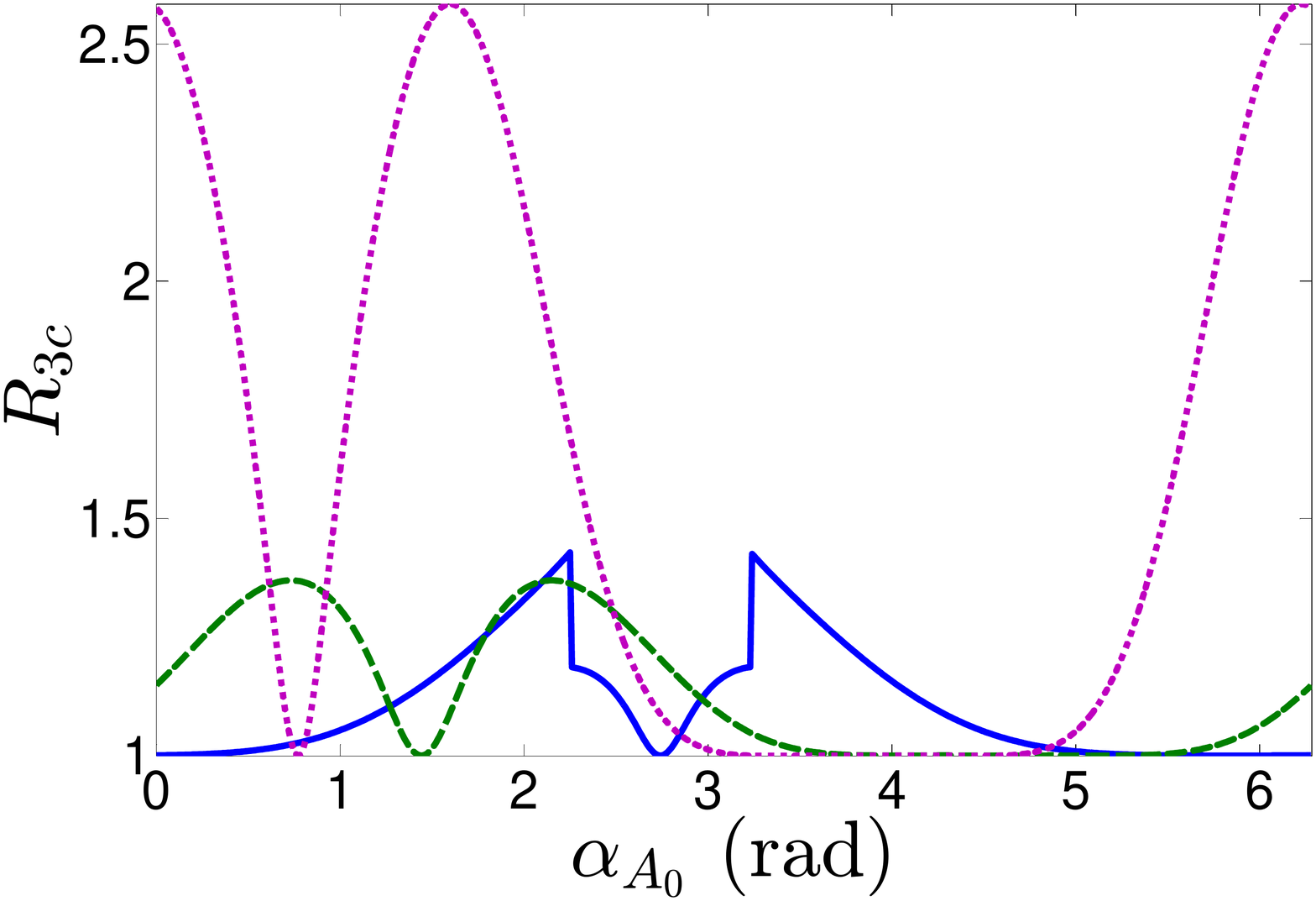}}\hglue5mm}}
\caption{Left panel: Variation of the $R_{3b}$ versus $\theta_{\mu}$ for the case with the contributions of the vectorlike generation.
Second  left panel: Variation of the $R_{3c}$ versus $\theta_{\mu}$ for the case with the contributions of the vectorlike generation.
The input corresponding to point 2 (solid curve), point 4 (dashed curve) and point 6 (dotted curve) in table~\ref{table:4}.
Second right panel: Variation of the $R_{3b}$ versus $\alpha_{A_0}$ ($\alpha_{A_0}=\alpha_{A^u_0}=\alpha_{A^d_0}$) for the case with the contributions of the vectorlike generation.
Right panel: Variation of the $R_{3c}$ versus $\alpha_{A_0}$  for the case with the contributions of the vectorlike generation.
The input corresponding to point 2 (solid curve), point 4 (dashed curve) and point 6 (dotted curve) in table~\ref{table:4}.}
\label{fig23}
\end{center}
\end{figure}

\section{Conclusion\label{sec6}}
An important phenomenon
in supersymmetric  models with inclusion of explicit CP violation relates to the mixing of CP even and CP
odd Higgs bosons. In this work we have investigated the implication of a vectorlike quark multiplet on the
CP even-CP odd mixing within an extended MSSM model. The sector brings with it new sources of CP violation
and our analysis shows that the vectorlike multiplet can generate substantial CP even-CP odd Higgs mixing
even in regions  where the mixing from the MSSM sector is small.
We have investigated the dependence of the mixings on the phases
and find that large mixings can occur in certain regions of the parameter space of CP phases.
The decays of the Higgs bosons into fermions are sensitive to new physics. We have investigated
these decays for the case of MSSM and for the case when one has in addition a vectorlike multiplet.
Further, for the latter case we have investigated the dependence of the Higgs decays widths into fermions
as a function of CP phases. These decays show a sharp dependence on the phase of $\mu$ and on the phase of the trilinear coupling. These results are of interest regarding the new
data expected from the LHC and the search for the heavy Higgs bosons.

\textbf{Acknowledgments: }
This research was supported in part by the NSF Grant PHY-1314774.\\

\section{Appendix: Squark mass  matrices}
In this Appendix we give further details of the model discussed in  section \ref{sec2}. As discussed in
section \ref{sec2} we allow for mixing between the vector generation and specifically the mirrors  and the standard three generations of quarks. The superpotential allowing such mixings is given by

\begin{align}
W&=\epsilon_{ij}  [y_{1}  \hat H_1^{i} \hat q_{1L} ^{j}\hat b^c_{1L}
 +y_{1}'  \hat H_2^{j}  \hat q_{1L} ^{i}\hat t^c_{1L}
+y_{2}  \hat H_1^{i} \hat Q^c{^{j}}\hat T_{L}
+y_{2}'  \hat H_2^{j} \hat Q^c{^{i}}\hat B_{L}\nonumber \\
 &+y_{3}  \hat H_1^{i} \hat q_{2L} ^{j}\hat b^c_{2L}
 +y_{3}'  \hat H_2^{j}  \hat q_{2L} ^{i}\hat t^c_{2L}
 +y_{4}  \hat H_1^{i} \hat q_{3L} ^{j}\hat b^c_{3L}
 +y_{4}'  \hat H_2^{j}  \hat q_{3L} ^{i}\hat t^c_{3L}
+y_{5}  \hat H_1^{i} \hat q_{4L} ^{j}\hat b^c_{4L}
+y_{5}'  \hat H_2^{j}  \hat q_{4L} ^{i}\hat t^c_{4L}
] \nonumber \\
&+ h_{3} \epsilon_{ij}  \hat Q^c{^{i}}\hat q_{1L}^{j}+
h_{3}' \epsilon_{ij}  \hat Q^c{^{i}}\hat q_{2L}^{j}+
h_{3}'' \epsilon_{ij}  \hat Q^c{^{i}}\hat q_{3L}^{j}+
h_{6} \epsilon_{ij}  \hat Q^c{^{i}}\hat q_{4L}^{j}
+h_4 \hat b_{1L}^c \hat B_{L} +h_5 \hat t_{1L}^c \hat T_{L}\nonumber\\
&+h_4'  \hat b_{2L}^c \hat B_{L} +h_5' \hat t_{2L}^c \hat T_{L}
+h_4''  \hat b_{3L}^c \hat B_{L} +h_5'' \hat t_{3L}^c \hat T_{L}
+h_7  \hat b_{4L}^c \hat B_{L} +h_8 \hat t_{4L}^c \hat T_{L}
  -\mu \epsilon_{ij} \hat H_1^i \hat H_2^j \ ,
 \label{7w}
\end{align}
Here the couplings are in general complex. Thus, for example,
 $\mu$ is the complex Higgs mixing parameter so that $\mu= |\mu| e^{i\theta_\mu}$.
The mass terms for the ups, mirror ups,  downs and  mirror downs arise from the term
\beq
{\cal{L}}=-\frac{1}{2}\frac{\partial ^2 W}{\partial{A_i}\partial{A_j}}\psi_ i \psi_ j+\text{h.c.},
\label{6}
\eeq
where $\psi$ and $A$ stand for generic two-component fermion and scalar fields.
After spontaneous breaking of the electroweak symmetry, ($\langle H_1^1 \rangle=v_1/\sqrt{2} $ and $\langle H_2^2\rangle=v_2/\sqrt{2}$),
we have the following set of mass terms written in the four-component spinor notation
so that
\beq
-{\cal L}_m= \bar\xi_R^T (M_u) \xi_L +\bar\eta_R^T(M_{d}) \eta_L +\text{h.c.},
\eeq
where the basis vectors are defined in  Eq. \ref{basis-xi} and Eq. \ref{basis-eta}.

  Next we  consider  the mixing of the down squarks and the charged mirror sdowns.
The mass squared  matrix of the sdown - mirror sdown comes from three sources:  the F term, the
D term of the potential and the soft {SUSY} breaking terms.
Using the  superpotential of  the mass terms arising from it
after the breaking of  the electroweak symmetry are given by
the Lagrangian
\beq
{\cal L}= {\cal L}_F +{\cal L}_D + {\cal L}_{\rm soft}\ ,
\eeq
where   $ {\cal L}_F$ is deduced from $F_i =\partial W/\partial A_i$, and $- {\cal L}_F=V_F=F_i F^{*}_i$ while the ${\cal L}_D$ is given by
\begin{align}
-{\cal L}_D&=\frac{1}{2} m^2_Z \cos^2\theta_W \cos 2\beta \{\tilde t_{ L} \tilde t^*_{ L} -\tilde b_L \tilde b^*_L
+\tilde c_{ L} \tilde c^*_{ L} -\tilde s_L \tilde s^*_L
+\tilde u_{ L} \tilde u^*_{ L} -\tilde d_L \tilde d^*_L
+\tilde t_{4 L} \tilde t^*_{4 L} -\tilde b_{4L} \tilde b^*_{4L}
 \nonumber \\
&+\tilde B_R \tilde B^*_R -\tilde T_R \tilde T^*_R\}
+\frac{1}{2} m^2_Z \sin^2\theta_W \cos 2\beta \{-\frac{1}{3}\tilde t_{ L} \tilde t^*_{ L}
 +\frac{4}{3}\tilde t_{ R} \tilde t^*_{ R}
-\frac{1}{3}\tilde c_{ L} \tilde c^*_{ L}
 +\frac{4}{3}\tilde c_{ R} \tilde c^*_{ R} \nonumber \\
&-\frac{1}{3}\tilde u_{ L} \tilde u^*_{ L}
 +\frac{4}{3}\tilde u_{ R} \tilde u^*_{ R}
+\frac{1}{3}\tilde T_{ R} \tilde T^*_{ R}
 -\frac{4}{3}\tilde T_{ L} \tilde T^*_{ L}
-\frac{1}{3}\tilde b_{ L} \tilde b^*_{ L}
 -\frac{2}{3}\tilde b_{ R} \tilde b^*_{ R}\nonumber\\
&-\frac{1}{3}\tilde s_{ L} \tilde s^*_{ L}
 -\frac{2}{3}\tilde s_{ R} \tilde s^*_{ R}
-\frac{1}{3}\tilde d_{ L} \tilde d^*_{ L}
 -\frac{2}{3}\tilde d_{ R} \tilde d^*_{ R}
+\frac{1}{3}\tilde B_{ R} \tilde B^*_{ R}\nonumber\\
&+\frac{2}{3}\tilde B_{ L} \tilde B^*_{ L}
-\frac{1}{3}\tilde t_{4 L} \tilde t^*_{4 L}
 +\frac{4}{3}\tilde t_{ 4R} \tilde t^*_{4 R}
-\frac{1}{3}\tilde b_{4 L} \tilde b^*_{4 L}
 -\frac{2}{3}\tilde b_{ 4R} \tilde b^*_{ 4R}
\}.
\label{12}
\end{align}
For ${\cal L}_{\rm soft}$ we assume the following form
\begin{align}
-{\cal L}_{\text{soft}}&= M^2_{\tilde 1 L} \tilde q^{k*}_{1 L} \tilde q^k_{1 L}
+ M^2_{\tilde 4 L} \tilde q^{k*}_{4 L} \tilde q^k_{4 L}
+ M^2_{\tilde 2 L} \tilde q^{k*}_{2 L} \tilde q^k_{2 L}
+ M^2_{\tilde 3 L} \tilde q^{k*}_{3 L} \tilde q^k_{3 L}
+ M^2_{\tilde Q} \tilde Q^{ck*} \tilde Q^{ck}
 + M^2_{\tilde t_1} \tilde t^{c*}_{1 L} \tilde t^c_{1 L} \nonumber \\
& + M^2_{\tilde b_1} \tilde b^{c*}_{1 L} \tilde b^c_{1 L}
+ M^2_{\tilde t_2} \tilde t^{c*}_{2 L} \tilde t^c_{2 L}
+ M^2_{\tilde b_4} \tilde b^{c*}_{4 L} \tilde b^c_{4 L}
+ M^2_{\tilde t_4} \tilde t^{c*}_{4 L} \tilde t^c_{4 L}\nonumber\\
&+ M^2_{\tilde t_3} \tilde t^{c*}_{3 L} \tilde t^c_{3 L}
+ M^2_{\tilde b_2} \tilde b^{c*}_{2 L} \tilde b^c_{2 L}
+ M^2_{\tilde b_3} \tilde b^{c*}_{3 L} \tilde b^c_{3 L}
+ M^2_{\tilde B} \tilde B^*_L \tilde B_L
 +  M^2_{\tilde T} \tilde T^*_L \tilde T_L \nonumber \\
&+\epsilon_{ij} \{y_1 A_{b} H^i_1 \tilde q^j_{1 L} \tilde b^c_{1L}
-y_1' A_{t} H^i_2 \tilde q^j_{1 L} \tilde t^c_{1L}
+y_5 A_{b_4} H^i_1 \tilde q^j_{4 L} \tilde b^c_{4L}
-y_5' A_{t_4} H^i_2 \tilde q^j_{4 L} \tilde t^c_{4L}
+y_3 A_{s} H^i_1 \tilde q^j_{2 L} \tilde b^c_{2L}\nonumber\\
&-y_3' A_{c} H^i_2 \tilde q^j_{2 L} \tilde t^c_{2L}
+y_4 A_{d} H^i_1 \tilde q^j_{3 L} \tilde b^c_{3L}
-y_4' A_{u} H^i_2 \tilde q^j_{3 L} \tilde t^c_{3L}
+y_2 A_{T} H^i_1 \tilde Q^{cj} \tilde T_{L}
-y_2' A_{B} H^i_2 \tilde Q^{cj} \tilde B_{L}
+\text{h.c.}\}\ .
\label{13}
\end{align}
Here $M_{\tilde 1 L},  M_{\tilde T}$, etc are the soft masses and $A_t, A_{b}$, etc are the trilinear couplings.
The trilinear couplings are complex  and we define their phases so that
\begin{gather}
A_b= |A_b| e^{i \alpha_{A_b}} \  ,
 ~~A_{t}=  |A_{t}|
 e^{i\alpha_{A_{t}}} \ ,
  \cdots \ .
\end{gather}
From these terms we construct the scalar mass squared matrices.
Thus we  define the scalar mass squared   matrix $M^2_{\tilde{d}}$  in the basis $(\tilde  b_L, \tilde B_L, \tilde b_R,
\tilde B_R, \tilde s_L, \tilde s_R, \tilde d_L, \tilde d_R,
\tilde b_{4L}, \tilde b_{4R}
)$. We  label the matrix  elements of these as $(M^2_{\tilde d})_{ij}= M^2_{ij}$ which is a hermitian matrix.
We can diagonalize this hermitian mass squared  matrix  by the
 unitary transformation
\begin{gather}
 \tilde D^{d \dagger} M^2_{\tilde d} \tilde D^{d} = diag (M^2_{\tilde d_1},
M^2_{\tilde d_2}, M^2_{\tilde d_3},  M^2_{\tilde d_4},  M^2_{\tilde d_5},  M^2_{\tilde d_6},  M^2_{\tilde d_7},  M^2_{\tilde d_8}
 M^2_{\tilde d_9},  M^2_{\tilde d_{10}}
 )\ .
\end{gather}
Similarly we write the   mass squared   matrix in the up squark sector in the basis $(\tilde  t_{ L}, \tilde T_L,$
$ \tilde t_{ R}, \tilde T_R, \tilde  c_{ L},\tilde c_{ R}, \tilde u_{ L}, \tilde u_{R}
,\tilde t_{4 L}, \tilde t_{4R}
 )$.
 Thus here we denote the up squark  mass squared matrix in the form
$(M^2_{\tilde u})_{ij}=m^2_{ij}$ which is also a hermitian matrix.
We can diagonalize this mass square matrix  by the  unitary transformation
\begin{equation}
 \tilde D^{u\dagger} M^2_{\tilde u} \tilde D^{u} = \text{diag} (M^2_{\tilde u_1}, M^2_{\tilde u_2}, M^2_{\tilde u_3},  M^2_{\tilde u_4},M^2_{\tilde u_5},  M^2_{\tilde u_6}, M^2_{\tilde u_7}, M^2_{\tilde u_8}
, M^2_{\tilde u_9}, M^2_{\tilde u_{10}}
)\ .
\end{equation}

We  label the matrix  elements of these as $(M^2_{\tilde d})_{ij}= M^2_{ij}$ where the elements of the matrix are given by
\begin{align}
M^2_{11}&= M^2_{\tilde 1 L}+\frac{v^2_1|y_1|^2}{2} +|h_3|^2 -m^2_Z \cos 2 \beta \left(\frac{1}{2}-\frac{1}{3}\sin^2\theta_W\right), \nonumber\\
M^2_{22}&=M^2_{\tilde B}+\frac{v^2_2|y'_2|^2}{2}+|h_4|^2 +|h'_4|^2+|h''_4|^2
+|h_7|^2
+\frac{1}{3}m^2_Z \cos 2 \beta \sin^2\theta_W, \nonumber\\
M^2_{33}&= M^2_{\tilde b_1}+\frac{v^2_1|y_1|^2}{2} +|h_4|^2 -\frac{1}{3}m^2_Z \cos 2 \beta \sin^2\theta_W, \nonumber\\
M^2_{44}&=  M^2_{\tilde Q}+\frac{v^2_2|y'_2|^2}{2} +|h_3|^2 +|h'_3|^2+|h''_3|^2
+|h_6|^2
 +m^2_Z \cos 2 \beta \left(\frac{1}{2}-\frac{1}{3}\sin^2\theta_W\right), \nonumber
\end{align}
\begin{align}
M^2_{55}&=M^2_{\tilde 2 L} +\frac{v^2_1|y_3|^2}{2} +|h'_3|^2 -m^2_Z \cos 2 \beta \left(\frac{1}{2}-\frac{1}{3}\sin^2\theta_W\right), \nonumber\\
M^2_{66}&= M^2_{\tilde b_2}+\frac{v^2_1|y_3|^2}{2}+|h'_4|^2  -\frac{1}{3}m^2_Z \cos 2 \beta \sin^2\theta_W,\nonumber\\
M^2_{77}&=M^2_{\tilde 3 L}+\frac{v^2_1|y_4|^2}{2}+|h''_3|^2-m^2_Z \cos 2 \beta \left(\frac{1}{2}-\frac{1}{3}\sin^2\theta_W\right),  \nonumber\\
M^2_{88}&= M^2_{\tilde b_3}+\frac{v^2_1|y_4|^2}{2}+|h''_4|^2   -\frac{1}{3}m^2_Z \cos 2 \beta \sin^2\theta_W\ . \nonumber\\
M^2_{99}&=M^2_{\tilde 4 L}+\frac{v^2_1|y_5|^2}{2}+|h_6|^2-m^2_Z \cos 2 \beta \left(\frac{1}{2}-\frac{1}{3}\sin^2\theta_W\right)\nonumber\\
M^2_{1010}&= M^2_{\tilde b_4}+\frac{v^2_1|y_5|^2}{2}+|h_7|^2   -\frac{1}{3}m^2_Z \cos 2 \beta \sin^2\theta_W\ . \nonumber\\
\end{align}

\begin{align}
M^2_{12}=M^{2*}_{21}=\frac{ v_2 y'_2h^*_3}{\sqrt{2}} +\frac{ v_1 h_4 y^*_1}{\sqrt{2}} ,
M^2_{13}=M^{2*}_{31}=\frac{y^*_1}{\sqrt{2}}(v_1 A^*_{b} -\mu v_2),
M^2_{14}=M^{2*}_{41}=0,\nonumber\\
 M^2_{15} =M^{2*}_{51}=h'_3 h^*_3,
 M^{2}_{16}= M^{2*}_{61}=0,  M^{2}_{17}= M^{2*}_{71}=h''_3 h^*_3,  M^{2}_{18}= M^{2*}_{81}=0,
M^{2}_{19}=M^{2*}_{91}=h^*_3 h_6,
\nonumber\\
M^{2}_{110}=M^{2*}_{101}=0,
M^2_{23}=M^{2*}_{32}=0,
M^2_{24}=M^{2*}_{42}=\frac{y'^*_2}{\sqrt{2}}(v_2 A^*_{B} -\mu v_1),  M^2_{25} = M^{2*}_{52}= \frac{ v_2 h'_3y'^*_2}{\sqrt{2}} +\frac{ v_1 y_3 h^*_4}{\sqrt{2}} ,\nonumber\\
 M^2_{26}=M^{2*}_{62}=0,  M^2_{27} =M^{2*}_{72}=  \frac{ v_2 h''_3y'^*_2}{\sqrt{2}} +\frac{ v_1 y_4 h''^*_4}{\sqrt{2}},  M^2_{28} =M^{2*}_{82}=0, \nonumber\\
 M^2_{29} =M^{2*}_{92}=  \frac{ v_1 h^*_7y_5}{\sqrt{2}} +\frac{ v_2 y'^*_2 h_6}{\sqrt{2}},
M^{2}_{210}=M^{2*}_{102}=0,\nonumber\\
M^2_{34}=M^{2*}_{43}= \frac{ v_2 h_4 y'^*_2}{\sqrt{2}} +\frac{ v_1 y_1 h^*_3}{\sqrt{2}}, M^2_{35} =M^{2*}_{53} =0, M^2_{36} =M^{2*}_{63}=h_4 h'^*_4,\nonumber\\
 M^2_{37} =M^{2*}_{73} =0,  M^2_{38} =M^{2*}_{83} =h_4 h''^*_4,\nonumber\\
M^{2}_{39}=M^{2*}_{93}=0,
M^{2}_{310}=M^{2*}_{103}=h_4 h^*_7,\nonumber\\
M^2_{45}=M^{2*}_{54}=0, M^2_{46}=M^{2*}_{64}=\frac{ v_2 y'_2 h'^*_4}{\sqrt{2}} +\frac{ v_1 h'_3 y^*_3}{\sqrt{2}}, \nonumber\\
 M^2_{47} =M^{2*}_{74}=0,  M^2_{48} =M^{2*}_{84}=  \frac{ v_2 y'_2h''^*_4}{\sqrt{2}} +\frac{ v_1 h''_3 y^*_4}{\sqrt{2}},\nonumber\\
M^{2}_{49}=M^{2*}_{94}=0,
 M^2_{410} =M^{2*}_{104}=  \frac{ v_2 y'_2h^*_7}{\sqrt{2}} +\frac{ v_1 h_6 y^*_5}{\sqrt{2}},\nonumber\\
M^2_{56}=M^{2*}_{65}=\frac{y^*_3}{\sqrt{2}}(v_1 A^*_{s} -\mu v_2),
 M^2_{57} =M^{2*}_{75}=h''_3 h'^*_3,  \nonumber\\
 M^2_{58} =M^{2*}_{85}=0,
M^2_{59} =M^{2*}_{95}=h'^*_3 h_6,
 M^2_{510} =M^{2*}_{105}=0,
  M^2_{67} =M^{2*}_{76}=0,\nonumber\\
 M^2_{68} =M^{2*}_{86}=h'_4 h''^*_4,
 M^2_{69} =M^{2*}_{96}=0,
 M^2_{610} =M^{2*}_{106}=h'_4 h^*_7,
  M^2_{78}=M^{2*}_{87}=\frac{y^*_4}{\sqrt{2}}(v_1 A^*_{d} -\mu v_2)\ . \nonumber\\
 M^2_{79} =M^{2*}_{97}=h''^*_3 h_6,
 M^2_{710} =M^{2*}_{107}=0\nonumber\\
 M^2_{89} =M^{2*}_{98}=0,
 M^2_{810} =M^{2*}_{108}=h''_4 h^*_7,
  M^2_{910}=M^{2*}_{109}=\frac{y^*_5}{\sqrt{2}}(v_1 A^*_{b_4} -\mu v_2)\ . \nonumber
\label{14}
\end{align}

We can diagonalize this hermitian mass$^2$  matrix of the scalar downs  by the
 unitary transformation
\begin{gather}
 \tilde D^{d \dagger} M^2_{\tilde d} \tilde D^{d} = diag (M^2_{\tilde d_1},
M^2_{\tilde d_2}, M^2_{\tilde d_3},  M^2_{\tilde d_4},  M^2_{\tilde d_5},  M^2_{\tilde d_6},  M^2_{\tilde d_7},  M^2_{\tilde d_8},
M^2_{\tilde d_9},
M^2_{\tilde d_{10}}
 )\ .
\end{gather}


 Next we write the   mass$^2$  matrix in the sups sector the basis $(\tilde  t_{ L}, \tilde T_L,$
$ \tilde t_{ R}, \tilde T_R, \tilde  c_{ L},\tilde c_{ R}, \tilde u_{ L}, \tilde u_{R},
 \tilde t _{4 L}, \tilde t_{4R}
 )$.
 Thus here we denote the sups mass$^2$ matrix in the form
$(M^2_{\tilde u})_{ij}=m^2_{ij}$ where

\begin{align}
m^2_{11}&= M^2_{\tilde 1 L}+\frac{v^2_2|y'_1|^2}{2} +|h_3|^2 +m^2_Z \cos 2 \beta \left(\frac{1}{2}-\frac{2}{3}\sin^2\theta_W\right), \nonumber\\
m^2_{22}&=M^2_{\tilde T}+\frac{v^2_1|y_2|^2}{2}+|h_5|^2 +|h'_5|^2+|h''_5|^2
+|h_8|^2
 -\frac{2}{3}m^2_Z \cos 2 \beta \sin^2\theta_W, \nonumber\\
m^2_{33}&= M^2_{\tilde t_1}+\frac{v^2_2|y'_1|^2}{2} +|h_5|^2 +\frac{2}{3}m^2_Z \cos 2 \beta \sin^2\theta_W, \nonumber\\
m^2_{44}&=  M^2_{\tilde Q}+\frac{v^2_1|y_2|^2}{2} +|h_3|^2 +|h'_3|^2+|h''_3|^2
+|h_6|^2
 -m^2_Z \cos 2 \beta \left(\frac{1}{2}-\frac{2}{3}\sin^2\theta_W\right), \nonumber
\end{align}
\begin{align}
m^2_{55}&=M^2_{\tilde 2 L} +\frac{v^2_2|y'_3|^2}{2} +|h'_3|^2 +m^2_Z \cos 2 \beta \left(\frac{1}{2}-\frac{2}{3}\sin^2\theta_W\right), \nonumber\\
m^2_{66}&= M^2_{\tilde t_2}+\frac{v^2_2|y'_3|^2}{2}+|h'_5|^2  +\frac{2}{3}m^2_Z \cos 2 \beta \sin^2\theta_W,\nonumber\\
m^2_{77}&=M^2_{\tilde 3 L}+\frac{v^2_2|y'_4|^2}{2}+|h''_3|^2+m^2_Z \cos 2 \beta \left(\frac{1}{2}-\frac{2}{3}\sin^2\theta_W\right),  \nonumber\\
m^2_{88}&= M^2_{\tilde t_3}+\frac{v^2_2|y'_4|^2}{2}+|h''_5|^2   +\frac{2}{3}m^2_Z \cos 2 \beta \sin^2\theta_W,\nonumber\\
m^2_{99}&=M^2_{\tilde 4 L}+\frac{v^2_2|y'_5|^2}{2}+|h_6|^2+m^2_Z \cos 2 \beta \left(\frac{1}{2}-\frac{2}{3}\sin^2\theta_W\right),  \nonumber\\
m^2_{1010}&= M^2_{\tilde t_4}+\frac{v^2_2|y'_5|^2}{2}+|h_8|^2   +\frac{2}{3}m^2_Z \cos 2 \beta \sin^2\theta_W.\
\nonumber
\end{align}

\begin{align}
m^2_{12}&=m^{2*}_{21}=-\frac{ v_1 y_2h^*_3}{\sqrt{2}} +\frac{ v_2 h_5 y'^*_1}{\sqrt{2}} ,
m^2_{13}=m^{2*}_{31}=\frac{y'^*_1}{\sqrt{2}}(v_2 A^*_{t} -\mu v_1),
m^2_{14}=m^{2*}_{41}=0,\nonumber\\
 m^2_{15} &=m^{2*}_{51}=h'_3 h^*_3,
 m^{2}_{16}= m^{2*}_{61}=0,  m^{2*}_{17}= m^{2*}_{71}=h''_3 h^*_3,  m^{2*}_{18}= m^{2*}_{81}=0,\nonumber\\
m^2_{23}&=m^{2*}_{32}=0,
m^2_{24}=m^{2*}_{42}=\frac{y^*_2}{\sqrt{2}}(v_1 A^*_{T} -\mu v_2),  m^2_{25} = m^{2*}_{52}= -\frac{ v_1 h'_3y^*_2}{\sqrt{2}} +\frac{ v_2 y'_3 h'^*_5}{\sqrt{2}} ,\nonumber\\
 m^2_{26} &=m^{2*}_{62}=0,  m^2_{27} =m^{2*}_{72}=  -\frac{ v_1 h''_3y^*_2}{\sqrt{2}} +\frac{ v_2 y'_4 h''^*_5}{\sqrt{2}},  m^2_{28} =m^{2*}_{82}=0, \nonumber\\
m^2_{34}&=m^{2*}_{43}= \frac{ v_1 h_5 y^*_2}{\sqrt{2}} -\frac{ v_2 y'_1 h^*_3}{\sqrt{2}}, m^2_{35} =m^{2*}_{53} =0, m^2_{36} =m^{2*}_{63}=h_5 h'^*_5,\nonumber\\
 m^2_{37} &=m^{2*}_{73} =0,  m^2_{38} =m^{2*}_{83} =h_5 h''^*_5,\nonumber\\
m^2_{45}&=m^{2*}_{54}=0, m^2_{46}=m^{2*}_{64}=-\frac{y'^*_3 v_2 h'_3}{\sqrt{2}}+\frac{v_1 y_2 h'^*_5}{\sqrt{2}},
\nonumber\\
m^2_{47}&=m^{2*}_{74}=0,
m^2_{48}=m^{2*}_{84}=\frac{v_1 y_2 h''^*_5}{\sqrt{2}}-\frac{v_2 y'^*_4 h''_3}{\sqrt{2}},\nonumber\\
 m^2_{56}&=m^{2*}_{65}=\frac{y'^*_3}{\sqrt{2}}(v_2 A^*_{c}-\mu v_1), \nonumber\\
m^2_{57}&=m^{2*}_{75}= h''_3 h'^*_3, m^2_{58}=m^{2*}_{85}=0, \nonumber\\
m^2_{67}&=m^{2*}_{76}=0, m^2_{68}=m^{2*}_{86}= h'_5 h''^*_5, \nonumber\\
m^2_{78}&=m^{2*}_{87}=\frac{y'^*_4}{\sqrt{2}}(v_2 A^*_{u}-\mu v_1),\nonumber\\
m^2_{19}&=m^{2*}_{91}=h_6 h^*_3, m^2_{110}=m^{2*}_{101}=0, \nonumber\\
m^2_{29}&=m^{2*}_{92}=-\frac{y^*_2 v_1 h_6}{\sqrt{2}}+\frac{v_2 y^*_5 h_8}{\sqrt{2}},\nonumber\\
m^2_{210}&=m^{2*}_{102}=0, m^2_{39}=m^{2*}_{93}=0,\nonumber\\
m^2_{310}&=m^{2*}_{103}=h_5 h^*_8,\nonumber\\
m^2_{49}&=m^{2*}_{94}=0, m^2_{410}=m^{2*}_{104}=
-\frac{y'^*_5 v_2 h_6}{\sqrt{2}}+\frac{v_1 y_2 h^*_8}{\sqrt{2}},\nonumber\\
m^2_{59}&=m^{2*}_{95}=h_6 h'^*_3, m^2_{510}=m^{2*}_{105}=0\nonumber\\
m^2_{69}&=m^{2*}_{96}=0, m^2_{610}=m^{2*}_{106}= h'_5 h^*_8 \nonumber\\
m^2_{79}&=m^{2*}_{97}=h_6 h''^*_3, m^2_{710}=m^{2*}_{107}=0, \nonumber\\
m^2_{89}&=m^{2*}_{98}=0, m^2_{810}=m^{2*}_{108}=h''_5 h^*_8, \nonumber\\
 m^2_{910}&=m^{2*}_{109}=\frac{y'^*_5}{\sqrt{2}}(v_2 A^*_{t_4}-\mu v_1)
\end{align}

We can diagonalize the scalar up mass$^2$ matrix  by the  unitary transformation
\begin{equation}
 \tilde D^{u\dagger} M^2_{\tilde u} \tilde D^{u} = \text{diag} (M^2_{\tilde u_1}, M^2_{\tilde u_2}, M^2_{\tilde u_3},  M^2_{\tilde u_4},M^2_{\tilde u_5},  M^2_{\tilde u_6}, M^2_{\tilde u_7}, M^2_{\tilde u_8}
 M^2_{\tilde u_9}, M^2_{\tilde u_{10}}
)\ .
\end{equation}

\end{document}